\documentclass[showpacs,amsmath,amssymb,prd]{revtex4}
\usepackage{ulem}
\usepackage{amssymb,amsmath}
\newcommand{\beqs}{\begin{equation*}}
\newcommand{\eeqs}{\end{equation*}}
\newcommand{\beq}{\begin{equation}}
\newcommand{\eeq}{\end{equation}}
\newcommand{\beqn}{\begin{eqnarray}}
\newcommand{\eeqn}{\end{eqnarray}}
\newcommand{\beqns}{\begin{eqnarray*}}
\newcommand{\eeqns}{\end{eqnarray*}}
\newcommand{\bspl}{\begin{split}}
\newcommand{\espl}{\end{split}}

\newcommand{\vv}[1]{\mathbf{#1}}
\newcommand{\bkt}[1]{\langle #1 \rangle}

\newcommand{\sg}{\sqrt{\gamma}\,}

\usepackage{graphicx}
\usepackage{graphicx}
\usepackage{dcolumn}
\usepackage{bm}
\usepackage{epsf}

\begin{document}

\title{General Relativistic Simulations of Slowly and Differentially
  Rotating Magnetized Neutron Stars}
Version: \today

\author{Zachariah B. Etienne}
\author{Yuk Tung Liu}
\author{Stuart L. Shapiro}
\altaffiliation{Also at the Department of Astronomy and NCSA,
  University of Illinois at Urbana-Champaign, Urbana, IL 61801}
\affiliation{Department of Physics, University of Illinois at 
Urbana-Champaign, Urbana, IL 61801, USA}

\begin{abstract}
We present long-term ($\sim 10^4M$) axisymmetric simulations of
differentially rotating, magnetized neutron stars in the
slow-rotation, weak magnetic field limit using a perturbative metric
evolution technique.  Although this approach yields results
comparable to those obtained via nonperturbative (BSSN) evolution
techniques, simulations performed with the perturbative metric solver
require about $1/4$ the computational resources at a given resolution.
This computational efficiency enables us to observe and analyze the
effects of magnetic braking and the magnetorotational instability
(MRI) at very high resolution.  Our simulations
demonstrate that
(1) MRI is not observed unless the fastest-growing mode wavelength is
resolved by $\gtrsim 10$ gridpoints; (2) as resolution is improved,
the MRI growth {\it rate} converges, but due to the small-scale
turbulent nature of MRI, the maximum growth {\it amplitude} increases,
but does not exhibit convergence, even at the highest resolution; and
(3) independent of resolution, magnetic braking drives the star toward
uniform rotation as energy is sapped from differential rotation by
winding magnetic fields.
\end{abstract}


\maketitle

\section{Introduction \& Motivation}

In differentially rotating neutron stars, an initially weak magnetic
field will be amplified by processes such as magnetic braking and the
magnetorotational instability (MRI)~\cite{MRI0,MRI}, causing a
redistribution of angular momentum.  Such differentially rotating
stars may arise from the merger of binary neutron
stars~\cite{BinMerg-Rasio,BinMerg-ShapShib,BinMerg-Shib}, or from
collapse of massive stellar cores, even if the cores spin uniformly at
the outset~\cite{Zwerg} (see also~\cite{liu01}).

To better understand how magnetic braking affects differentially
rotating configurations, Shapiro performed a purely Newtonian,
magnetohydrodynamic (MHD) calculation~\cite{StusFirst} in which the
star is idealized as a differentially rotating, infinite cylinder of
homogeneous, incompressible, perfectly conducting gas (see
also~\cite{mouschpaleo}).  The magnetic field is taken to be radial
initially and is allowed to evolve 
according to the ideal MHD (flux-freezing) equations. This calculation
demonstrates that differential rotation generates a toroidal magnetic
field, which reacts back on the fluid flow. Without viscous
dissipation, the toroidal field energy and rotational kinetic energy
in differential motion undergo periodic exchange and oscillations on
the Alfv\'en timescale. The magnitude of these oscillations, and the
maximum field strength, are independent of the initial magnetic field
strength; only the growth and oscillation timescale depend on the
magnitude of the seed field. If viscosity is present, or if some of
the Alfv\'en waves are allowed to propagate out of the star and into
an ambient plasma atmosphere, the oscillations are damped,
driving the star to uniform rotation. 

Cook, Shapiro, and Stephens~\cite{JamesCookPaper} later generalized
Shapiro's calculations for compressible stars. In their model, the
star is idealized as a differentially rotating, infinite cylinder
supported by a polytropic equation of state. They performed Newtonian
MHD simulations for differentially rotating stars with various
polytropic indices and different initial values of $T/|W|$, where $T$
is the rotational kinetic energy and $W$ is the gravitational
potential energy. They found that when $T/|W|$ is below the upper
(mass-shedding) limit for uniform rotation, $\beta_{\rm max}$,
magnetic braking results in oscillations of the induced toroidal
fields and angular velocities, and the star pulsates stably. However,
when $T/|W|$ exceeds $\beta_{\rm max}$, their calculations suggest
that the core contracts significantly and shock waves are generated, 
while the outer layers are ejected to large radii to form a wind or 
an ambient disk.

Liu and Shapiro~\cite{LiuShap} carried out both Newtonian and general
relativistic MHD (GRMHD) simulations on slowly and differentially
rotating incompressible stars. They considered the situation in
which $T \ll \mathcal{M} \ll |W|$, where $\mathcal{M}$ is the magnetic
energy. Due to the assumptions of slow rotation and weak magnetic
field, the star is well approximated as a sphere. They found that
toroidal fields are generated by magnetic braking, and the toroidal
fields and angular velocities oscillate independently along each
poloidal field line. The incoherent oscillations on different field
lines stir up turbulent-like motion on tens of Alfv\'en timescales, a
phenomenon called phase mixing (see~\cite{spruit99} and references
therein). In the presence of viscosity, the stars eventually are
driven to uniform rotation, with the energy contained in the initial
differential rotation going into heat. 

Most recently, Duez et al.~\cite{MHDLett,BigPaper} and Shibata et
al.~\cite{GRB2} performed GRMHD simulations on rapidly, differentially
rotating magnetized neutron stars in axisymmetry using two, newly
developed GRMHD codes~\cite{MHDIntro,SS}. They found that if a star is
hypermassive (i.e.\ the star's mass is larger than the maximum mass 
of a uniformly rotating star), magnetic braking and MRI will 
eventually induce collapse to a rotating black hole surrounded by a hot,
massive torus with a collimated magnetic field aligned along the spin
axis. This system provides a promising central engine for
short-duration gamma ray bursts~\cite{GRB2}.  They found the behavior
of nonhypermassive, differentially rotating neutron stars to be quite
different.  If a star initially spins at a rate exceeding the limit
for uniform rotation (``ultraspinning'' case), then instead of
collapsing, such a star settles to an equilibrium state consisting of
a nearly uniformly rotating core surrounded by a differentially
rotating torus.  Although this torus maintains differential rotation,
the angular velocity is constant along the magnetic field lines, so
further magnetic braking will not occur.  However, if a magnetized
star initially exhibits rapid, differential rotation at a spin
below the limit for uniform rotation (``normal'' case), the star
settles into a uniformly rotating configuration.

The purpose of this paper is to study the same magnetic effects
examined by Duez et al.~\cite{MHDLett,BigPaper}, but in the slow
rotation, weak magnetic field limit (i.e., $\mathcal{M} \ll T \ll
|W|$).  Analyzing the behavior of stars with such weak, but
astrophysically realistic, magnetic fields on the magnetic braking
(Alfv\'en) timescale requires simulations spanning $\sim 10^4M$ for
the models we consider.  Thus the primary challenge of this 
work is its exorbitant computational expense.  To overcome this
difficulty, we adopt a {\it perturbative} metric approach similar to
the one developed by Hartle~\cite{Hartle,Hartle2}, valid to first
order in the angular velocity $\Omega$.  The second computational
challenge is to resolve the wavelength of the dominant MRI mode, which
is small for the weak initial fields we wish to treat
($\lambda_{\text{MRI}} \propto B$). Solving the metric equations via
perturbation theory allows us to adopt sufficiently high spatial
resolution to track MRI.

Two aspects of our perturbative approach make simulations
significantly less costly than a nonperturbative, metric evolution at
a given resolution and ultimately allow us to perform simulations at
roughly $1/4$ the total computational cost.  First, the perturbed
metric is time independent except for the $\phi$-component of the
shift, $\beta^{\phi}$, which gives rise to frame-dragging.  The shift
$\beta^{\phi}$ varies with $\Omega$ on the Alfv\'en 
timescale, equivalent to many {\it thousands} of (Courant) timesteps
in a typical simulation.  Our perturbative metric solver uses a
simple, ordinary differential equation (ODE) solver to compute the
shift and allows us to skip many timesteps between matter evolution
updates.  Second, nonperturbative metric schemes incorporate
approximate asymptotic outer boundary conditions, which cause problems
if the outer boundary is moved too close to the star.  The perturbed 
metric on the other hand depends only on quantities defined within the
(spherical) star, so the outer boundary of the MHD evolution grid may
be moved significantly closer to the stellar radius with significant
reduction in computational expense.

The validity of our slow-rotation perturbative code is tested to
$\sim 2$ Alfv\'en timescales ($\sim 10^4M$).  We evolve a star with
both the perturbative and the nonperturbative
Baumgarte-Shapiro-Shibata-Nakamura (BSSN)~\cite{BSSN} gravitational
field scheme, and compare the results. We validate our simulations
self-consistently by checking that the evolution data satisfy the slow
rotation, weak magnetic field assumptions.

We study resolution-dependent MRI effects using two techniques.
First, we vary the grid spacing $\Delta$ at fixed initial magnetic
field strength, and second, we vary the initial magnetic field
strength ($\propto \lambda_{\text{MRI}}$) at fixed $\Delta$.  We
observe MRI-driven rapid growth of the poloidal fields when
$\lambda_{\text{MRI}}/\Delta \gtrsim 10$ (consistent
with~\cite{MHDLett,BigPaper}) and convergence in the growth rate if
$\lambda_{\text{MRI}}/\Delta \gtrsim 25$. However, convergence in the
maximum amplitude of these fields is not achieved even when
$\lambda_{\text{MRI}}$ is resolved to $\approx 41$ points. This is due
to small-scale turbulence intrinsic to MRI and to axisymmetry.

In addition to an analysis of MRI, we also study magnetic braking.  
Our simulations indicate that the winding of magnetic fields due to
magnetic braking saps a considerable fraction of the energy
associated with differential rotation in roughly one Alfv\'en 
timscale $t_A$, regardless
of resolution or metric evolution technique. Once the rotation profile
becomes more uniform, the magnetic fields begin to {\it unwind},
pumping differential rotation energy back into the star.

In the above simulations, we choose a differentially rotating star
with an angular velocity profile that initially decreases with
increasing distance from the rotation axis.  In our final simulation,
we evolve a differentially rotating star possessing an angular
velocity profile that initially {\it increases} with distance from the
rotation axis.  Magnetic braking should occur both models, but MRI
should not occur in the later case~\cite{MRI,MRIrev}. Our code yields
the expected result. 

The remainder of this paper is structured as follows. 
In Section~II, magnetic winding and MRI are explained qualitatively.
Section~III presents the mathematical and numerical framework for our
simulations. Section~IV outlines our initial data and numerical input
parameters, Section~V analyzes the validity of our perturbative metric
solver, Section~VI discusses our simulation results, and Section~VII
summarizes our conclusions. We adopt geometrized units in which
$G=c=1$.

\section{Qualitative Overview}
\subsection{Magnetic braking}
In an infinitely conducting plasma (MHD limit), magnetic field lines
are ``frozen-in'' to the fluid elements they connect.  We evolve
differentially rotating stars in this limit with initially purely
poloidal magnetic fields.  Differential rotation causes the magnetic
fields to wind toroidally on the Alfv\'en
timescale~\cite{BSS,StusFirst}
\begin{equation}
  t_A \sim \frac{R}{v_A} \sim 80~{\rm s} \left( \frac{B}{10^{12}~{\rm G}}
\right)^{-1}
\left( \frac{R}{15~{\rm km}}\right)^{-1/2}\left( \frac{M}{1.4 M_{\odot}}
\right)^{1/2} \ ,
\end{equation} 
where $B$ is the magnetic field strength, $v_A \sim B/\sqrt{4\pi
  \rho}\sim B\sqrt{R^3/3M}$ is the Alfv\'en speed, and $R$ and $M$ are
the radius and mass of the star, respectively. As shown
  in~\cite{BigPaper}, in the early stage of the magnetic braking, the
toroidal magnetic field $B^{\phi}$ increases linearly with time
according to
\begin{equation}
  B^T(t;\varpi,z) \equiv \varpi B^{\phi} \approx 
t\varpi B^i(0;\varpi,z)\partial_i\Omega(0;\varpi,z)
\ \ \ \ \ (i=\varpi,z) \ ,
\label{lineargrowth}
\end{equation}
where $\Omega=v^{\phi}$ is the angular velocity, $v^i=u^i/u^t$ is the 
matter three-velocity, and $\varpi=\sqrt{x^2+y^2}$ is the cylindrical
radius.  The coordinates are set up so that the rotation axis is along
the $z$-direction. 

The increase of $B^T$ with time adds energy to the
magnetic fields and saps the energy available in differential rotation
$T_{\text{DR}}$, conserving total angular momentum.  After roughly one
Alfv\'en time $t_A$, $T_{\text{DR}}$ is exhausted
\cite{StusFirst,JamesCookPaper,LiuShap} and $B^T$ reaches a maximum,
so eventually the growth of $B^T$ must deviate from the linear
relation~(\ref{lineargrowth}).  Although $t_A$ depends on the initial
magnetic field amplitude, the amplitude at which the field saturates
{\it does not}~\cite{StusFirst}.

\subsection{MRI}
MRI-induced turbulence occurs in a weakly-magnetized, gravitating body
if the angular velocity decreases with increasing distance from the
rotation axis \cite{MRI,MRIrev}.  According to a local, linearized
perturbation analysis in the Newtonian limit
\cite{MRI,MRIrev,BigPaper}, this instability causes the poloidal
field magnitude to increase exponentially with an $e$-folding time
$\tau_{\text{MRI}}$ independent of the seed field strength before
saturating:
\beq
\tau_{\text{MRI}} = -2 \left(\frac{\partial \Omega}{\partial \ln
  \varpi} \right)^{-1} \ .
\label{tauMRI}
\eeq
The wavelength of the fastest-growing mode,
$\lambda_{\text{MRI}}$, is given by
\beq
\lambda_{\text{MRI}} \approx \frac{8\pi\Omega}{\sqrt{16
    \Omega^4-\kappa^4}} v_A \ ,
\label{lamMRI}
\eeq
where 
\beq 
\kappa=\frac{1}{\varpi^3}\frac{d}{d\varpi}(\varpi^4 \Omega^2)
\eeq
is the epicyclic frequency of Newtonian theory. For a star
with $\partial \ln \Omega/\partial \ln \varpi \sim -1$, we have
$\tau_{\text{MRI}} \sim 2/\Omega \sim P_c$, where $P_c$ is the
central rotation timescale, and $\lambda_{\text{MRI}} \sim 2\pi
v_A/\Omega$. 

For configurations we consider, $\lambda_{\text{MRI}}/R \approx 1/6$,
where $R$ is the stellar radius. Therefore resolution on the order of
$R/\Delta \sim 20$ (where $\Delta$ is the grid spacing) is necessary
to resolve the fastest growing MRI wave mode $\lambda_{\text{MRI}}$.
However, the onset of MRI results in a buildup of small-scale MHD
turbulence, so the actual resolution requirements for fine-scale MRI
modelling are much higher.

\subsection{Upper Bound on Magnetic Energy}
The magnetic energy $\mathcal{M}$ in a magnetized star undergoing
magnetic braking increases as differential rotation is destroyed.
In the slow rotation limit, the gravitational potential energy $W$ and
internal energy $U$ of the star do not change significantly, so the
maximum possible value of $\mathcal{M}$ for such a star,
$\mathcal{M}_{\text{max}}$, is given by
\beq
\mathcal{M}_{\text{max}} \sim T_{\text{DR}},
\label{Bmaxest}
\eeq
where $T_{\text{DR}}$ is the kinetic energy associated with
differential rotation.
Note that $T_{\text{DR}}$ may be estimated at $t=0$ by constructing a
rigidly rotating star with the same total angular momentum $J$ as the
differentially rotating star and computing the difference in kinetic
energy between the two configurations.  Eq.~(\ref{Bmaxest}) therefore
provides us a way to estimate the maximum allowed magnetic energy {\it
  a priori}.

\section{Basic Equations and Numerical Technique}

In this section, we describe two methods to evolve the metric: the
perturbative approach and the nonperturbative BSSN 
scheme. The perturbative approach (Sec.~\ref{sec:perturb}) takes
advantage of the fact that the system is nearly spherically
symmetric. With this scheme, the evolution of the metric can be
simplified considerably. The nonperturbative metric evolution approach
(Sec.~\ref{sec:nonperturb}) is the same as that used
in~\cite{MHDIntro}.  The Maxwell and GRMHD equations are discussed in
Sec.~\ref{sec:MHD-Maxwell}. They are evolved with the same
high-resolution shock-capturing technique as in~\cite{MHDIntro}.

\subsection{Perturbative metric evolution scheme}
\label{sec:perturb}
For a slowly rotating, quasi-stationary axisymmetric star, the
rest-mass density $\rho_0$ and pressure $P$ differ from those of a
spherical star to second order in rotation frequency
$\Omega$. Further, if the stress-energy  tensor satisfies the
circularity conditions (see Eq.~(\ref{cond:circular})
and~\cite{circular}), we can choose a coordinate system so that the
only off-diagonal component of the metric is the frame-dragging term
$g_{t\phi}$.  In this approximation, the line element may be written
to first order in rotation frequency $\Omega$ and magnetic field
strength $|B|$ as
\beq
ds^2 = -e^{\nu(r)} dt^2 + e^{\lambda(r)} dr^2 - 2 \omega(t,r,\theta)
r^2 \sin^2{\theta} dt d\phi + r^2 (\sin^2\theta d\phi^2 + d\theta^2),
\label{metric}
\eeq
where $r$ is the areal radius.  Thus full determination of the metric
requires expressions for $e^{\nu(r)}$, $e^{\lambda(r)}$, and
$\omega(t,r,\theta)=-\beta^{\phi}(t,r,\theta)$.  The first two
quantities comprise the time-independent components of metric,
computed once for all time using the initial spherical $P$ and
$\rho_0$.  However, $\omega$ is a dynamical quantity that depends on
the rotation profile of the star. It must therefore be recomputed as
the star evolves in a quasi-stationary fashion.

As stated before, the metric~(\ref{metric}) is valid only if the 
stress-energy tensor satisfies the circularity conditions. To 
simplify our calculation, we also require that the azimuthal 
momentum energy density associated with the electromagnetic 
field be small compared with those associated with the fluid. 
These conditions are satisfied if (1) the meridional components 
of the fluid's velocity are much smaller than the rotational 
velocity, and (2) the energy density of the poloidal magnetic 
fields is much smaller than the energy density of the fluid.
Table~\ref{tab:slow_rot_assump} summarizes these conditions, which are
derived in Appendix~\ref{app:slow-rot-der}.

\subsubsection{Computing the time-independent metric components}
For small $\Omega$, the equilibrium star is spherical (the deviation
from sphericity is of order $\Omega^2$),
and the metric components $e^{\nu(r)}$ and $e^{\lambda(r)}$ are
independent of time (to order $\Omega$). They can be computed by 
solving the Oppenheimer-Volkoff (OV) equations~\cite{TOV}:
\beqn
\frac{dm(r)}{dr} &=& 4\pi r^2 \rho(r) \label{eq:ov1} \\
\frac{dP(r)}{dr} &=& -\frac{[\rho(r) + P(r)] [m(r) + 4 \pi r^3
  P(r)]}{r [r-2 m(r)]},  \label{eq:ov2} \\ 
  e^{\lambda(r)} &=& [1-2 m(r)/r]^{-1}, \label{eq:ov3} \\  
  \frac{d\nu(r)}{dr} &=& \frac{2[m(r) + 4\pi r^3 P(r)]} {r[r-2 m(r)]},
\label{eq:ov4}
\eeqn
with boundary conditions
\beqn
\rho_0(0) &=& \rho_c = \text{constant} \ , \\
m(0) &=& 0 \ , \\ 
 \lim_{r\to \infty} \nu(r) &=& 0 \ .
\eeqn
We close the above set of equations via a polytropic equation of
state:
\beq
P = K \rho_0^\Gamma,\ \ \ \Gamma = 1+1/n,
\eeq
where $K$ is the polytropic constant, $\Gamma$ is the adiabatic index,
and $n$ is the polytropic index. Note that $\rho$ is related to
$\rho_0$ by $\rho = \rho_0 (1+\epsilon)$ where $\epsilon=P/[(\Gamma-1)
  \rho_0]$ is the specific internal energy.  In our perturbative
scheme, $\nu(r)$ and $m(r)$ are frozen to their initial values and are
not evolved with time.  Note that outside the star, the diagonal
metric components describe the Schwarzschild spacetime, with mass
$m(r>R) = M$.

\subsubsection{The time-dependent shift term $\omega(t,r,\theta)$}
What remains is to compute time-dependent quantity
$\omega(t,r,\theta)=-\beta^{\phi}(t,r,\theta)$.
Given the slow rotation assumptions
summarized in Table~\ref{tab:slow_rot_assump}, the momentum 
constraint equation in the Arnowitt-Deser-Misner (ADM)
formalism~(Eq.~24 in \cite{ADM}) yields the following partial
differential equation for $\omega(t,r,\theta)$ (see
Appendix~\ref{app:shiftEq} for further details): 
\beqn
\frac{1}{r^4} \frac{\partial}{\partial r} \left[ r^4 j(r) 
\frac{\partial \omega}{\partial r} \right] +
e^{\frac{\lambda-\nu}{2}} \frac{1}{r^2 \sin^3{\theta}} 
\frac{\partial}{\partial \theta} \left[
  \sin^3{\theta} \frac{\partial \omega}{\partial \theta} \right] +
\frac{4}{r} j'(r) \omega &=& \frac{4}{r} j'(r) \Omega, 
\label{finalpde}
\eeqn
where $j(r) = \exp \{-[\lambda(r)+\nu(r)]/2\}$ and $j'(r)=d j(r)/dr$.

Following~\cite{Hartle2}, we solve Eq.~(\ref{finalpde}) by expanding
$\omega$ and $\Omega$ in terms of associated Legendre polynomials:
\beqn
\omega(t,r,\theta) &=& \sum_{l=1}^{\infty}{P_l'(\cos\theta)
  \omega_l(t,r)},
\label{littomeg_exp} \\ 
\Omega(t,r,\theta) &=& \sum_{l=1}^{\infty}{P_l'(\cos\theta)
  \Omega_l(t,r)}.
\label{bigomeg_exp}
\eeqn
Due to the assumption of equatorial symmetry (in addition to
axisymmetry), all even terms in the above expansions vanish.
Substituting Eqs.~(\ref{littomeg_exp}) and~(\ref{bigomeg_exp}) into
Eq.~(\ref{finalpde}), we obtain the same radial equation for each $l$
as Hartle (Eq.~(30) of \cite{Hartle2}):
\beq
\frac{1}{r^4} \frac{d}{dr} \left[r^4 j(r) \frac{d\omega_l}{dr}\right] +
\left[\frac{4}{r} \frac{dj}{dr} - e^{(\lambda-\nu)/2} \frac{l(l+1) -
    2}{r^2}\right] \omega_l = \frac{4}{r} \frac{dj}{dr} \Omega_l,
\label{finalode}
\eeq
where $\Omega_l(r>R)=0$. Our analysis of this equation in the limits
$r \to 0$ and $r\to \infty$  (see Appendix~\ref{app:shiftEq}) yields
the following boundary conditions 
\beqn
\omega_l(t,r)|_{r\to0} &=&
\begin{cases}
\Omega_1(t,0) + A_1(t), & \text{if } l=1\cr \cr
A_3(t) r^2 - \frac{16\pi}{21}[4 \rho(0) + 3 P(0)] \Omega_3(t,0) r^2
\ln{r}, & \text{if } l=3\cr  \cr
A_l(t) r^{l-1} +\displaystyle\frac{16\pi[4\rho(0) + 3P(0)]
  \Omega_l(t,0) r^2}{3[l(l+1) - 12]}, & 
\text{otherwise.}
\end{cases}
\label{rtozero_soln}
\\
\omega_l(t,r)|_{r\to \infty} &=& C_l(t) r^{-l-2},
\label{rtoinf_soln}
\eeqn
where $C_l(t)$ and $A_l(t)$ are determined (using the shooting method) 
at a given time $t$ by matching the interior ($r<R$) and exterior
($r>R$) solutions at the stellar surface $r=R_{\pm}$:
\beqn
\omega_l(t,R_+) &=& \omega_l(t,R_-) 
\label{match1} \\
\frac{d}{dr} \omega_l(t,r)|_{R_+} &=& \frac{d}{dr}\omega_l(t,r)|_{R_-}
\label{match2}
\eeqn

For the models we consider in this paper, we find that contributions
from modes above $l=5$ are negligible, so we only calculate modes up
to and including $l=5$.

\begin{table}
\caption{Assumptions made in the slow-rotation approximation (see
Appendix~\ref{app:slow-rot-der} for a derivation).} 
\label{tab:slow_rot_assump}
\begin{tabular}{|c||cc|c|c|c|}
  \hline
  Orthonormal Component & Velocity\ \ \ \ \ \ \ [max. average]$^{\star}$ &&
  Magnetic Field\ \ \ \ [max. average]$^{\star}$ \\
  \hline
  $\hat{\phi}$ & 
  $v^{\hat{\phi}} = \Omega r \sin \theta $\ \ \ \ \ \ \ \ \ \ \ \ \ \
  \ \ \ \ \ \ &  &
  $\displaystyle\frac{ (B^{\hat{\phi}})^2}{4\pi \rho_0 h}
  \lesssim 1$\ \ \ \ \ \ \ \  $[6\times 10^{-5}]$ \\ 
  \hline
  $\hat{\theta}$ & $\displaystyle\left|\frac{v^{\hat{\theta}}}{\Omega
    r}\right| \ll 1$\ \ \ \ \ \ \   $[0.03]$ &  &
  $\displaystyle\frac{(B^{\hat{\theta}})^2}{4\pi \rho_0h}
   \ll  1$\ \ \ \ \ \ \  $[4\times 10^{-7}]$ \\
  \hline
  $\hat{r}$ & $\displaystyle\left|\frac{v^{\hat{r}}}{\Omega
    r}\right|  \ll 1$\ \ \ \ \ \ \   $[0.04]$ &
  & $\displaystyle\frac{(B^{\hat{r}})^2}{4\pi \rho_0 h} 
  \ll 1$\ \ \ \ \ \ \ $[3\times 10^{-7}]$ \\
  \hline
\end{tabular}
\begin{minipage}[b]{5in}
\raggedright
${}^{\star}$ The above nondimensional ratios are local in space and in
time, so we compute a mass-density weighted average of these
quantities at various times.  We denote the maximum value (in time)
observed in our simulations ``max. average''.
\end{minipage}
\end{table}

\subsection{BSSN metric evolution scheme}
\label{sec:nonperturb}
The line element for a generic spacetime is written in the standard
3+1 form as follows:
\beq
ds^2 = -\alpha^2 dt^2 + \gamma_{ij} (dx^i + \beta^i dt)(dx^j + \beta^j
dt), 
\label{metric2}
\eeq
where $\alpha$ is the lapse, $\beta^i$ is the shift and 
$\gamma_{ij}$ is the three-dimensional spatial metric.
We evolve the metric $\gamma_{ij}$ and the extrinsic curvature 
$K_{ij}$ using the BSSN formalism~\cite{BSSN}. The BSSN evolution variables are:
\beqn
\phi &=& \frac{1}{12}\ln[\det(\gamma_{ij})]\ , \\
  \tilde\gamma_{ij} &=& e^{-4\phi}\gamma_{ij}\ , \\
  K &=& \gamma^{ij}K_{ij}\ , \\
  \tilde A_{ij} &=& e^{-4\phi}(K_{ij} - \frac{1}{3}\gamma_{ij}K)\ , \\
  \tilde\Gamma^i &=& -\tilde\gamma^{ij}{}_{,j} \ .
\eeqn 
The equations for evolving these variables are given in~\cite{BSSN}. 
For the gauges, we use the hyperbolic driver 
conditions~(\cite{abpst01,HJYo}) to evolve
the lapse and shift.

We adopt the Cartoon method~\cite{cartoon} to impose axisymmetry and
use a Cartesian grid. In this scheme, the coordinate $x$ is identified
with the cylindrical radius $\varpi$, the $y$-direction
corresponds to the azimuthal direction, and $z$ lies along the
rotation axis. For example, for any 3-vector $V^i$, $V^x \equiv
V^{\varpi}$, and $V^y \equiv \varpi V^{\varphi}$. 

\subsection{Maxwell and MHD Equations}
\label{sec:MHD-Maxwell}
In terms of the Faraday tensor $F^{\mu\nu}$, the MHD condition is
given by
\beq
F^{\mu\nu} u_{\nu} = E^{\mu}_{(u)} = 0,
\eeq
where $E^{\mu}_{(u)}$ is the electric field measured by an observer
comoving with the fluid.  As in \cite{BigPaper}, we evolve the
following set of variables:
\beqn
\rho_{\star} &=& \alpha \sqrt{\gamma}\, \rho_0 u^0, \\
\tilde{\tau} &=& \alpha^2 \sqrt{\gamma}\, T^{00} - \rho_{\star}\ , \\
\tilde{S}_i &=& \alpha\sqrt{\gamma}\, T^0{}_i \ , \\
\tilde{B}^i &=& \sqrt{\gamma}\, B^i,
\eeqn
where $\gamma = \det(\gamma_{ij})$, $B^{\mu}=\frac{1}{2}
\epsilon^{\mu\nu\alpha\beta} F_{\alpha\beta} n_{\mu}$ denotes the
magnetic field measured by a normal observer, 
and $n^{\mu}$ is the unit normal vector orthogonal to the time 
slice. These variables satisfy the following evolution equations: 
\beqn
 \partial_t \vv{U} + \nabla \cdot \vv{F} &=& \vv{S}, \text{ where}
\label{eq:MHD-Induction}
\eeqn
\beqn
\partial_t \vv{U} &=& \partial_t \left[
  \begin{array}{ c }
    \rho_{\star} \\
    \tilde{\tau} \\
    \tilde{S_i} \\
    \tilde{B^i} 
  \end{array} \right], \\
\nabla \cdot \vv{F} &=& \partial_j \left[
  \begin{array}{ c }
    \rho_{\star} v^j \\
    \alpha^2 \sqrt{\gamma}\, T^{0j} - \rho_{\star} v^j \\
    \alpha \sqrt{\gamma}\, T^j{}_i \\
    v^j \tilde{B}^i - v^i \tilde{B}^j 
  \end{array} \right], \text{ and}\\
\vv{S} &=& \left[
  \begin{array}{ c }
    0 \\
    \alpha \sqrt{\gamma}\,[(T^{00} \beta^i \beta^j + 2 T^{0i}\beta^j +
      T^{ij})K_{ij} - (T^{00}\beta^i+T^{0i})\partial_i \alpha] \\
    \frac{1}{2}\alpha\sqrt{\gamma}\, T^{\alpha\beta} g_{\alpha\beta , i} \\
    0 
  \end{array} \right].
\eeqn
The stress-energy tensor $T^{\mu\nu}$ for a magnetized, infinitely
conducting, perfect fluid is given by 
\beqn
T^{\mu\nu} &=& (\rho_0 h + b^2) u^{\mu}u^{\nu} + (P+\frac{b^2}{2})
g^{\mu\nu} - b^{\mu}b^{\nu}.
\eeqn
Here, $h=1+\epsilon+P/\rho_0$ is the specific enthalpy, and $\sqrt{4
  \pi} b^{\mu}=B^{\mu}_{(u)}$ is the magnetic field measured by an 
observer comoving with the fluid, which is related to
$B^{\mu}$ by
\beq
\sqrt{4 \pi} b^{\mu} = - \frac{P^{\mu\nu}
  B_{\nu}}{n^{\alpha}u_{\alpha}} = B^{\mu}_{(u)}, 
\eeq
where $P^{\mu\nu}=g^{\mu\nu}+u^{\mu}u^{\nu}$. 

We evolve Eq.~(\ref{eq:MHD-Induction}) using a high-resolution 
shock-capturing scheme as in~\cite{MHDIntro}. Specifically, we use 
the piecewise parabolic method (PPM)~\cite{PPM} algorithm for data
reconstruction and the Harten-Lax-Van Leer (HLL) flux
formula~\cite{HLL} for the approximate Riemann solver.

\subsection{Diagnostics}
\label{sec:diagnostics}

During the simulations, we monitor the following conserved quantities: 
rest mass $M_0$, angular momentum $J$. We also monitor the ADM mass $M$, 
which is nearly conserved, as the energy emitted as gravitational
radiation is negligible. We also compute the rotational kinetic energy
$T$, magnetic energy $\mathcal{M}$, internal energy $U$, and
gravitational potential energy $W$. All of these global quantities are
calculated using the formulae given in~\cite{BigPaper}.

\section{Initial Data and Numerical Parameters}
\label{sec:ID}

To understand the behavior of slowly-rotating, weakly magnetized
neutron stars, we perform four studies.  First, in our ``MRI Resolution
Study,'' we start with a differentially rotating, poloidally
magnetized configuration in which the angular velocity decreases away
from the rotation axis.  We then evolve this star at various
resolutions, with the goal of uncovering the detailed,
resolution-dependent behavior of MRI.  In our second study, the
``$B$ Variation Study,'' we evolve the same star as in the first
study at lowest resolution, varying only the strength of the initial
poloidal fields.  This study also examines the resolution-dependent
nature of the observed MRI by varying $B$ and hence
$\lambda_{\text{MRI}}$ at fixed spatial resolution.  Finally, in the
``Rotation Profile Study,'' we evolve the same star as with
our ``MRI Resolution Study,'' changing the angular velocity
distribution so that it initially increases with distance from the
rotation axis.  In this study, we expect to observe magnetic  winding,
but not MRI (out to $\sim 1 t_A$).  As a code test, we also perform the
``Rigid Rotation Profile Study,'' where we explore the same
configuration as with the first study, only with solid body rotation
at the same total angular momentum $J$.  We expect that the magnetic
field will not change in time and have no effect on the
star. Tables~\ref{tab:Bstrengths} and~\ref{tab:parameters} present a
summary of initial parameters for the stars we consider in these
studies.

For simulations using the BSSN metric solver, we construct initial
data for a differentially rotating, relativistic star in equilibrium
using the code of Cook et al.~\cite{CookPaper} with the following
rotation law:
\beq
u^0 u_{\phi} = A R^2 (\Omega_c-\Omega) \ ,
\label{realrotlaw}
\eeq
where $R$ is the equatorial coordinate radius, $\Omega_c$ 
is the central angular velocity, and $A$ is a constant 
parameter which determines the degree of differential rotation. 
In the Newtonian limit, this rotation law reduces to the so-called 
``j-constant'' law: 
\beq
  \Omega=\frac{\Omega_c}{1+ \frac{\varpi^2}{A R^2}} \ .
\label{rotlaw}
\eeq
For a slowly rotating star, the spatial metric $\gamma_{ij}$ is nearly
conformally flat $\tilde{\gamma}_{ij}\approx
f^{\text{flat}}_{ij}$. The Cook et al.~code uses spherical isotropic
coordinates, so to obtain the desired $\tilde{\gamma}_{ij}$, we only
need to transform the Cook et al.~initial data to Cartesian
coordinates.

For simulations with the perturbative metric solver, we set up the 
initial data by first computing the diagonal components of the metric 
and the hydrodynamic quantities by solving the OV
equations~(\ref{eq:ov1})--(\ref{eq:ov4}). Then the shift
$\beta^{\phi}$ is computed via the perturbative technique described in
Section~\ref{sec:perturb}, with the angular velocity distribution
computed by either solving Eq.~(\ref{realrotlaw}) in the slow rotation
limit (the ``Rotation Profile Study''), or using the solution of
Eq.~(\ref{realrotlaw}) as computed by the Cook et al.~code (``MRI
Resolution Study'' and ``$B$ Variation Study'').  Note that since the
initial data computed by the perturbative technique is only accurate
to order $\Omega$, the resulting star will undergo small amplitude
oscillations [due to $\mathcal{O}(\Omega^2)$ effects]. Further, to
more easily compare perturbative simulation results with those using
the BSSN scheme, we perform the coordinate transformations necessary
to facilitate evolution of the Maxwell and MHD equations in the same
Cartesian coordinates as in the BSSN evolution scheme.

In our ``MRI Resolution'' and ``$B$ Variation'' studies, we consider
an $A=1$ differentially rotating star which satisfies the $n=1$
polytropic equation of state (EOS).  Other parameters are set so that
the equilibrium star possesses the following properties: the ratio of
equatorial to polar radii $R_p/R = 0.98$, central rotation period
$P_c=2\pi/\Omega_c = 264.7M$, compactness $M/R=0.182$, ratio of
angular velocity at the equator to that at the center
$\Omega_{\text{eq}}/\Omega_c \approx 0.3$, and $T/|W|=4.88\times
10^{-3}$. The mass of this star is determined by the polytropic
constant $K$, which we set to unity. However, our results can be
easily rescaled to any values of $K$ (see~\cite{CookPaper}), and hence
to any values of the mass. For example, the model we just described
has $R = 9.2~(M/1.4M_{\odot}){\rm km}$, $\rho_c=1.54\times
10^{15}~(1.4M_{\odot}/M)^2{\rm g/cm^3}$, and $P_c =
1.8~(M/1.4M_{\odot}){\rm ms}$. 

The ``Rotation Profile Study'' involves the same star as in the ``MRI
Resolution Study'', but with rotation profile parameters set so that
$A=-1$, which corresponds to $\Omega(R)/\Omega_c = 2.8$.

Next, we add a small seed magnetic field to the stellar models above by 
specifying the vector potential $A_i = A_{\phi} \delta_i{}^{\phi}$ as
\beq
  A_{\phi} = \varpi^2 \max[ A_b (P-P_{\rm cut}),0]\ ,
\eeq
where the pressure cutoff $P_{\rm cut}$ is set to 4\% of the maximum
pressure ($P_{\rm cut}=0.04P_{\rm max}$). The strength of the initial
seed field is determined by the constant $A_b$ and may be
characterized by the parameter $C$, the maximum value of $b^2/P$ at
$t=0$.

\begin{table}
\caption{Initial models: Magnetic field-related parameters}
\begin{tabular}{c |c|c|c|c|c|c|c}
  \hline \hline
  Study & $C^{\dagger}$ &  $\bkt{t_A}^{\ddagger}/M$ &
  $\bkt{t_A}/\left(\frac{M}{1.4 M_\odot}\right)$ & $\bkt{t_A}/P_c$ &
  $(|B|^{t=0}_{\text{max}})^*/\left(\frac{1.4 M_\odot}{M}\right)$ &
  $(\mathcal{M}/|W|)^{**}$ & $T/|W|^{\dagger\dagger}$ \\
  \hline
  Rigid Rotation Profile & $6.1 \times 10^{-5}$ & 4800 & 33ms & 10.2 &
  $4.9 \times 10^{14}$G & $4.4 \times 10^{-6}$ & $4.55 \times 10^{-3}$ \\ 
  \hline
  MRI Resolution & $6.1 \times 10^{-5}$ &  4800 & 33ms & 17.9 &
  $4.9 \times 10^{14}$G & $4.4 \times 10^{-6}$ & $4.85 \times 10^{-3}$ \\
  \hline
  & $4.97 \times 10^{-6}$ & 16200 & 112ms & 61.1 & $1.4 \times
  10^{14}$G &  $3.8 \times 10^{-7}$ & $4.85 \times 10^{-3}$ \\

  $B$ Variation  & $1.96 \times 10^{-5}$ & 8100 & 56ms & 30.7 & $2.8
  \times 10^{14}$G & $1.5 \times 10^{-6}$  & $4.85 \times 10^{-3}$ \\
   & $6.1 \times 10^{-5}$ &  4800 & 33ms & 17.9 & $4.9 \times
  10^{14}$G & $4.4 \times 10^{-6}$ & $4.85 \times 10^{-3}$ \\
  \hline
  Rotation Profile& $6.1 \times 10^{-5}$ & 4800  & 33ms &
  7.2 & $4.9 \times 10^{14}$G & $4.4 \times 10^{-6}$ & $4.90 \times 10^{-3}$ \\
  \hline \hline
\end{tabular}
\label{tab:Bstrengths}
\begin{minipage}[b]{5in}
\raggedright
  ${}^{\dagger}$ $C$ is the maximum value of $b^2/P$ at $t=0$. \\
  ${}^{\ddagger}$ $\bkt{t_A}$ is the mass density-weighted Alfv\'en
  time, given by Eq.~(\ref{tAdef}). \\
  ${}^{*}$ $|B|^{t=0}_{\text{max}}$ is the maximum magnitude of the
  magnetic field at $t=0$. \\
  ${}^{**}$ $\mathcal{M}/|W|$ is the initial ratio of magnetic energy
  to gravitational potential energy. \\
  ${}^{\dagger\dagger}$ $T/|W|$ is the initial ratio of kinetic energy
  to gravitational potential energy.
\end{minipage}
\end{table}

The strength of the magnetic field can also be measured by the mass 
density-averaged Alfv\'en time $\bkt{t_A}$ defined as
\beq
  \bkt{t_A} \equiv \left( \frac{\langle v_A \rangle}{R}\right)^{-1}
 = \left[ \frac{1}{RM_0} \int v_A \rho_* d^3x \right]^{-1} \ ,
\label{tAdef}
\eeq
where $v_A = \sqrt{b^2/(\rho_0 h + b^2)}$ is the Alfv\'en speed.
Since $\lambda_{\text{MRI}}$ is a local quantity, we define the
magnetic energy density-weighted average of $\lambda_{\text{MRI}}$ as
follows:
\beq
  \langle \lambda_{\rm MRI} \rangle \equiv 
\frac{\int \tilde{\lambda}_{\rm MRI} b^2 \sg d^3x}{\int b^2 \sg d^3x} \ ,
\eeq
where
\beq
  \tilde{\lambda}_{\rm MRI}(r) = \begin{cases}
  \lambda_{\text{MRI}}(r) & \text{ if
    $0< \tau_{\text{MRI}}$(r) $< t_A$} \\
  0 & \text{ otherwise} 
\end{cases} \ .
  \label{lambMRI2}
\eeq
Here $\lambda_{\rm MRI}$ and $\tau_{\rm MRI}$ are calculated by
Eqs.~(\ref{lamMRI}) and (\ref{tauMRI}), respectively. The cutoff in
Eq.~(\ref{lambMRI2}) is set so that we only consider the region where
the MRI is present ($\tau_{\text{MRI}}>0$) and where the MRI timescale
is less than the Alfv\'en time (where $\tau_{\text{MRI}}>t_A$, magnetic 
braking is expected to dominate).

All our nonperturbative simulations are performed on a square grid
with outer boundary at $2.0R$ $(= 11M)$.  Our perturbative simulations
on the other hand use an outer boundary of $1.2R$, with metric
updates every 8-10 timesteps.  We have verified that if the
outer boundary is set to $1.5R$ instead, all quantities we studied are
the same to within $\sim 1\%$ until the magnetic field hits the $1.2R$
outer boundary.  Due to this loss of accuracy, we stop our simulations
soon after this boundary crossing.  The time at which each
simulation was stopped, $t_{\text{stop}}$, is listed in
Table~\ref{tab:parameters}.  By $\sim 2t_A$, both magnetic winding and
MRI are fully developed.

\begin{table}
\begin{center}
\caption{Initial models: Parameters related to resolution and
  numerical evolution}

\begin{tabular}{c|c|c|c|c|c}
  \hline \hline
  Study & Method & $(R/\Delta){ }^*$ & $C$ & 
  $\bkt{\lambda_{\text{MRI}}}/\Delta$ & $(t_{\text{stop}}/P_c){
  }^\dagger$ \\ 
  \hline
  Rigid Rotation Profile & Perturbative & 100 & $6.1\times 10^{-5}$ &
  -- & 10.2 (ns) \\
  \hline
   &Perturbative& 75 & $6.1\times 10^{-5}$ &   12.2  & 35.8 (ns) \\
  &  & 100 &  &   16.3 & 15.5 \\
  MRI Resolution &  & 150 &  & 24.4 & 13.1 \\
  &  & 200 &  & 32.6 & 13.3 \\
  &  & 250 &  & 40.7 & 12.3 \\
  \cline{2-6}
  & Nonperturbative & 75 & $6.1\times 10^{-5}$ &  12.2 & 35.8 (ns)\\
  &  & 100 & &  16.3 & \\
  \hline
  & Perturbative & 75 & $4.97\times 10^{-6}$ & 3.6 & 61.1 (ns)\\
  &              &    & $1.96\times 10^{-5}$ & 7.2 & 30.7 (ns) \\
  $B$ Variation &  & & $6.1\times 10^{-5}$ &  12.2  & 17.9 (ns) \\
  \cline{2-6}
  & Nonperturbative & 75 & $4.97\times 10^{-6}$ & 3.6 & 61.1 (ns)\\
  &  &  &  $1.96\times 10^{-5}$ & 7.2 & 30.7 (ns)\\
  &  & & $6.1\times 10^{-5}$ & 12.2 & 17.9 (ns)\\
  \hline
  Rotation Profile & Perturbative & 100 & $6.1\times
  10^{-5}$ & -- &  7.2 (ns)\\
  \hline \hline
\end{tabular}
\label{tab:parameters}
\end{center}
\begin{minipage}[b]{4.5in}
\raggedright
  ${}^*$ $R$ is the equatorial coordinate radius of the star, and
  $\Delta$ is the grid spacing. \\
  ${}^{\dagger}$ $t_{\text{stop}}$ is the time at which the simulation
  was stopped due to loss of accuracy, which happens soon after the magnetic
  field hits the outer boundary. (ns) indicates that the magnetic fields 
  have not hit the outer boundary at $2 \langle t_A \rangle$,
  and the simulation was terminated at the indicated time.
\end{minipage}
\end{table}

We specify resolution by the quantity $R/\Delta$, where $R$ is the
stellar radius and $\Delta$ is the grid spacing.  Thus a simulation
with $150^2$ points and outer boundary at $2R$ has $R/\Delta=75$, and
a simulation with $90^2$ points and outer boundary at $1.2R$ has
$R/\Delta=75$ as well.  In these simulations, MRI does not become
evident until $t\approx 6 P_c$ (e.g., see Figure~\ref{MRIresstudy1}).
Thus for computational efficiency in our highest resolution run,
$R/\Delta=250$, we evolve the star at resolution $R/\Delta=100$ until
$t=5P_c$ and then regrid to $R/\Delta=250$.
Table~\ref{tab:parameters} summarizes the resolutions chosen in our
simulations.

\section{Code Tests}

\subsection{Test of the perturbative shift solver}
\begin{figure}
  \begin{center}
    \epsfxsize=2.7in
    \leavevmode
    \epsffile{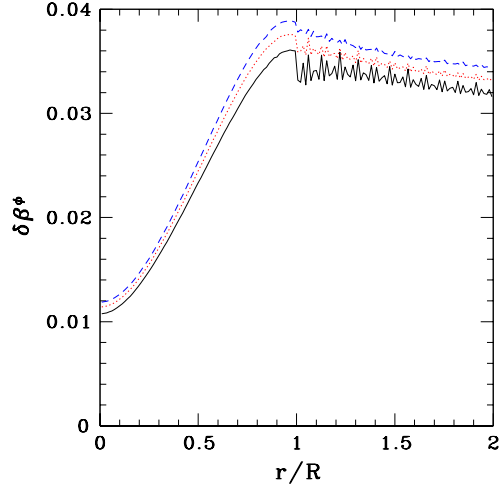}
    \caption{Relative error in $\beta^{\phi}$ along equatorial
      plane of star at $t=0$.  Exact solutions from Cook et
      al.~code~\cite{CookPaper} are compared to perturbative shift
      solver results at $T/|W|= 2.43 \times 10^{-3}$ (solid black
      lines), $4.9 \times 10^{-3}$ (dashed blue lines), and $9.9
      \times 10^{-3}$ (dotted red lines). Here $\delta \beta^{\phi}
      \equiv (\beta^{\phi}_{\text{Cook}} - \beta^{\phi}_{\text{Pert}})
      / [(\beta^{\phi}_{\text{Cook}} +
      \beta^{\phi}_{\text{Pert}})/2]$. To demonstrate the approximate
      scaling $\delta \beta^{\phi} \propto T/|W|$, we have multiplied
      the value of $\delta \beta^{\phi}$ a factor of 4 for the case
      $T/|W|= 2.43 \times 10^{-3}$, and by a factor of 2 for the case
      $T/|W|= 4.9\times 10^{-3}$.}
    \label{err_shift}
  \end{center}
\end{figure}

To verify that our perturbative shift solver produces the shift
$\beta^{\phi}$ accurately to order $\Omega$, we compute $\beta^{\phi}$
for differentially rotating, equilibrium star models and compare the
results with those computed without approximation by the Cook et
al.~code~\cite{CookPaper}.  Figure~\ref{err_shift} shows the error,
$\delta\beta^{\phi}$, for three models with the same central density
($\rho_c=1.54\times 10^{15}~(1.4M_{\odot}/M)^2~{\rm g}/{\rm cm}^3$)
but with various $T/|W|$. We see that $\delta \beta^{\phi}$ decreases
as $\Omega^2 \propto T/|W|$, as expected. This shows that our
perturbative shift solver accurately calculates $\beta^{\phi}$ to
first order in $\Omega$.

\subsection{Rigid Rotation Profile Study}

\begin{figure}
  \begin{center}
    \epsfxsize=1.4in
    \leavevmode
    \hspace{-0.3cm}\epsffile{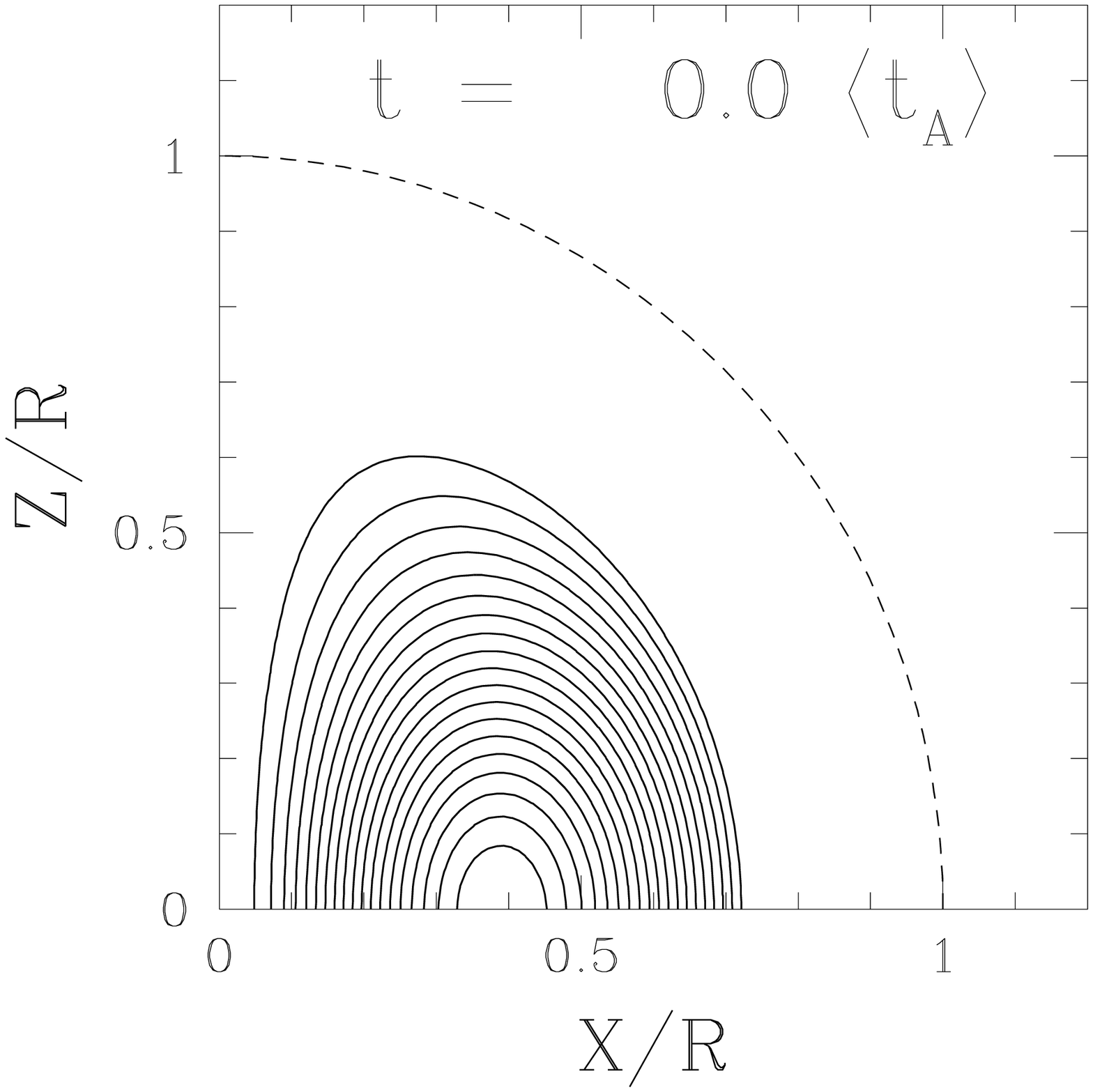}
    \epsfxsize=1.4in
    \leavevmode
    \hspace{-0.3cm}\epsffile{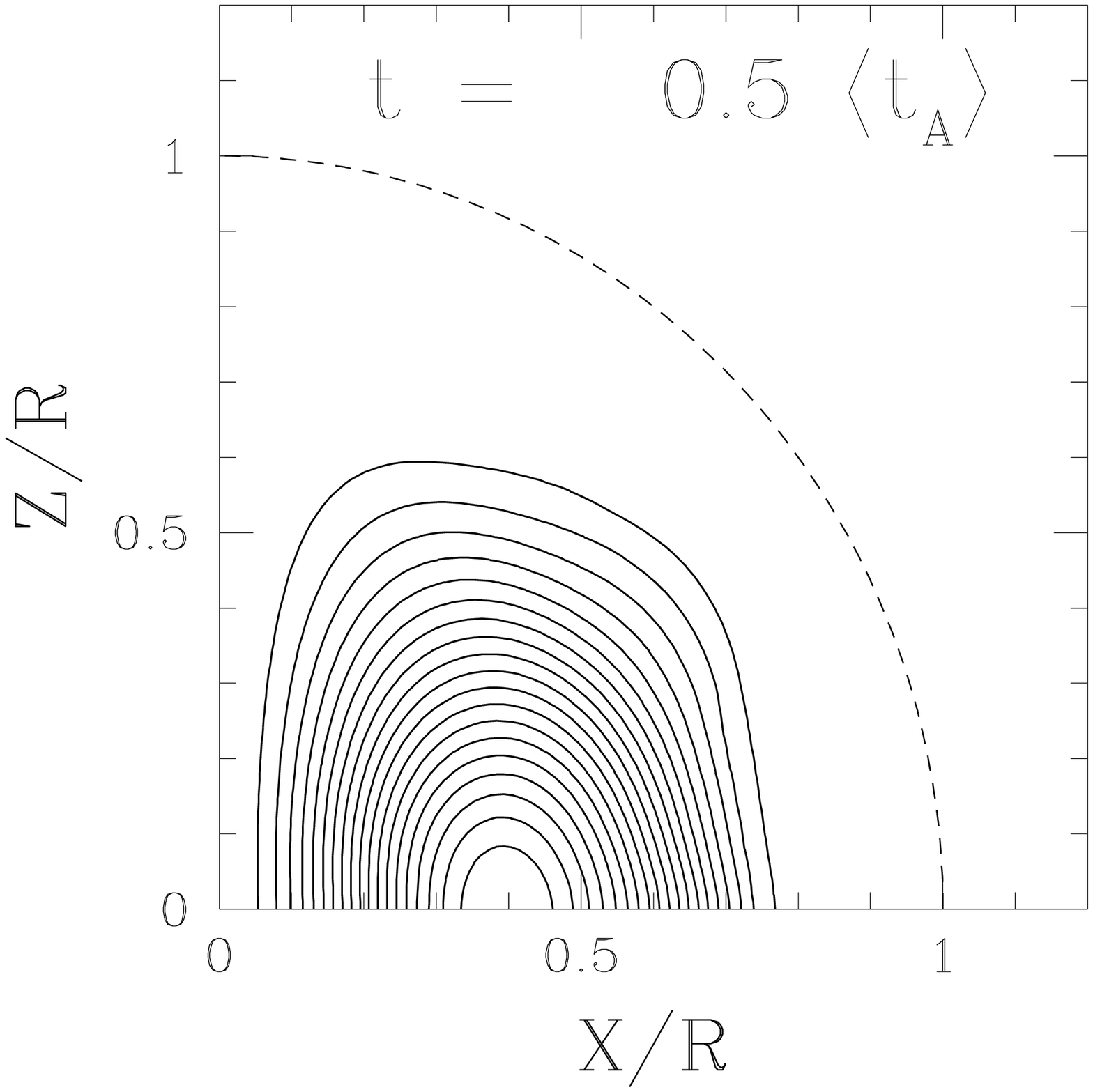}
    \epsfxsize=1.4in
    \leavevmode
    \hspace{-0.3cm}\epsffile{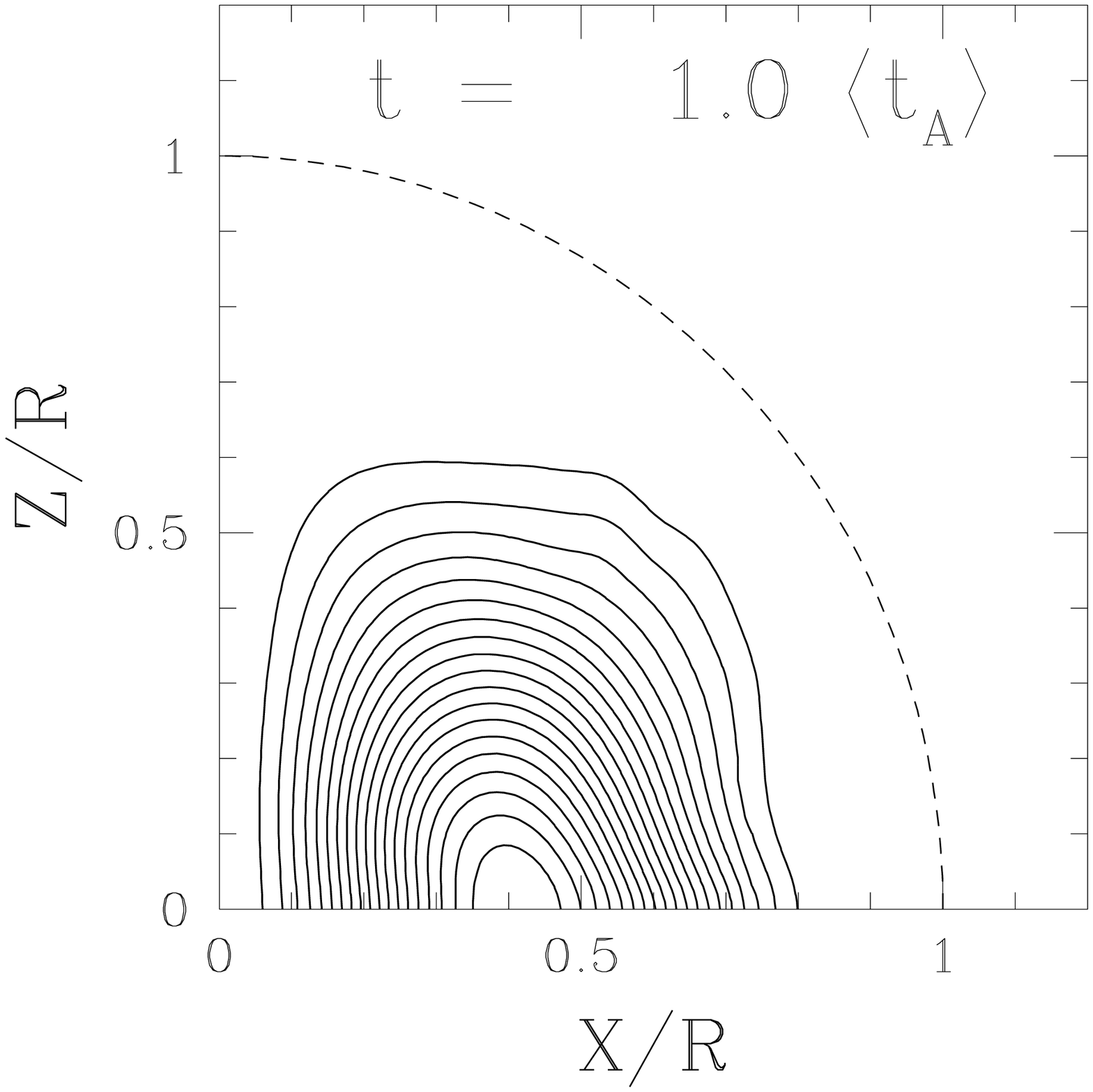} 
    \caption{Snapshots of the poloidal magnetic field lines
      (contours of $A_{\phi}$) at $t/\langle t_A \rangle = 0$,
      $0.5$, and $1.0$ for the uniformly rotating star. The field 
      lines are drawn for $A_{\phi}=A_{\phi,\rm min} + (A_{\phi,\rm
      max} - A_{\phi,\rm min}) i/20,~(i=1$--19), where $A_{\phi,\rm
      max}$ and $A_{\phi,\rm min}$ are the maximum and minimum values
      of $A_{\phi}$, respectively, at the given time.  The dashed line
      in each plot indicates the initial stellar surface.}
  \label{flatrot1}
  \end{center}
\end{figure}

\begin{figure}
  \begin{center}
    \epsfxsize=2.8in
    \leavevmode
    \hspace{-0.7cm}\epsffile{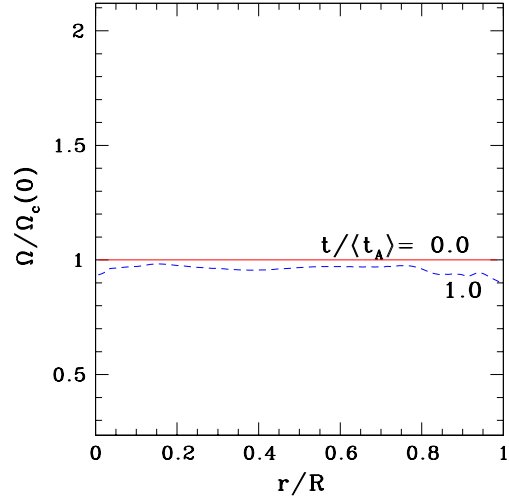}
    \caption{Snapshots of the rotation profile at the equatorial plane at
    $t/\langle t_A \rangle = 0$ and $1.0$ for the uniformly rotating
    star.}
  \label{flatrot2}
  \end{center}
\end{figure}

When the star is uniformly rotating, a (weak) poloidal magnetic field
should not change with time and it should have no effect on the
star. To test our code, we evolve a uniformly rotating star with the
physical parameters specified in Tables~\ref{tab:Bstrengths}
and~\ref{tab:parameters} (Rigid Rotation Profile Study). We follow the
star for one Alfv\'en time ($\langle t_A\rangle =10.2P_c=4800M$).
Figure~\ref{flatrot1} displays snapshots of the poloidal magnetic
field in time, and Figure~\ref{flatrot2} shows the evolution of the
rotational profile on the equatorial plane.  As expected, neither the
star's rotation profile nor its magnetic field change significantly
over the Alfv\'en timescale. 


\section{Numerical Results}

\subsection{MRI Resolution Study}

\begin{figure} 
  \begin{center}
    \epsfxsize=2.7in
    \leavevmode
    \hspace{-0.7cm}\epsffile{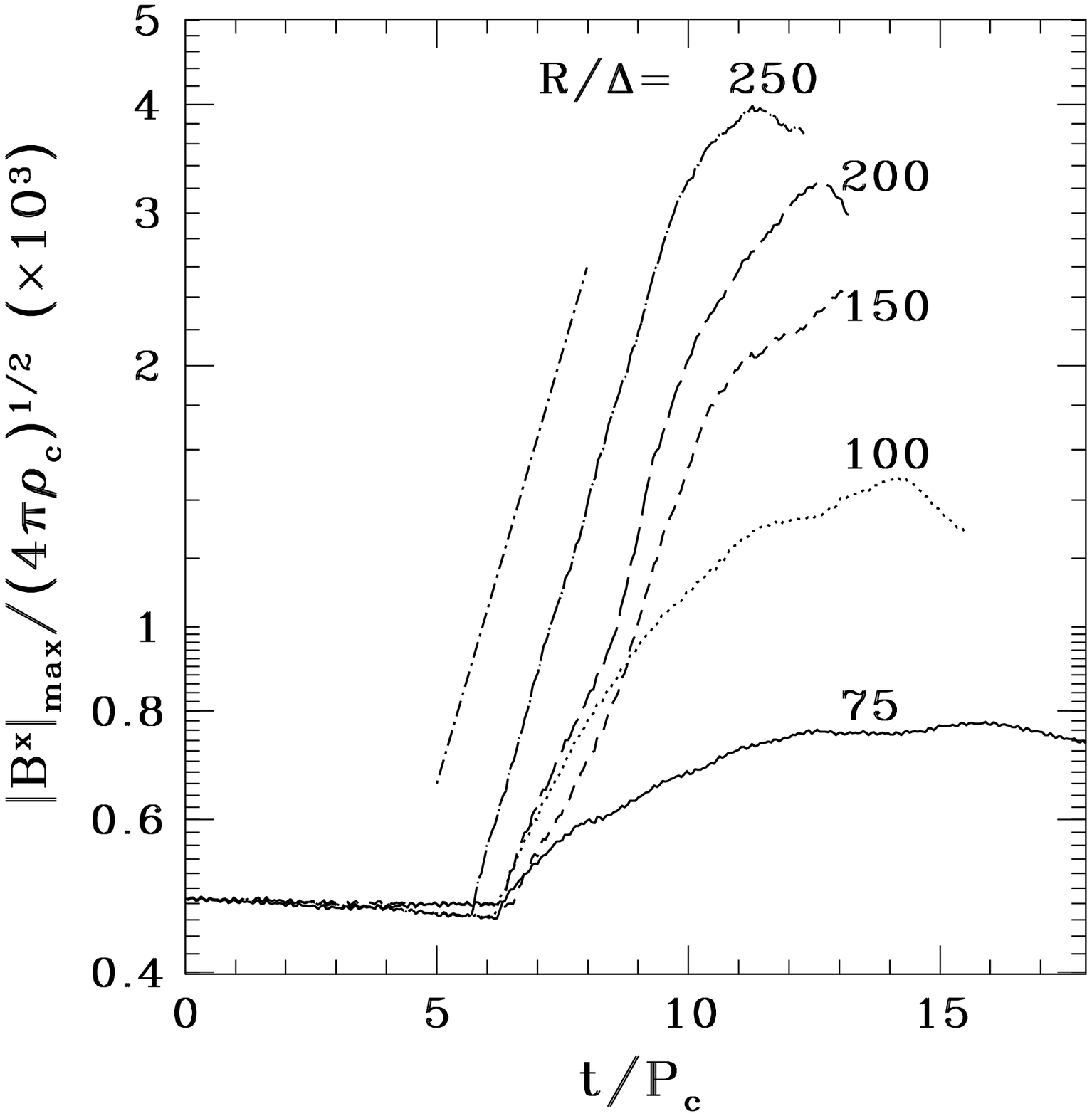}
    \epsfxsize=2.7in
    \leavevmode
    \hspace{-0.2cm}\epsffile{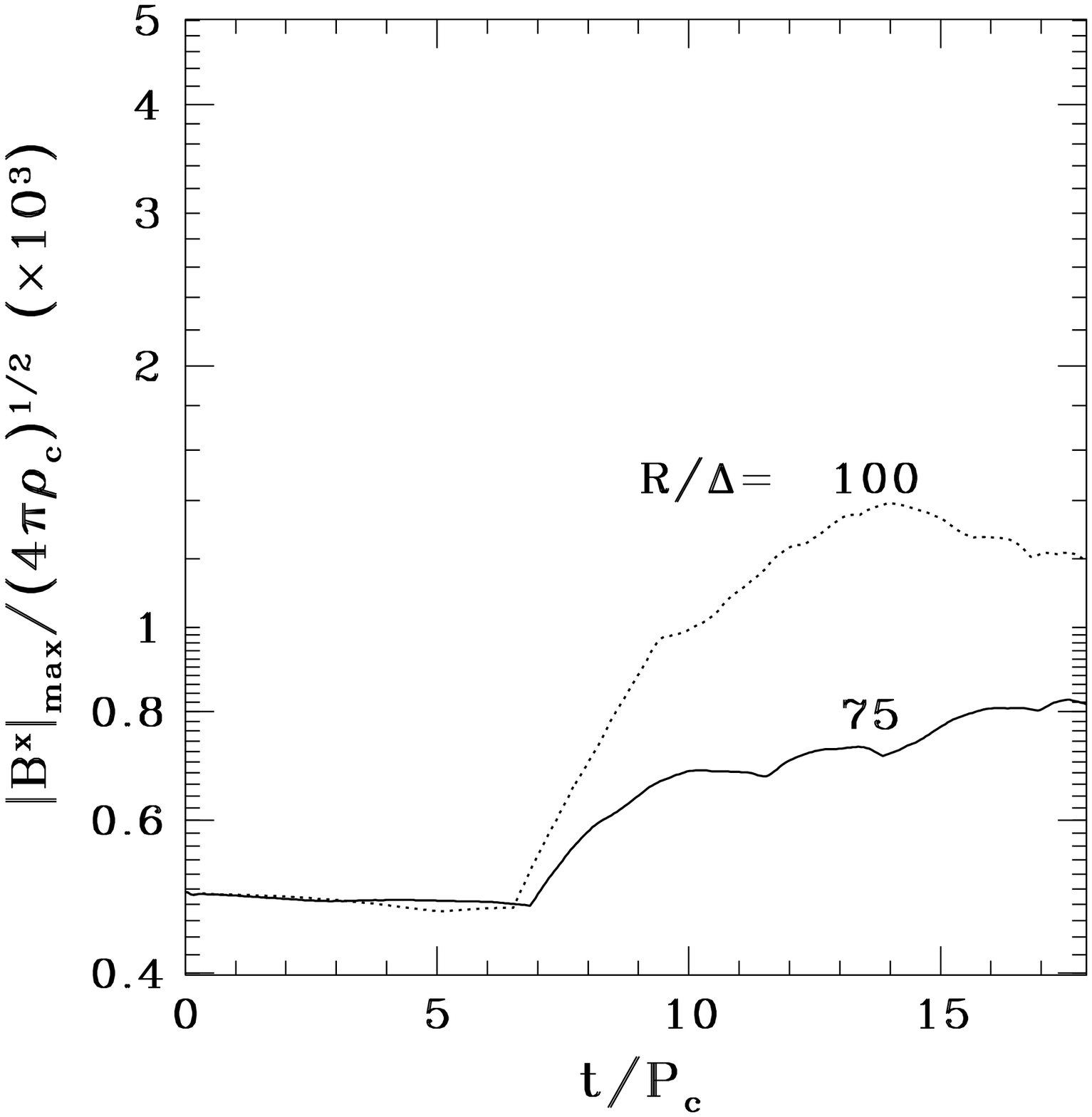}
    \caption{$|B^x|_{\text{max}}$ vs.~time, with perturbative results
      plotted on the left and nonperturbative (BSSN) on the right. 
      Resolutions of $R/\Delta=75$, $100$, $150$, $200$, and $250$ 
      are shown with solid, dotted, short dashed, long dashed, and
      long dash-dotted lines, respectively.  The short dash-dotted
      line represents an approximate slope $\gamma = 1/(5.0P_c)$ for
      the exponential growth rate of the MRI, $\delta B^x \propto
      e^{\gamma t}$.  Note that $\bkt{t_A}/P_c = 17.9$.}
  \label{MRIresstudy1}
  \end{center}
\end{figure}

In the MRI Resolution Study, we perform simulations on a magnetized
differentially rotating star (see Table~\ref{tab:parameters}) at
resolutions $R/\Delta=75$, 100, 150, 200, and 250.  The initial
magnetic field is set so that $C=6.1\times 10^{-5}$.  To demonstrate
that the two metric solvers yield the same results, we also perform
simulations with the BSSN metric solver at the two lowest resolutions
($R/\Delta=$75 and 100).  
Figure~\ref{MRIresstudy1} displays the maximum magnitude of $|B^x|$ as
a function of time. MRI causes the sudden increase of $|B^x|_{\rm
  max}$ at $t \approx 6P_c$. As resolution is increased, $|B^x|_{\rm
  max}$ saturates at larger values. Due to the turbulent nature of the
MRI, we do not achieve convergence, even at the highest resolutions
($R/\Delta=200$ and 250).  However the exponential growth time of the
MRI, $\tau_{\text{MRI}}$, does converge at the highest resolutions
($R/\Delta=150 \to 250$). The numerically determined value for
$\tau_{\text{MRI}}$ is $\approx 5.0P_c$, which does not significantly
deviate from the linearized, Newtonian theory estimate (obtained by
applying Eq.~(\ref{tauMRI}) at $t=0$) of $\tau_{\text{MRI,min}} =
5.7P_c$. We see that regardless of resolution or metric evolution
scheme, the MRI-induced amplification of the magnetic field above its
initial value becomes evident by $t\approx 6P_c$.  

\begin{figure}
  \begin{center}
    \epsfxsize=2.7in
    \leavevmode
    \hspace{-0.7cm}\epsffile{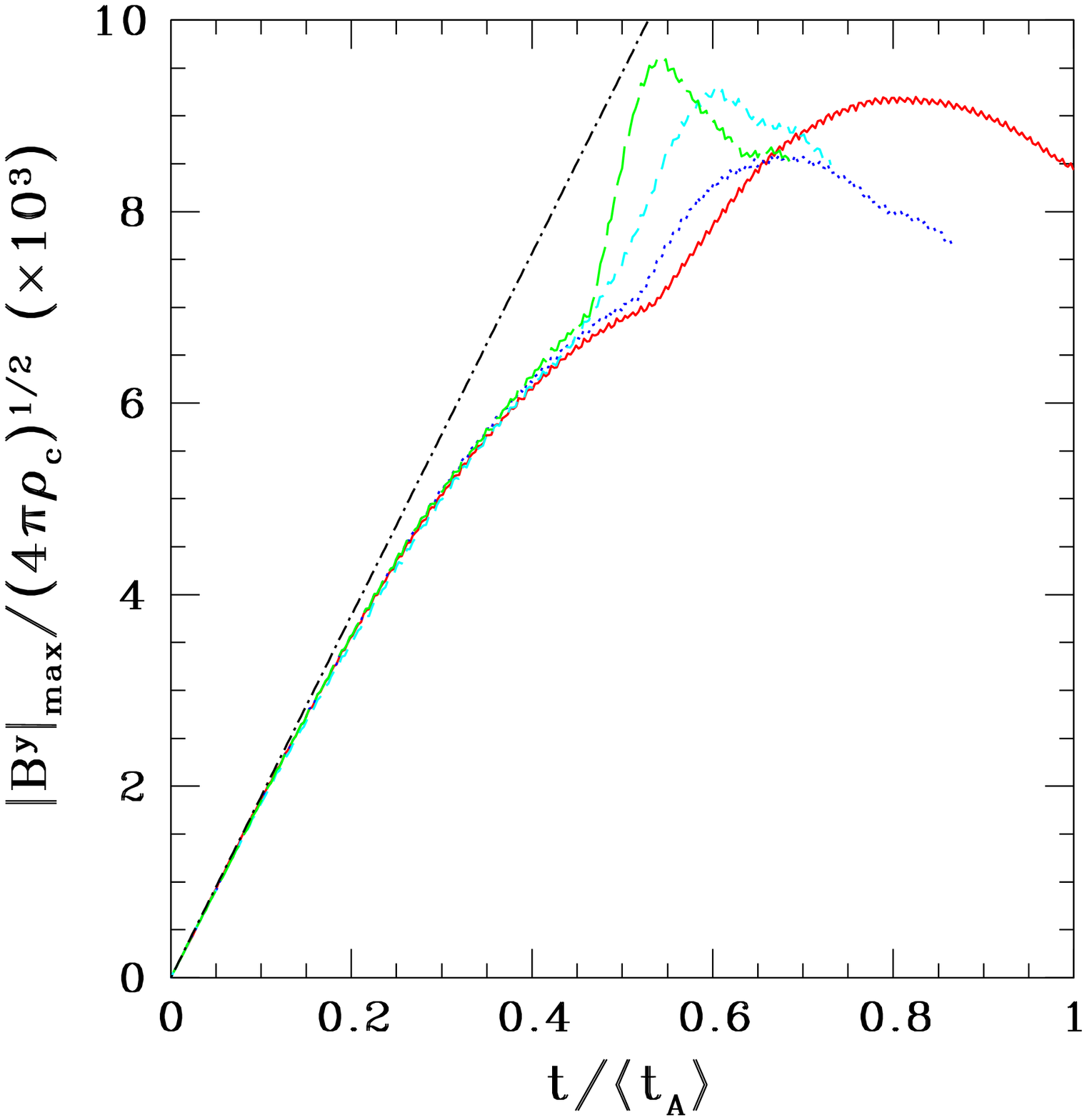}
    \epsfxsize=2.7in
    \leavevmode
    \hspace{-0.2cm}\epsffile{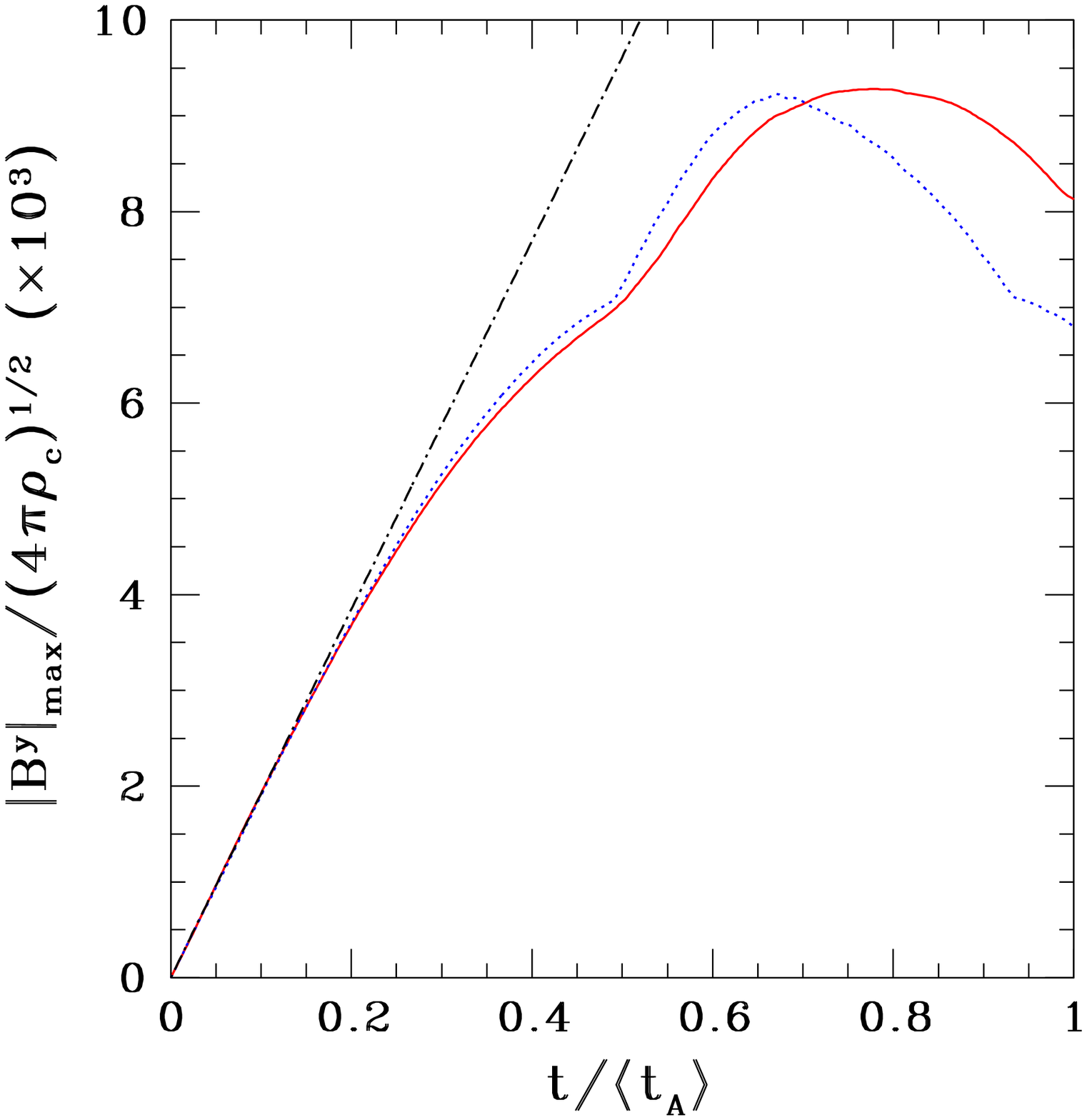}
    \caption{Evolution of $|B^y|_{\text{max}}(t)$. Simulations with
      resolution $R/\Delta=75$, $100$, $150$, and $250$ are shown with
      solid red, dotted blue, short dashed cyan, and long dashed
      green lines respectively, with perturbative results on the
      left and nonperturbative (BSSN) on the right.  The dash-dotted
      black line represents the expected early-time linear growth of
      $|B^y|_{\rm max}$, as predicted by Eq.~(\ref{lineargrowth}).}
  \label{MRIresstudy2}
  \end{center}
\end{figure}

Figure~\ref{MRIresstudy2} plots the maximum value of $|B^y|$ as a
function of time.  The straight lines in each plot indicate the
expected growth rate via magnetic braking in the linear regime,
according to Eq.~(\ref{lineargrowth}).  Notice that at early time,
$|B^y|_{\text{max}}$ agrees well with the expected linear growth.
However, the slope begins to flatten once the magnetic field becomes
strong enough to induce fluid back-reaction.  Later, the slope of
$|B^y|_{\text{max}}$ increases once again before flattening at the
point of saturation. As resolution is improved, the saturation point
in $|B^y|_{\text{max}}$ occurs at earlier time.  We find that the
sudden increase in the slope correlates well with the increase of
$|B^x|_{\text{max}}$ in the MRI plot of Figure~\ref{MRIresstudy1}. 
We also find that at the same time, the region at which $|B^y|_{\rm
  max}$ occurs shifts to the region where the MRI occurs. This
is further evidence that the sudden increase in $|B^y|_{\text{max}}$
results from MRI-induced magnetic field rearrangement.

\begin{figure} 
  \begin{center}
    \epsfxsize=1.4in
    \leavevmode
    \hspace{-0.3cm}\epsffile{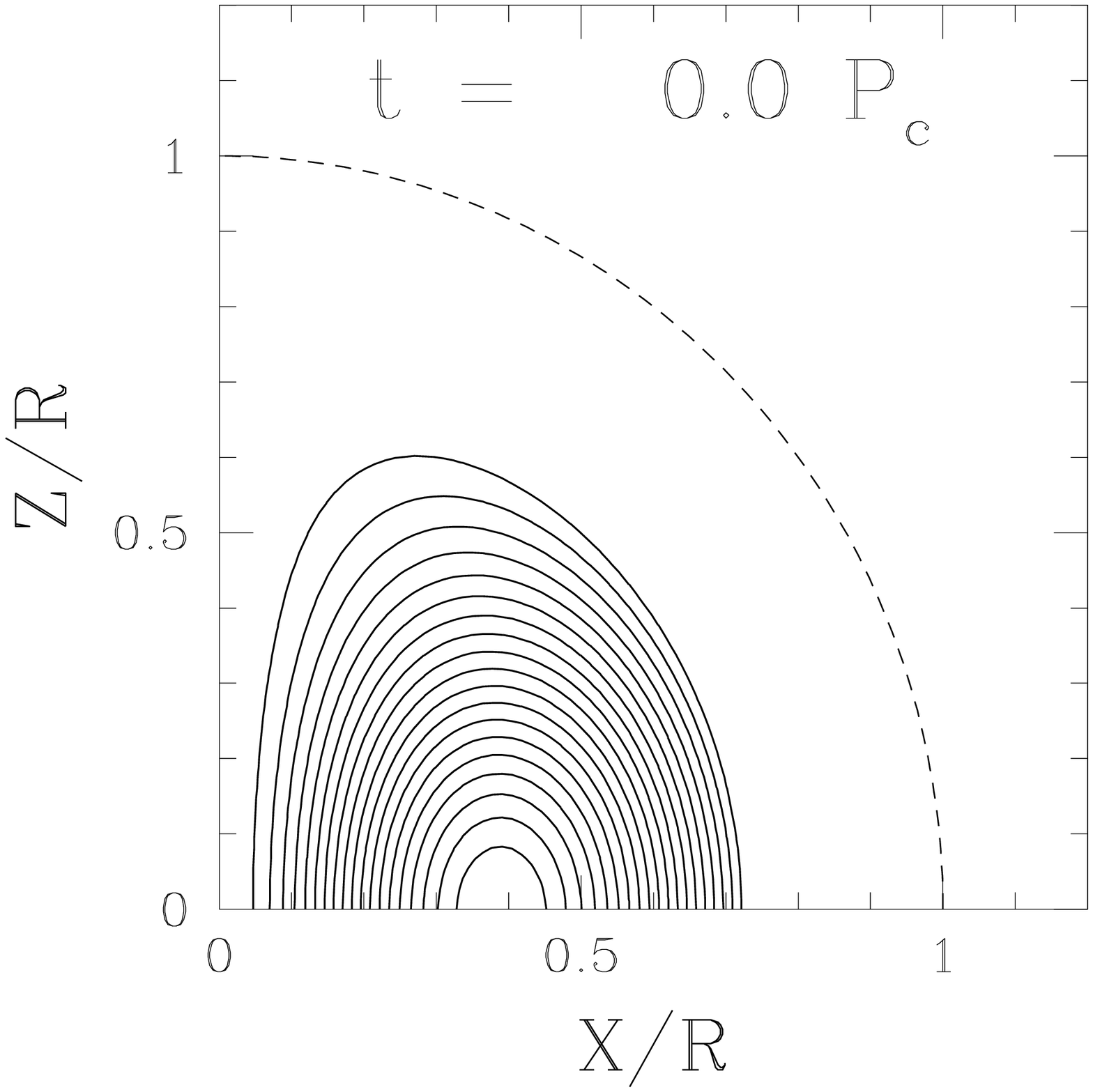}
    \epsfxsize=1.4in
    \leavevmode
    \hspace{-0.3cm}\epsffile{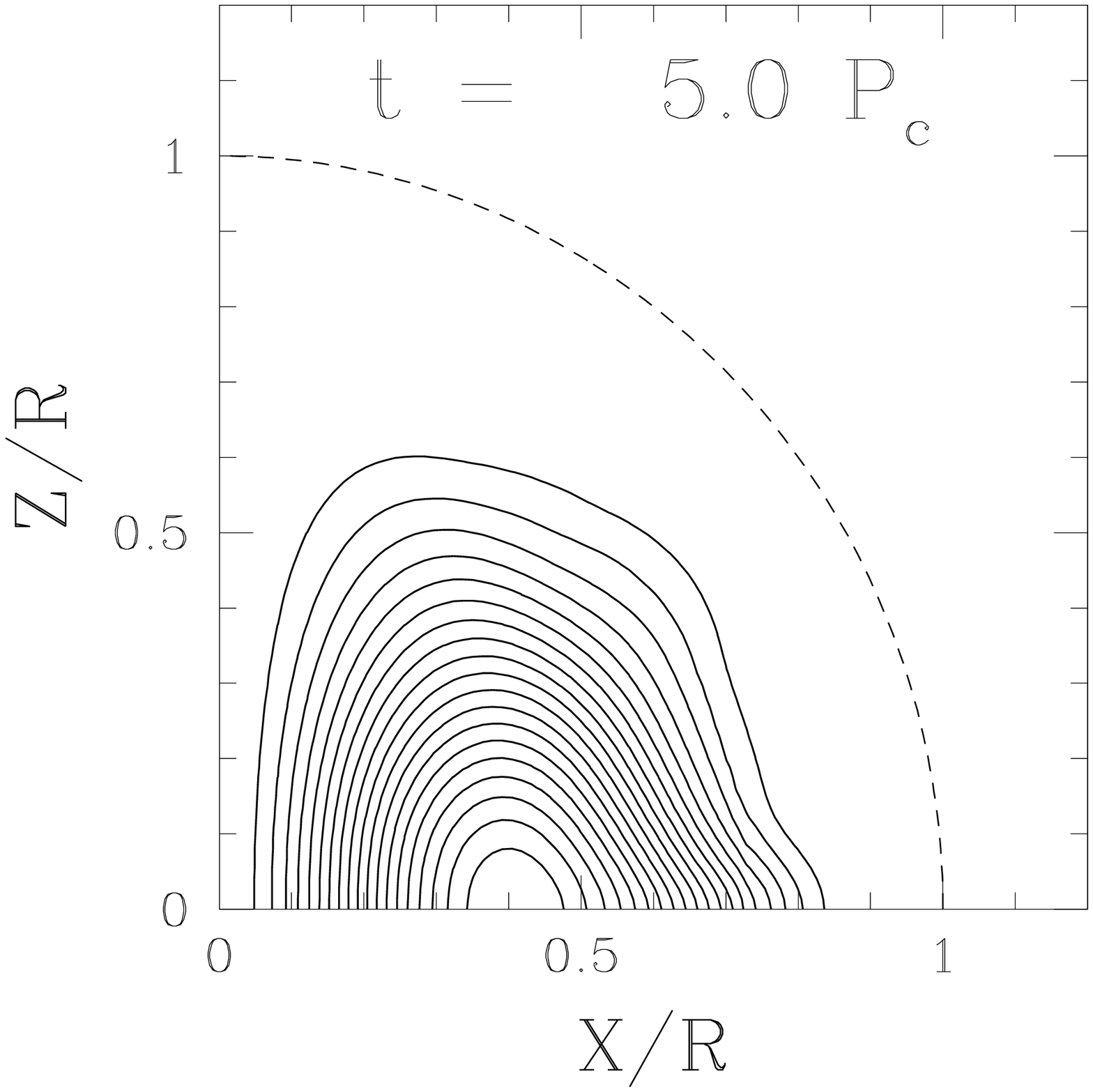}
    \epsfxsize=1.4in
    \leavevmode
    \hspace{-0.3cm}\epsffile{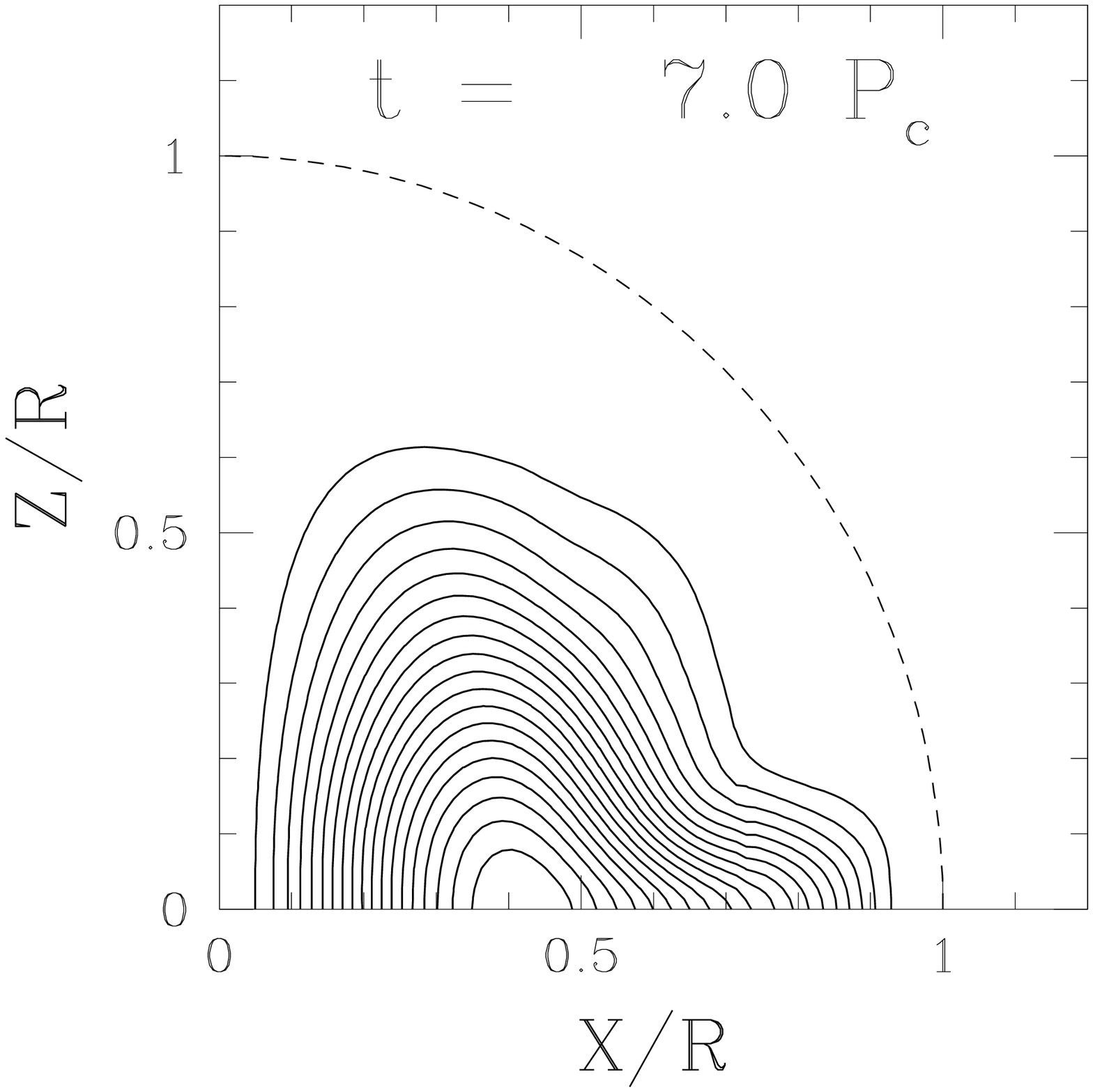}
    \epsfxsize=1.4in
    \leavevmode
    \hspace{-0.3cm}\epsffile{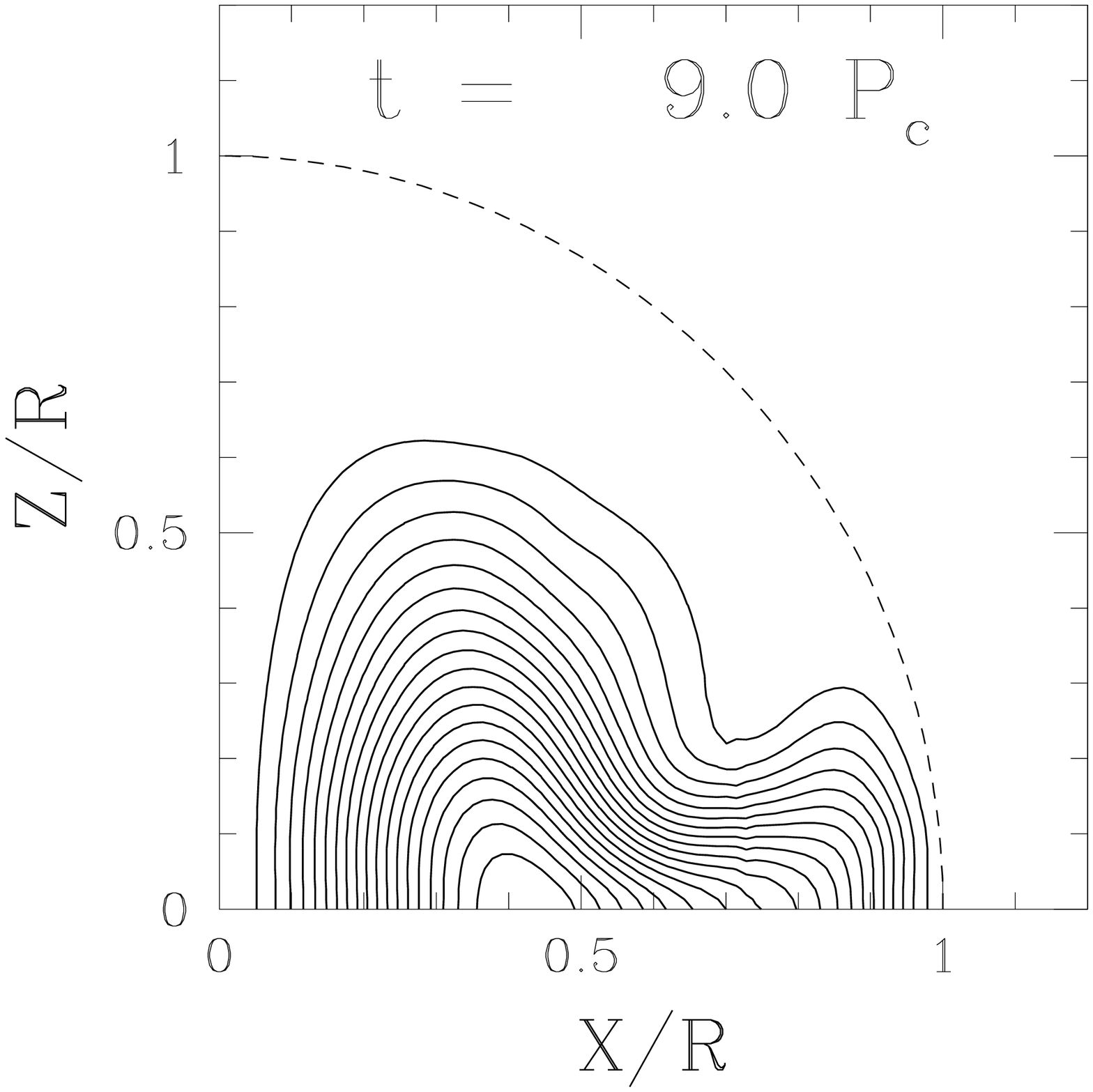}
    \epsfxsize=1.4in
    \leavevmode
    \hspace{-0.3cm}\epsffile{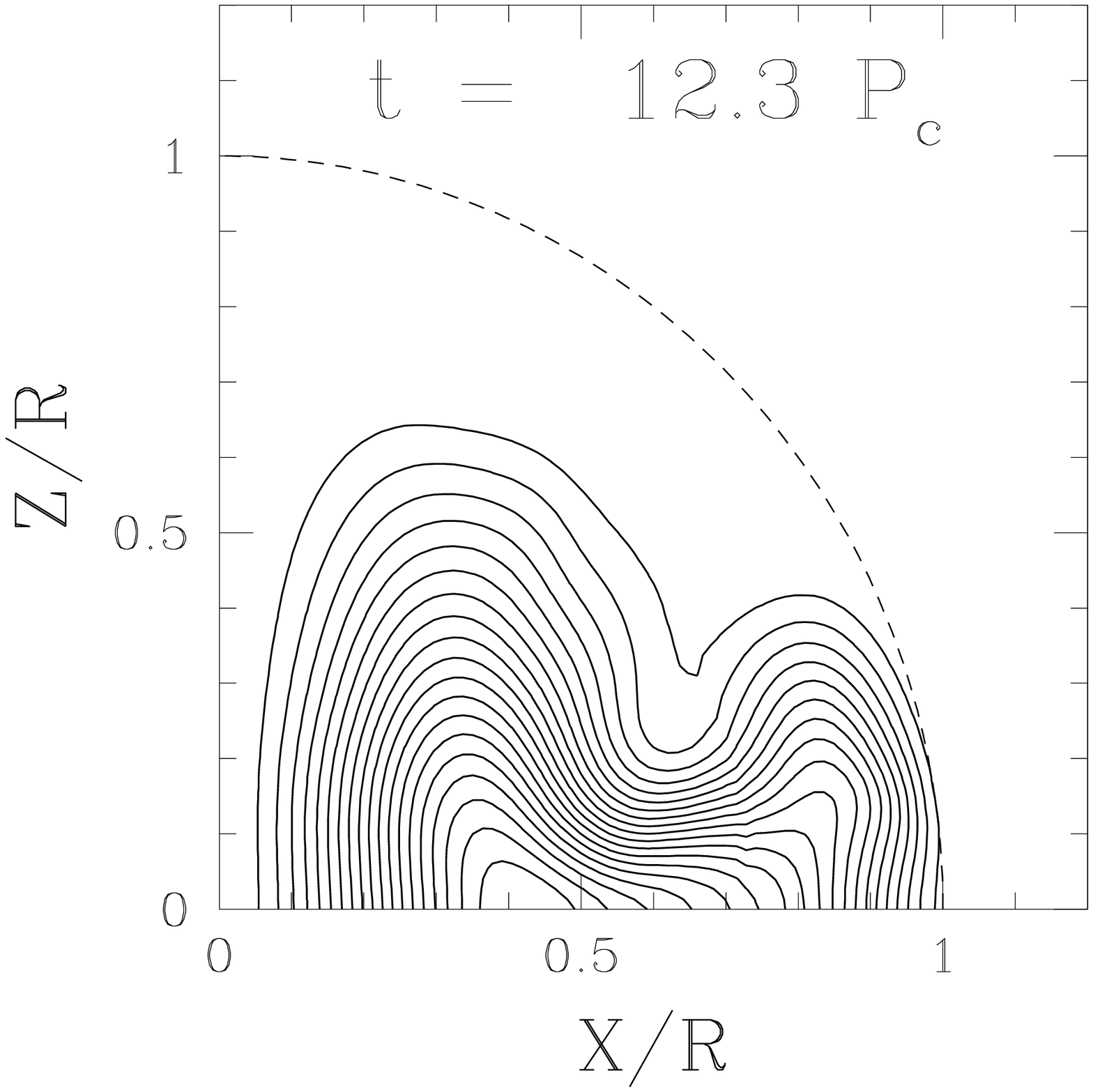}
    \epsfxsize=1.4in
    \leavevmode
    \epsfxsize=1.4in
    \leavevmode
    \epsfxsize=1.4in
    \leavevmode \\
    \epsfxsize=1.4in
    \leavevmode
    \hspace{-0.3cm}\epsffile{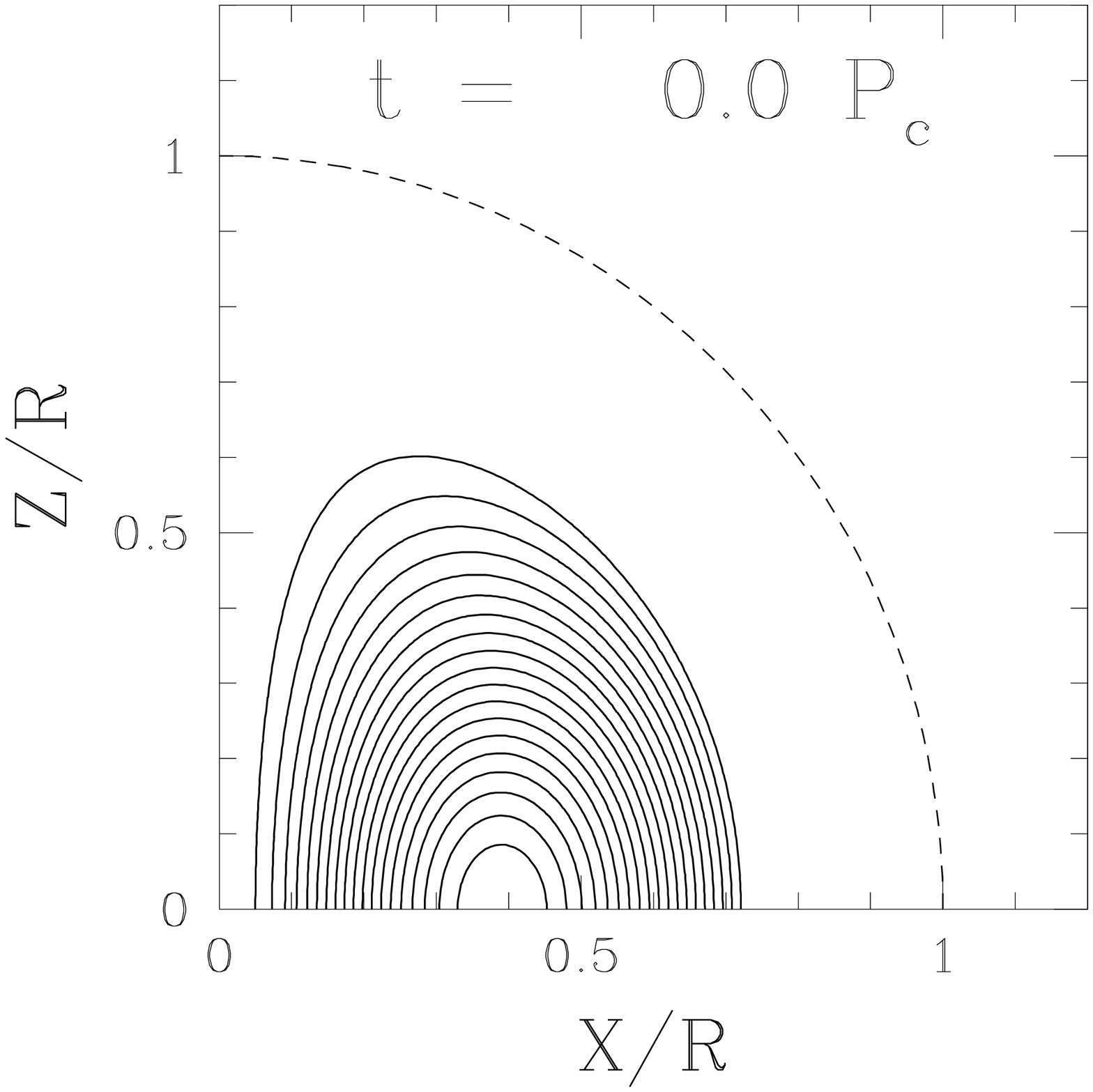}
    \epsfxsize=1.4in
    \leavevmode
    \hspace{-0.3cm}\epsffile{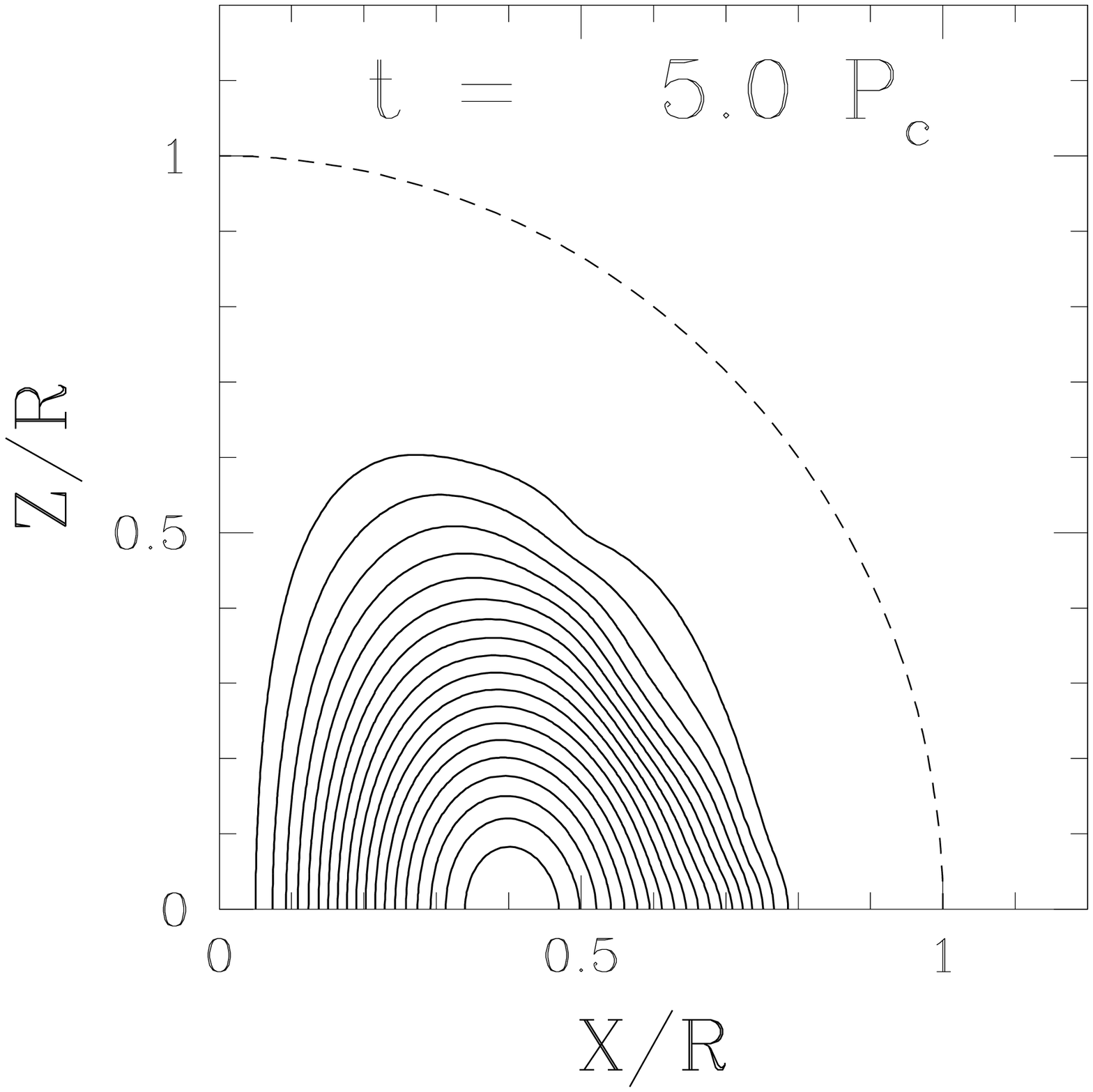}
    \epsfxsize=1.4in
    \leavevmode
    \hspace{-0.3cm}\epsffile{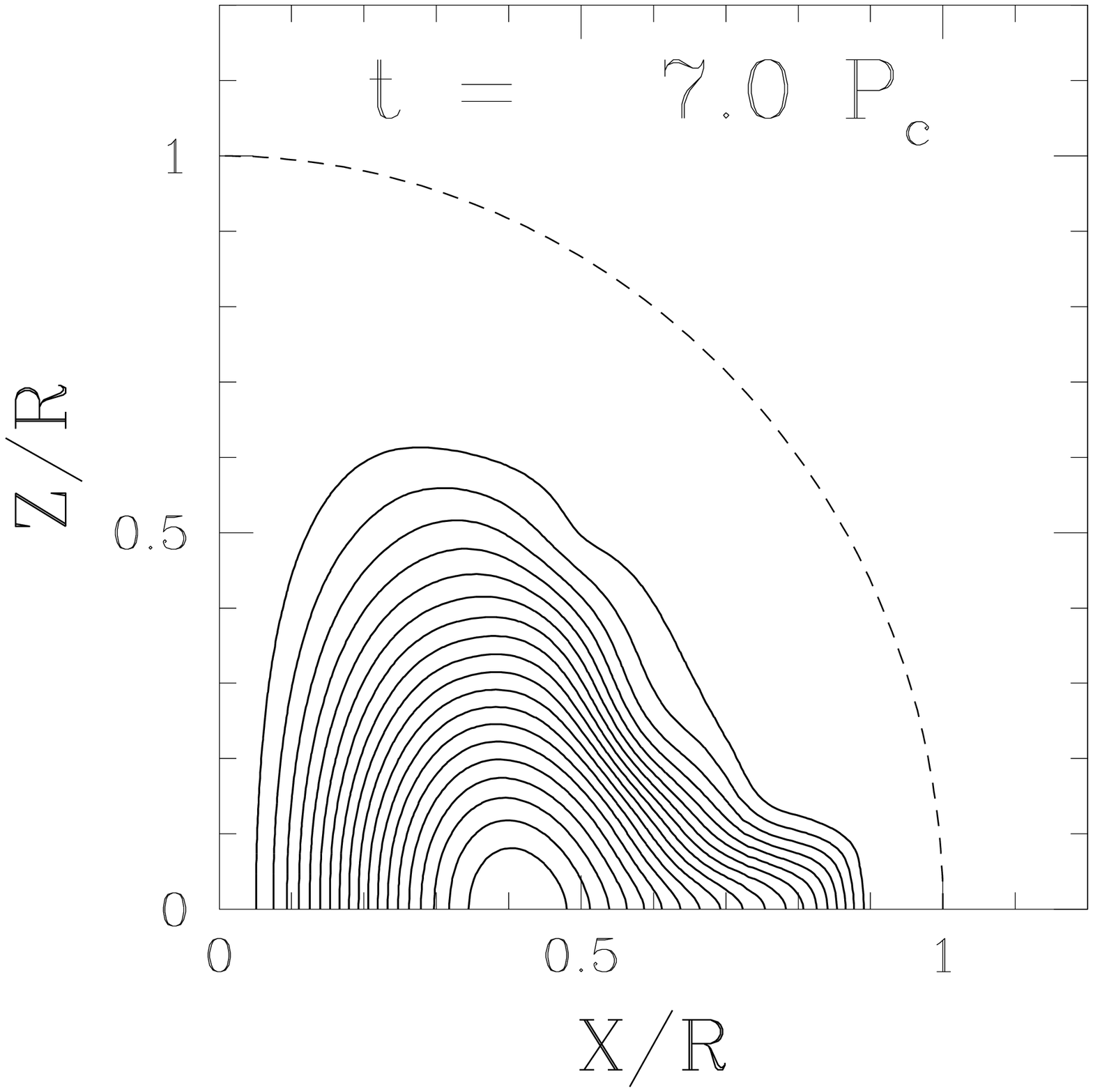}
    \epsfxsize=1.4in
    \leavevmode
    \hspace{-0.3cm}\epsffile{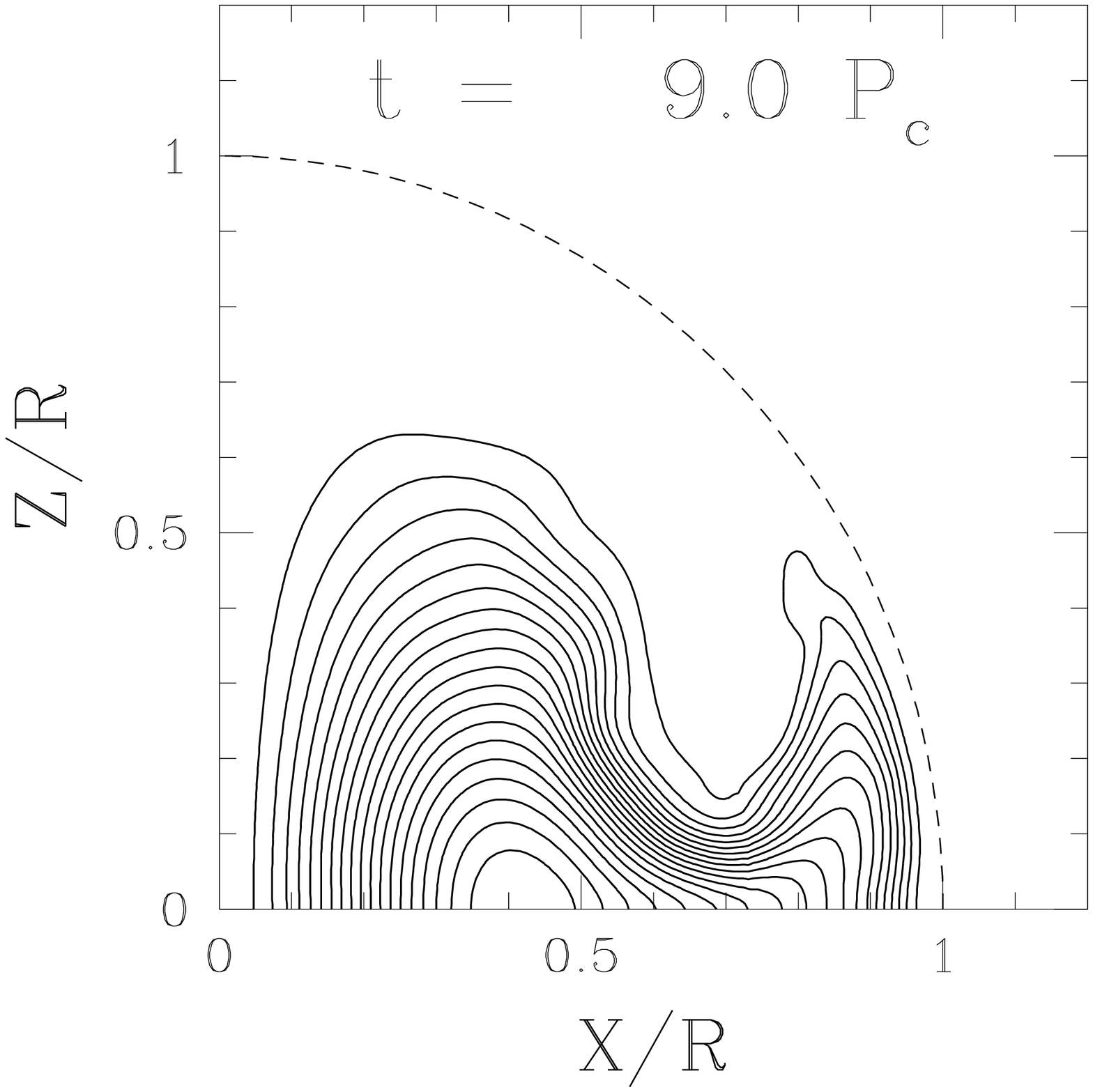}
    \epsfxsize=1.4in
    \leavevmode
    \hspace{-0.3cm}\epsffile{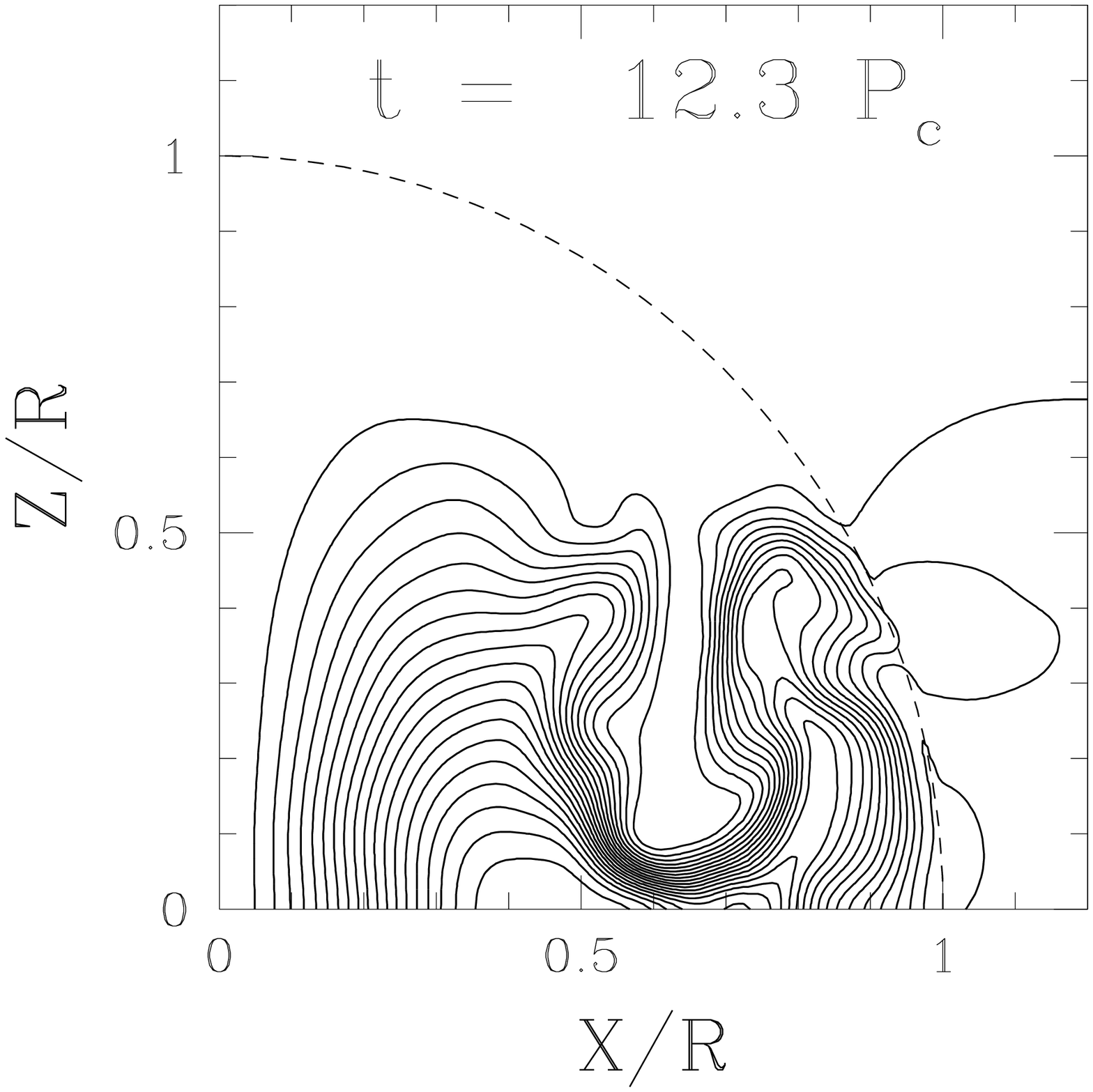}
    \epsfxsize=1.4in
    \leavevmode
    \epsfxsize=1.4in
    \leavevmode
    \epsfxsize=1.4in
    \leavevmode \\
    \hspace{-0.3cm}\epsffile{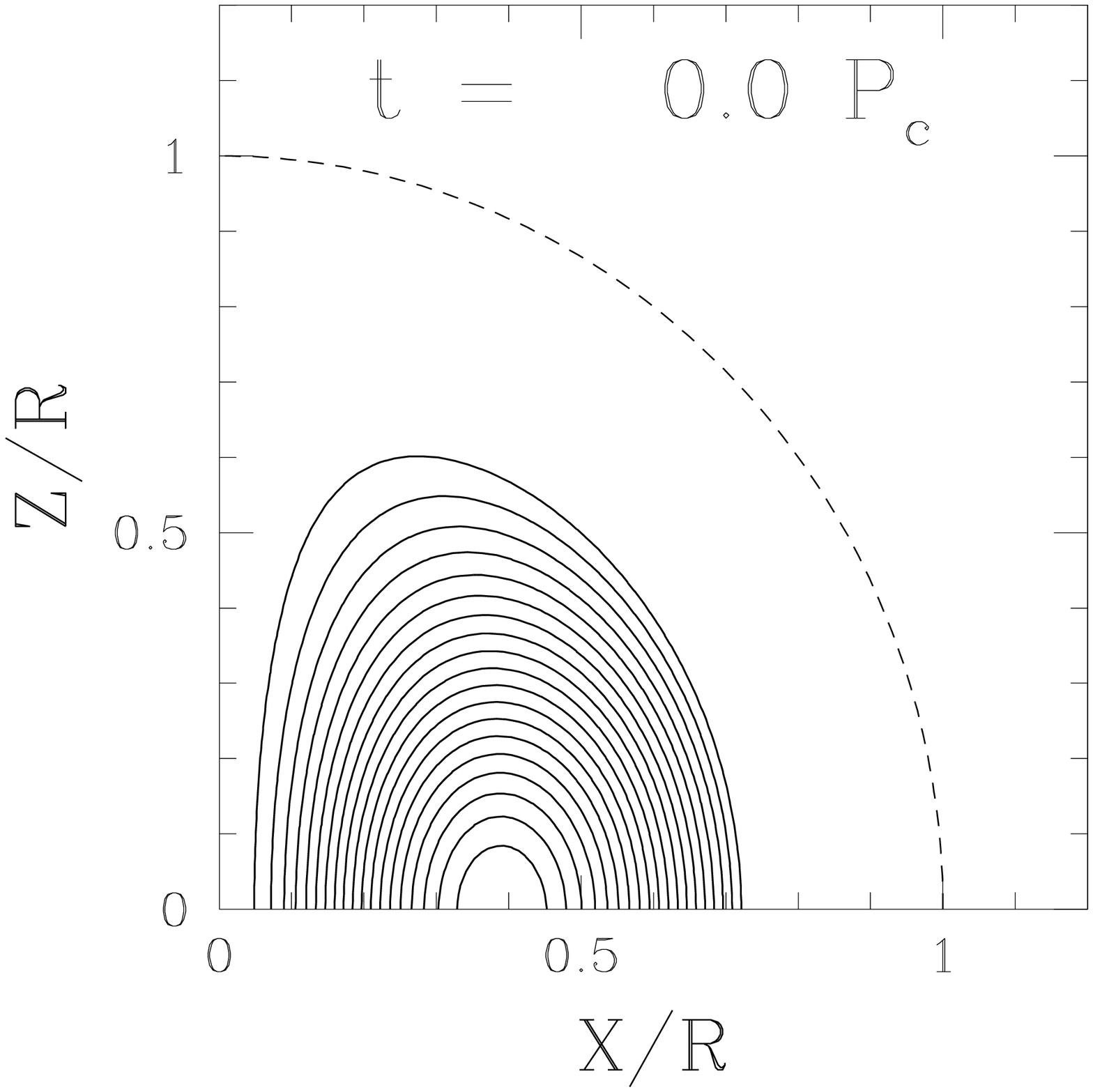}
    \epsfxsize=1.4in
    \leavevmode
    \hspace{-0.3cm}\epsffile{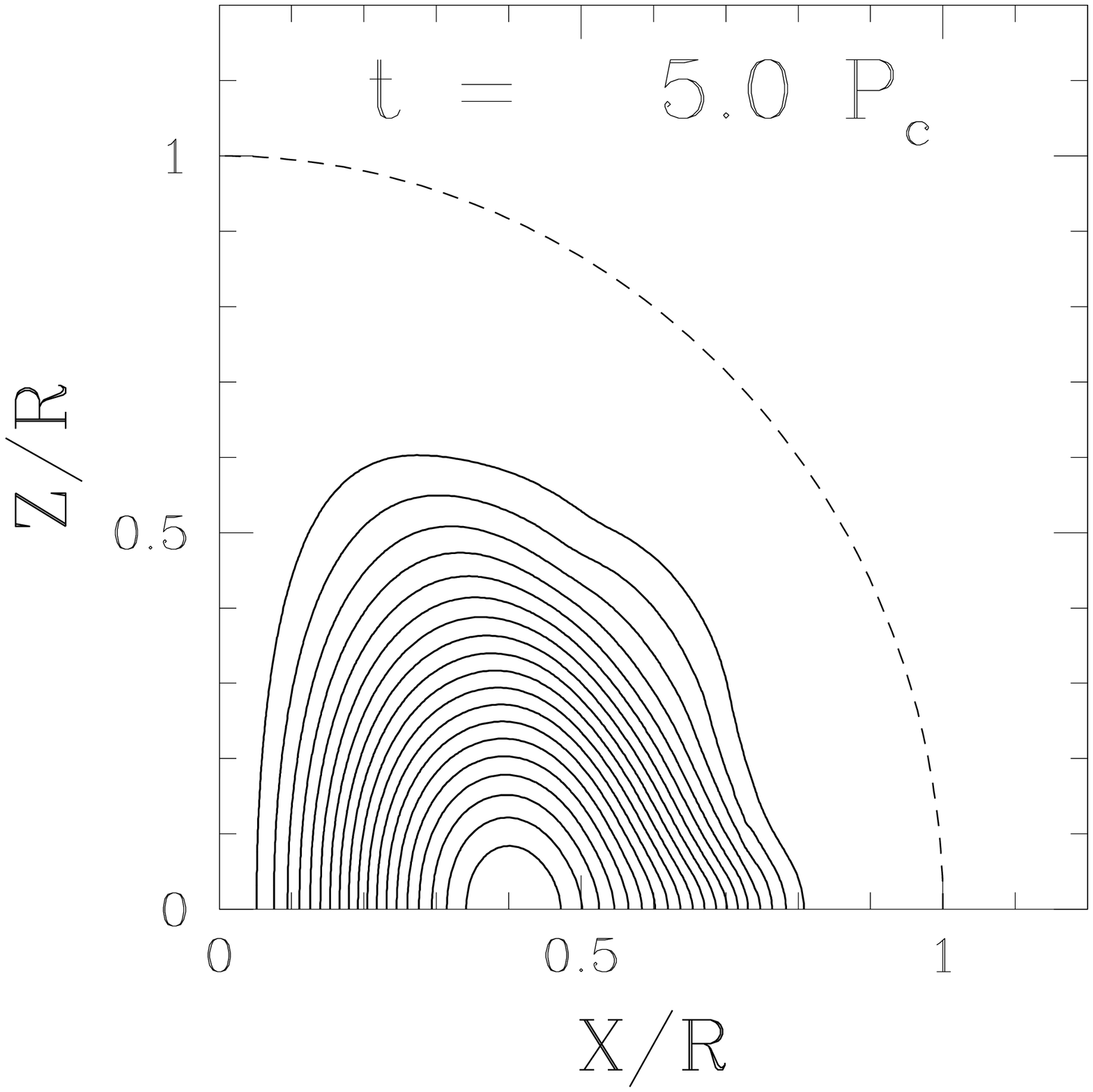}
    \epsfxsize=1.4in
    \leavevmode
    \hspace{-0.3cm}\epsffile{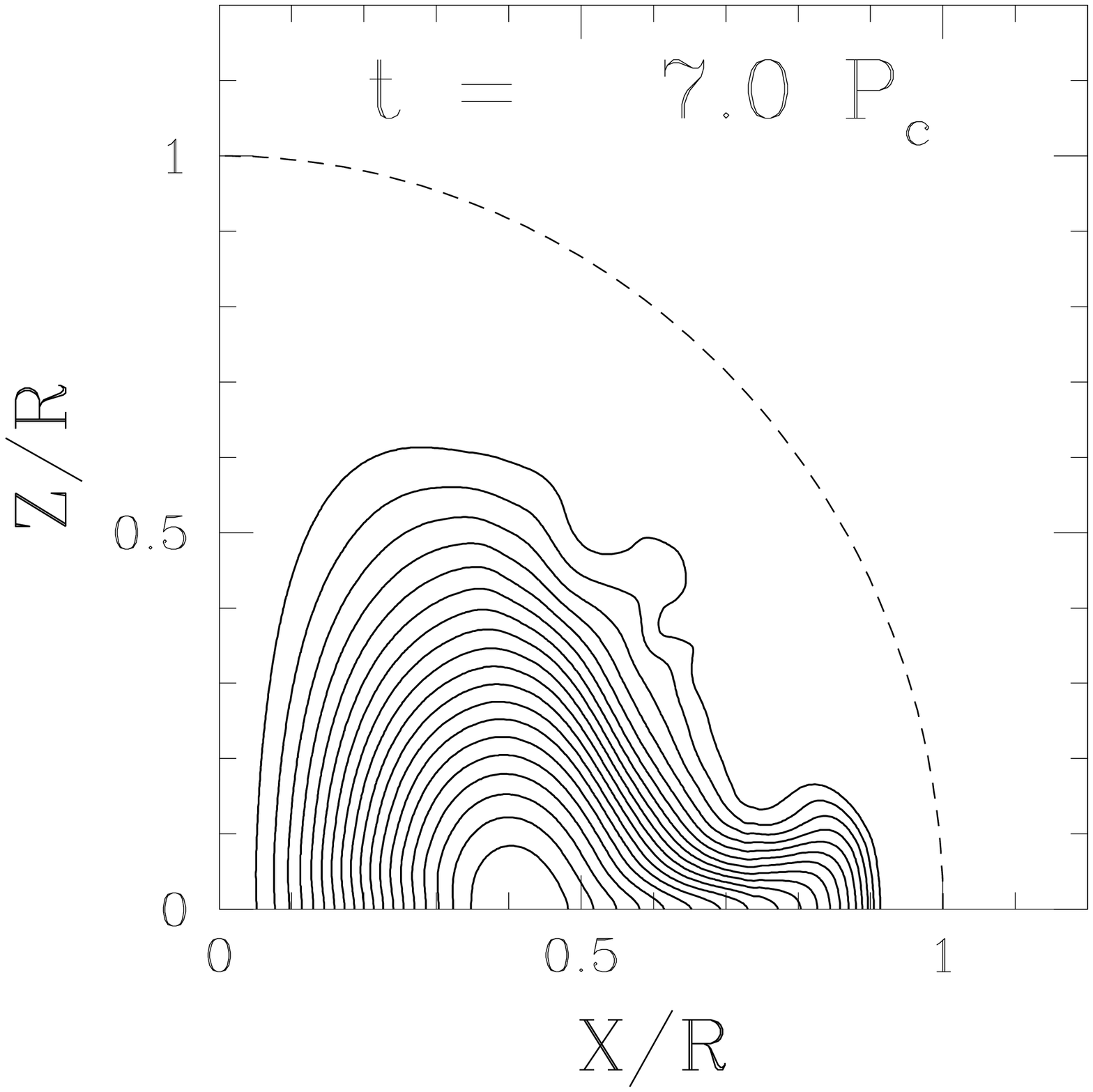}
    \epsfxsize=1.4in
    \leavevmode
    \hspace{-0.3cm}\epsffile{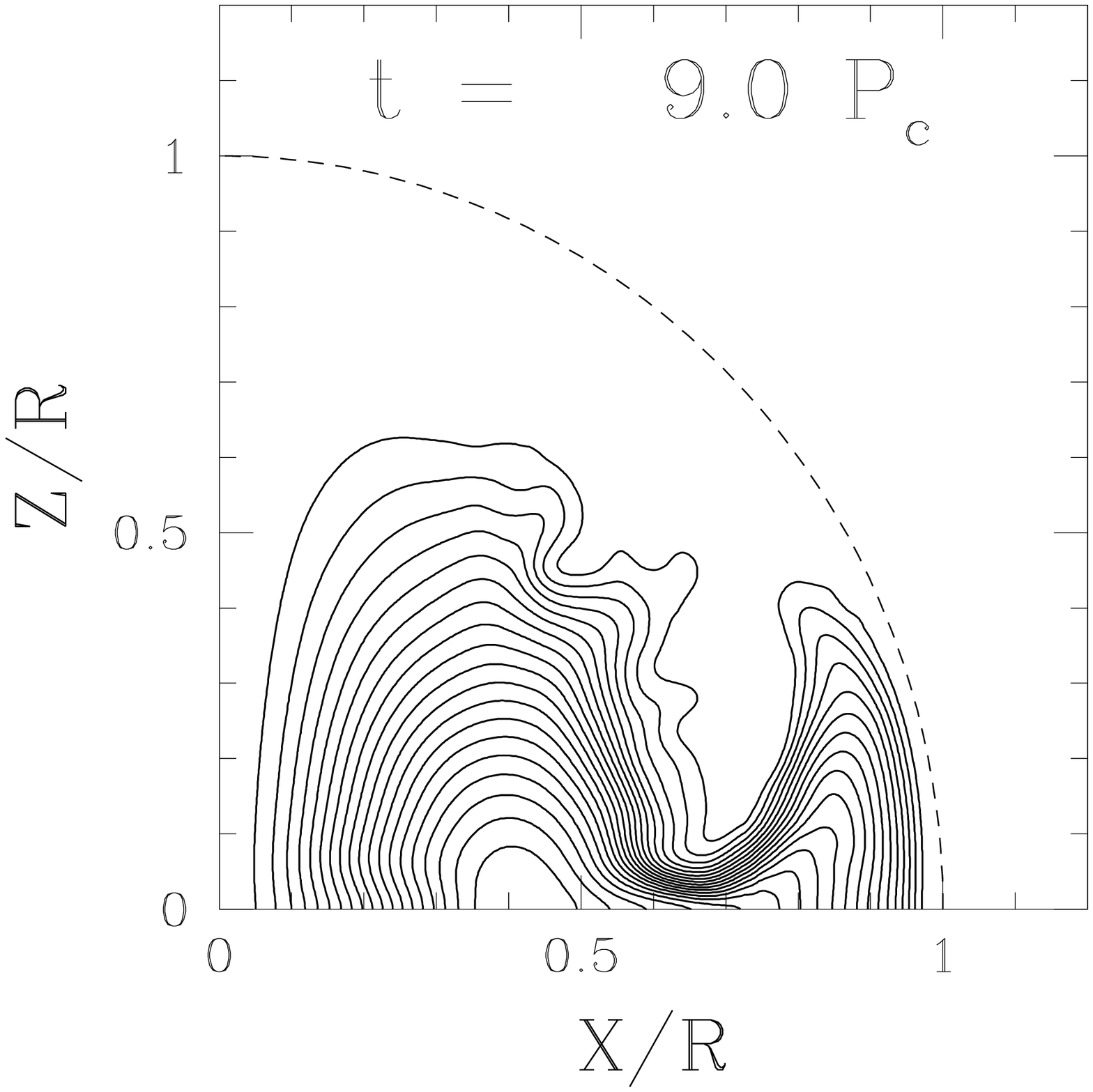}
    \epsfxsize=1.4in
    \leavevmode
    \hspace{-0.3cm}\epsffile{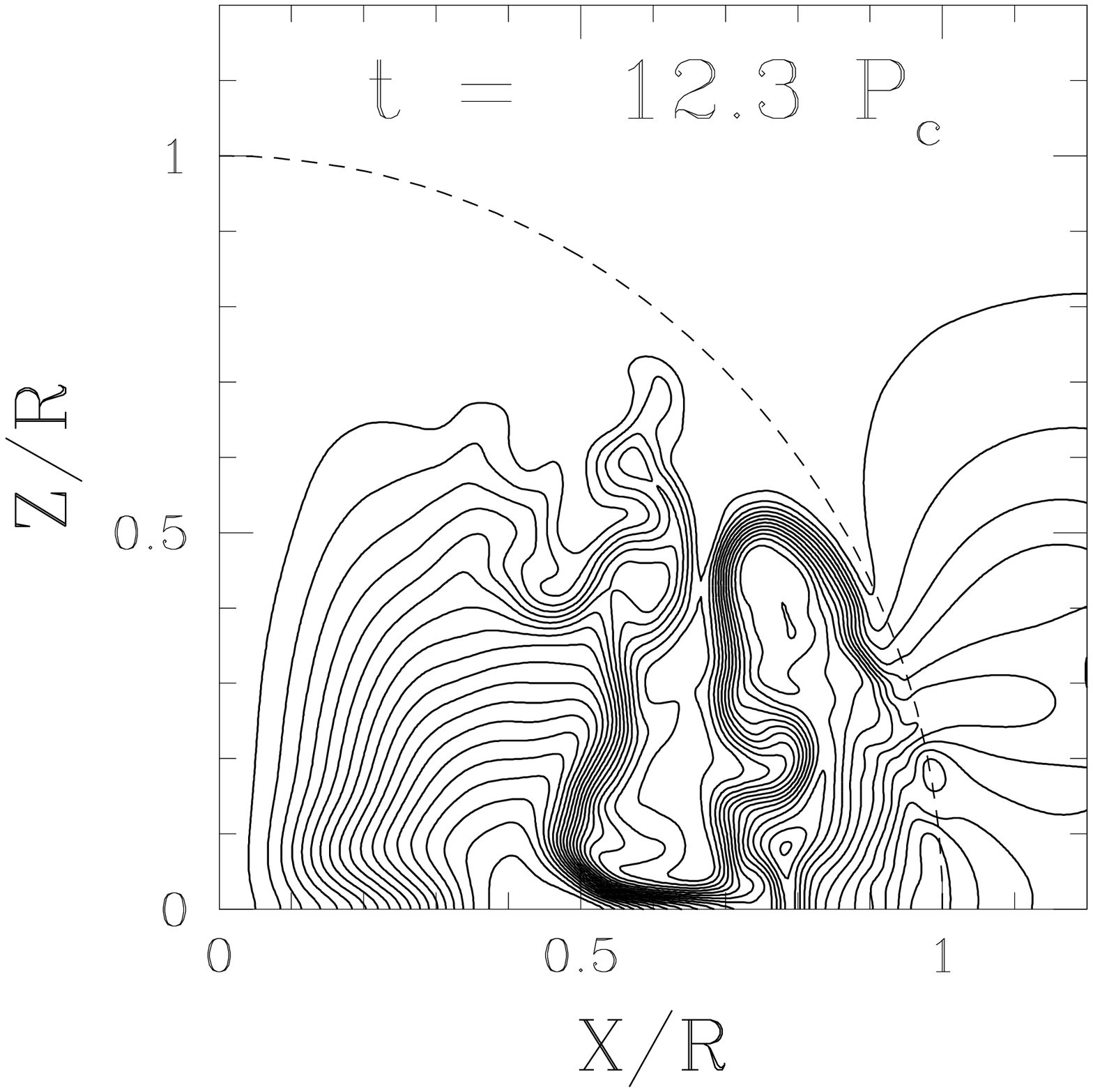}
    \epsfxsize=1.4in
    \leavevmode
    \epsfxsize=1.4in
    \leavevmode
    \epsfxsize=1.4in
    \leavevmode
    \caption{Snapshots of poloidal magnetic field lines at various times,
      with resolution $R/\Delta=75$ (top), $150$, (middle), and $250$
      (bottom). The field lines are drawn for $A_{\phi}=A_{\phi,\rm
      min} + (A_{\phi,\rm max} - A_{\phi,\rm min}) i/20,~(i=1$--19),
      where $A_{\phi,\rm max}$ and $A_{\phi,\rm min}$ are the maximum
      and minimum values of $A_{\phi}$, respectively, at the given
      time.  The dashed line in each plot indicates the initial stellar
      surface.}
  \label{MRIresstudy3}
  \end{center}
\end{figure}

To further explore magnetic field rearrangement in our star at
different resolutions, Figure~\ref{MRIresstudy3} provides snapshots 
of poloidal magnetic field lines (i.e., contours of the vector
potential $A_{\phi}$) at various times. Notice that MRI induces the
largest distortion of the field lines in the outer equatorial region
of the star. This is consistent with the linear analysis:
Eq.~(\ref{tauMRI}), together with the star's angular velocity profile,
gives a shorter  $\tau_{\rm MRI}$ near the outer part of the star.
Similar behavior has been observed in simulations of magnetized,
rapidly and differentially rotating stars~\cite{MHDLett,BigPaper}.
The distortion becomes more prominent at finer resolution, which
suggests that more and more small-scale MRI modes are being resolved
as resolution is improved.

\begin{figure} 
  \begin{center}
    \epsfxsize=2.8in
    \leavevmode
    \hspace{-0.7cm}\epsffile{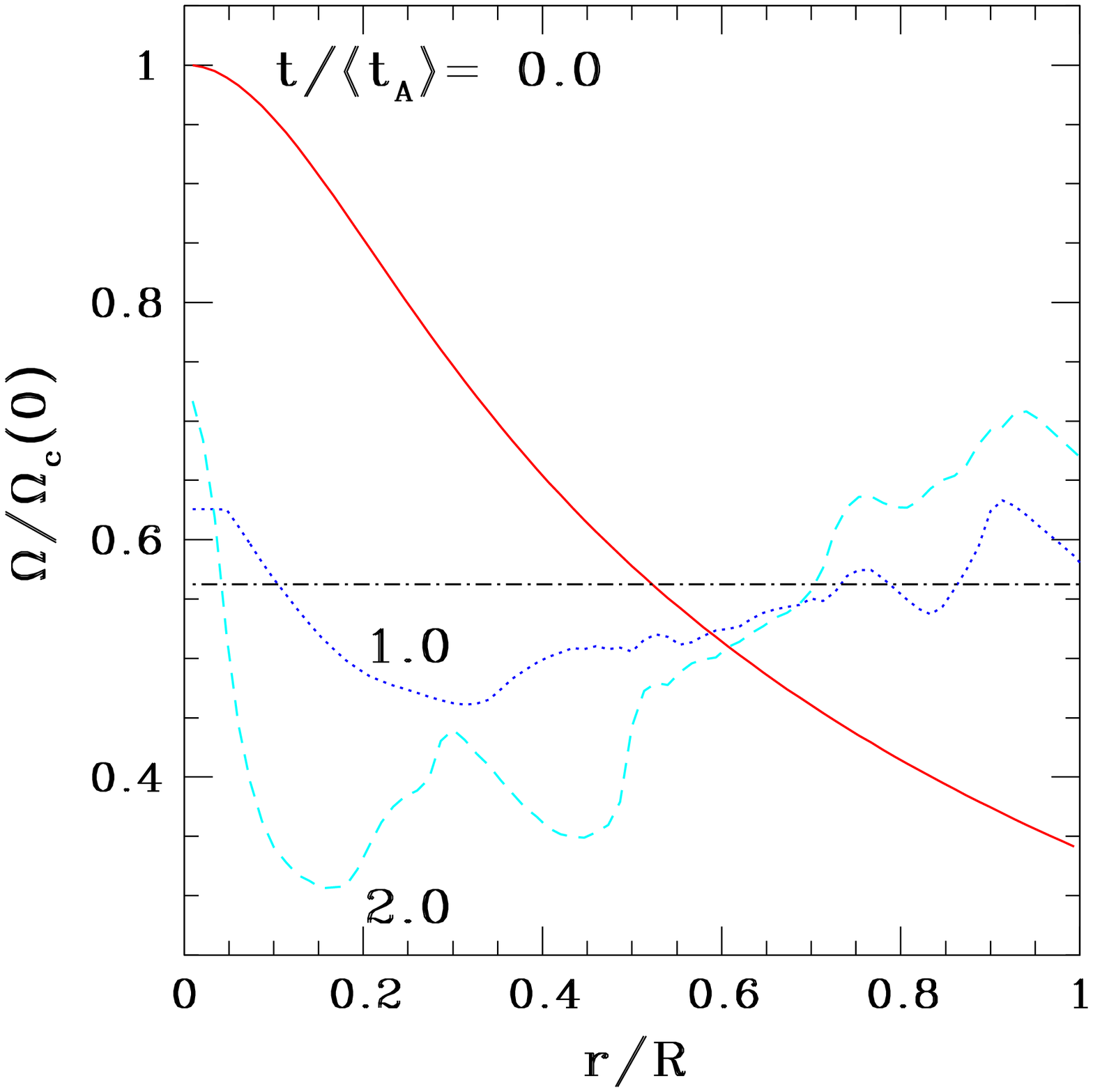}
    \epsfxsize=2.8in
    \leavevmode
    \hspace{-0.2cm}\epsffile{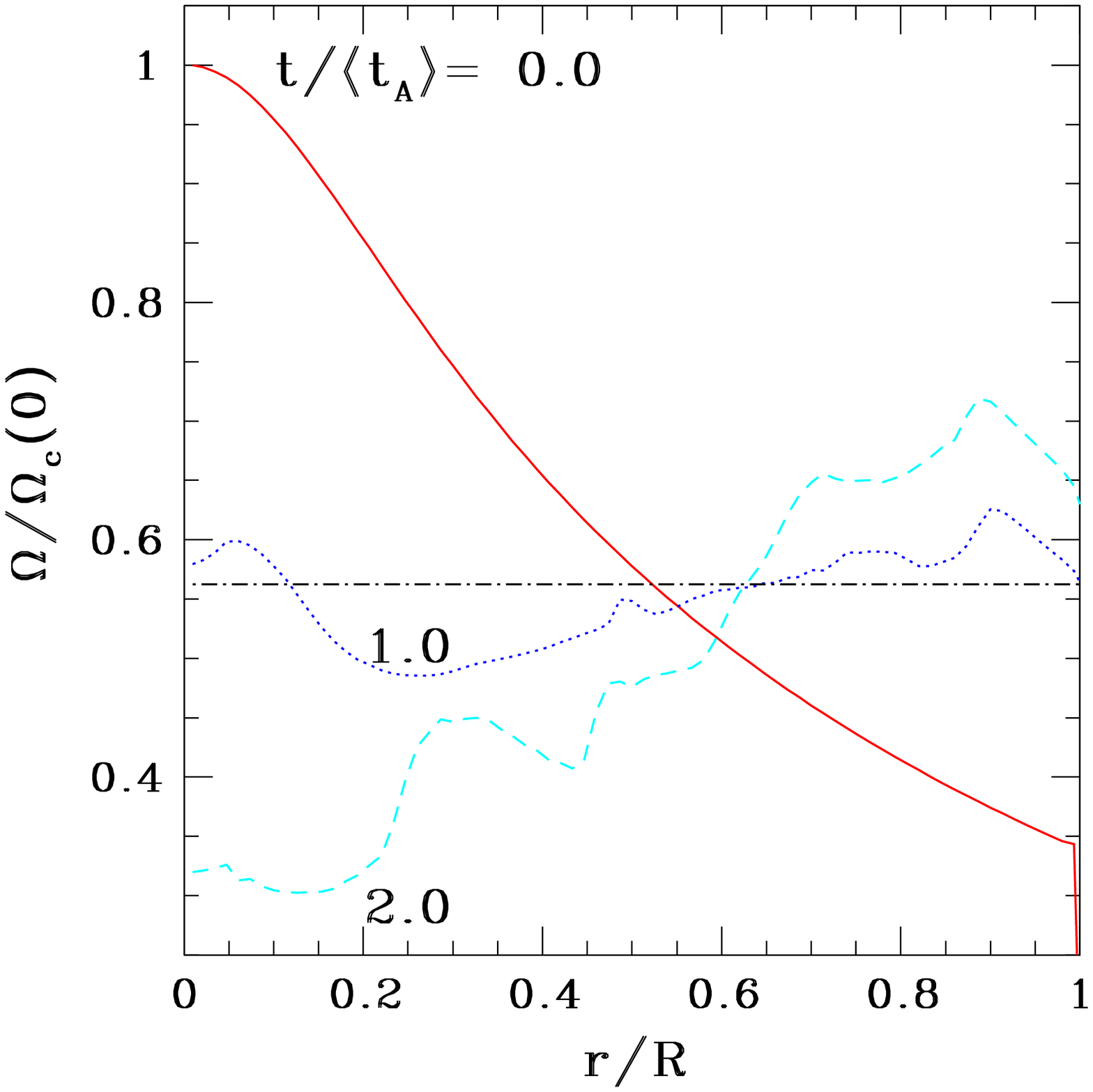}
    \caption{Rotation profile $\Omega$ measured in the equatorial
      plane at various times, comparing perturbative scheme (left)
      with nonperturbative (right) at $R/\Delta=75$ resolution.  The
      straight line indicates the solid body angular frequency
      $\Omega_{\text{const}}$  of a star with the same angular
      momentum $J$ and rest mass $M_0$.}
  \label{MRIresstudy4}
  \end{center}
\end{figure}

Next we analyze how the rotation profile of the star changes as a
result of its magnetic field.  Figure~\ref{MRIresstudy4} shows the
equatorial rotation profile at various times.
Consistent with the results of \cite{StusFirst}, we find that magnetic
winding destroys the differential rotation profile on the Alfv\'en
timescale, causing the star to rotate nearly as a solid body with
$\Omega = \Omega_{\text{const}}$ at $t \approx 1 \langle t_A\rangle$. 
Here $\Omega_{\text{const}}$ is the angular velocity of a uniformly 
rotating star with the same rest mass and angular momentum as the 
star under study. When the toroidal field saturates around 
$t \approx 1 \langle t_A\rangle$, the magnetic fields begin to unwind, 
eventually causing the rotation profile to increase with increasing 
radius at roughly $2 \langle t_A \rangle$. In addition, the MRI stirs
up a turbulent-like flow, causing the bumpy rotation profile seen at
later times.

\begin{figure}
  \begin{center}
    \epsfxsize=2.8in
    \leavevmode
    \hspace{-0.7cm}\epsffile{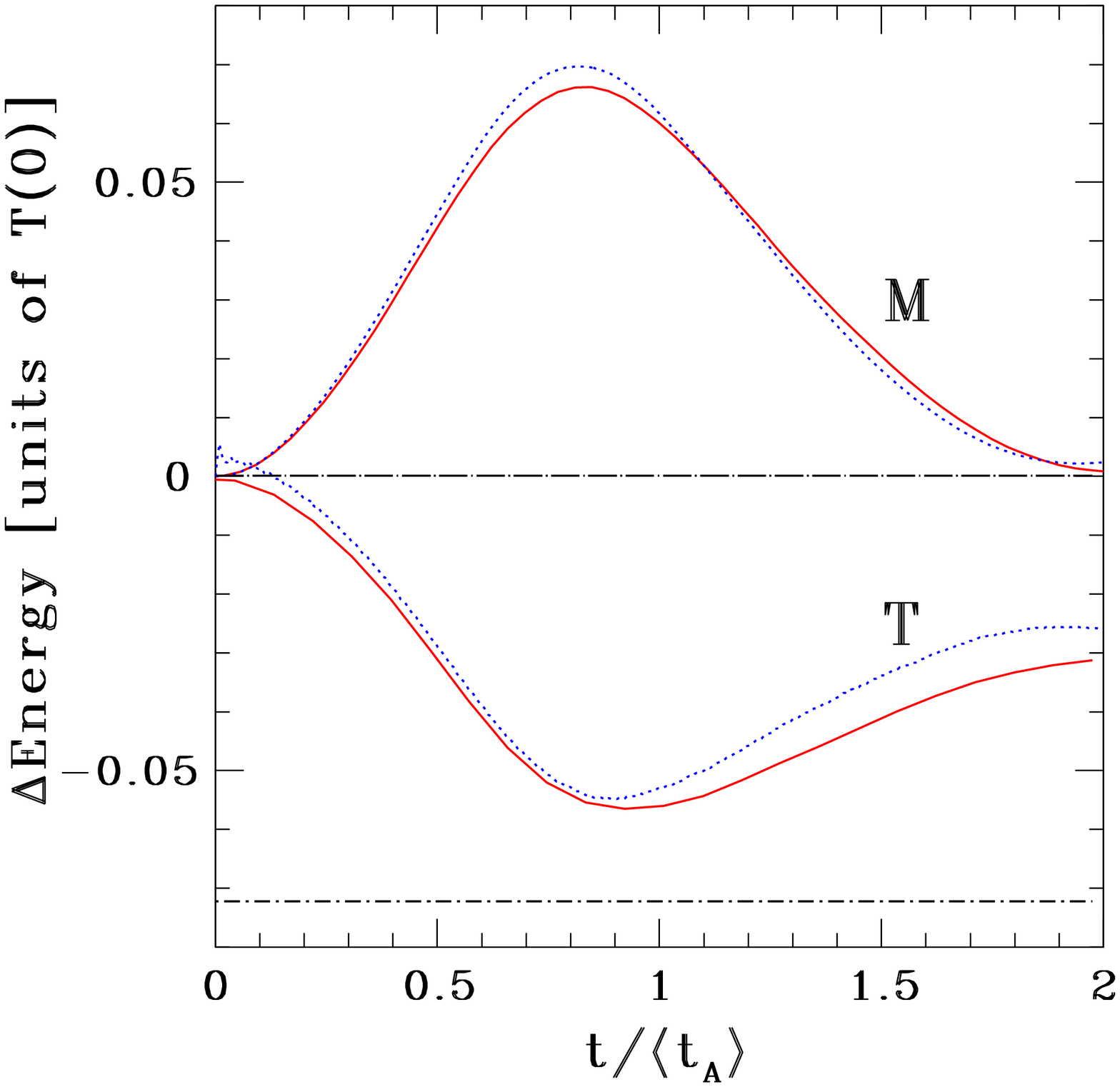}
    \epsfxsize=2.8in
    \leavevmode
    \hspace{-0.2cm}\epsffile{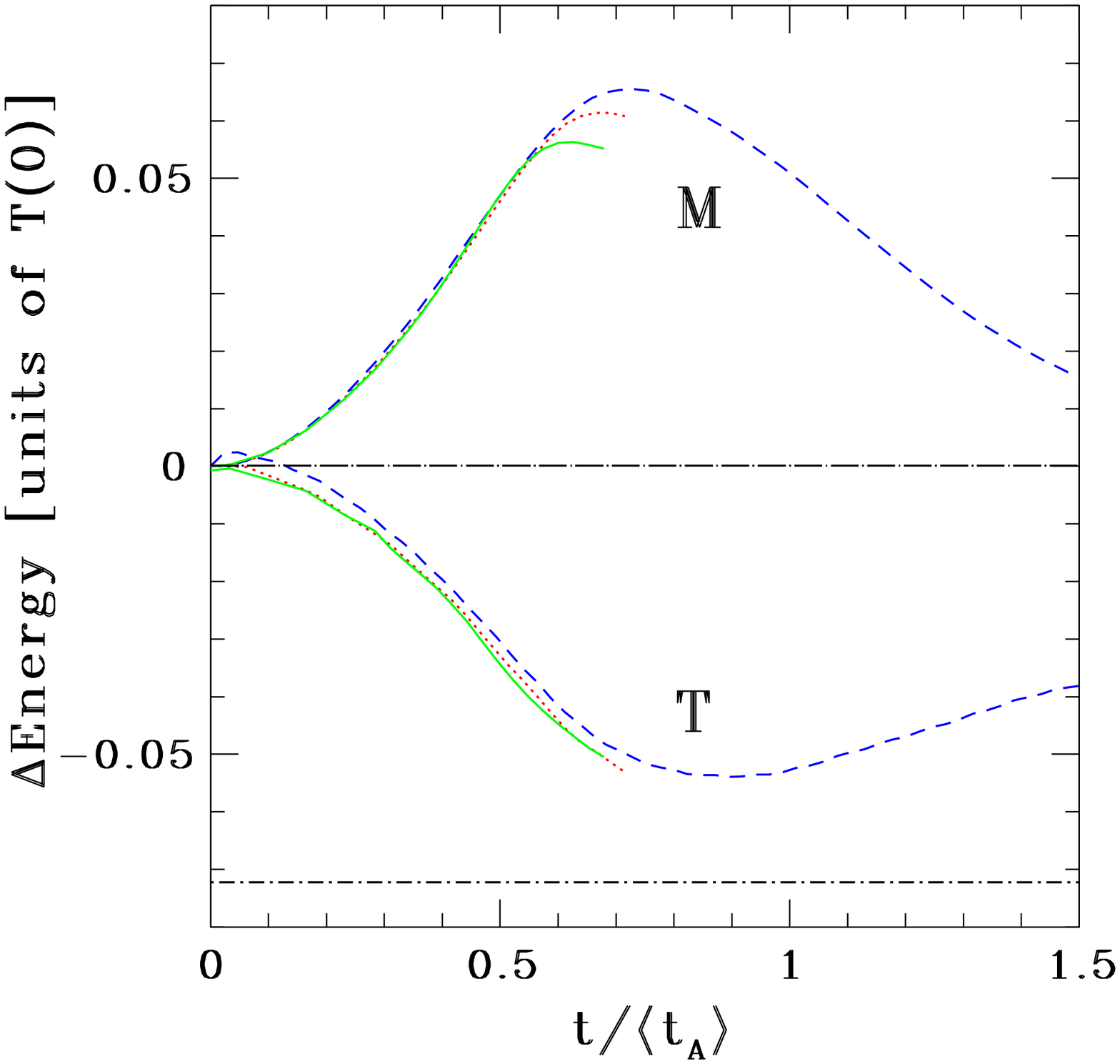}
    \caption{Rotational kinetic ($T$) and magnetic ($\mathcal{M}$) 
      energies vs.~time. 
      For a given energy $E$, we define $\Delta E = [E(t)-E(0)]/T(0)$.  
      The left plot compares results from the perturbative (solid red)
      and BSSN (dashed blue) metric algorithms at $R/\Delta=75$ out to
      $t=2 \langle t_A \rangle$, and the right plot shows the same at
      $R/\Delta=100$ (perturbative: dotted red, BSSN: dashed blue)
      and $250$ (perturbative only: solid green).  The rotational
      kinetic energy of rigid body rotation for the given $J$ is
      plotted at the bottom of each graph. The data for the
      perturbative runs have been smoothed to remove unphysical
      oscillations (see the text for details).}
  \label{MRIresstudy5}
  \end{center}
\end{figure}

As the magnetic fields are wound, the magnetic energy $\mathcal{M}$
saps kinetic energy $T$ associated with differential rotation
(Fig.~\ref{MRIresstudy5}) until the star rotates nearly as a solid
body.  At this point, we find that the star's kinetic energy sinks to
its minimum value shortly after $\mathcal{M}$ reaches maximum.  Note
that, consistent with Fig.~\ref{MRIresstudy1}, the maximum of
$\mathcal{M}$ occurs earlier and earlier as resolution is improved due
to an interplay between MRI and magnetic braking.  
After $\mathcal{M}$ reaches maximum, the fields unwind, pumping energy
back into differential rotation, as shown at lowest resolutions in
Fig.~\ref{MRIresstudy5}.  We speculate, based on the $\alpha$-disk 
model~\cite{MRIrev} and on our previous
work~\cite{StusFirst,JamesCookPaper,MHDLett,LiuShap}, that the
oscillations of $T$ and $\mathcal{M}$ will continue for many Alfv\'en
times until the rotational kinetic energy associated with differential
rotation is dissipated by phase mixing caused by MRI-induced
turbulence. However,
since the star is slowly rotating ($T/|W|=4.88\times 10^{-3}$) with
weak magnetic fields ($\mathcal{M} \ll T$), the Alfv\'en time is long,
$\langle t_A \rangle = 10.2P_c = 4800M$. It is therefore
computationally taxing to accurately evolve the star for many Alfv\'en
times, even if the perturbative metric solver is used.

As discussed in Section~\ref{sec:ID}, perturbative metric solver
initial data are only accurate to order $\Omega$. This causes
oscillations to arise in our perturbative $T$ data at the level of
$\Delta T\approx0.0065$, where $\Delta T = [T-T(0)]/T(0)$ is the
fractional deviation of the rotational kinetic energy from its initial
value. The oscillations are evident in raw Fig.~\ref{MRIresstudy5}
perturbative data.  A simple, local (in time) averaging technique is
used to smooth out these oscillations in Fig.~\ref{MRIresstudy5}.
Although they are smaller by an order of magnitude, we remove the
oscillations in our perturbative $\mathcal{M}$ data as well, 
for consistency.  Notice also in Fig.~\ref{MRIresstudy5} that the
value of $\mathcal{M}_{\text{max}}$ is comparable to that derived in
Eq.~(\ref{Bmaxest}).  

\begin{figure} 
  \begin{center}
    \epsfxsize=2.8in
    \leavevmode
    \epsffile{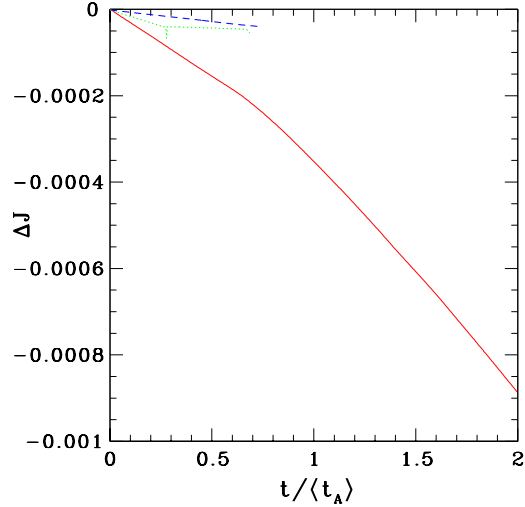}
    \caption{Relative change in angular momentum $J$ 
      ($\Delta J = [J-J(0)]/J(0)$) vs. time at
      resolutions $R/\Delta=75$ (solid red), $150$ (dashed blue), and
      $250$ (dotted green).  Recall that the $R/\Delta=250$ data is
      from a regridding run, where the data before $t/\langle t_A
      \rangle\approx 0.3$ is at $R/\Delta=100$, and $250$ after.  This
      explains the varying slope in the $R/\Delta=250$ data.  We plot
      results from the perturbative scheme only; the nonperturbative
      technique conserves $J$ to the same degree.}
  \label{MRIresstudy7}
  \end{center}
\end{figure}

Figure~\ref{MRIresstudy7} demonstrates that the angular momentum $J$
in our long-term simulations is well-conserved out to $2 \langle
t_A\rangle$.  We see that angular momentum is lost at nearly a
constant rate in our perturbative simulations, but the loss decreases
with increasing resolution.  In addition to angular momentum
conservation, the binding energy $M_0 - M_{\text{ADM}}$ is conserved
to within $\approx 0.5\%$ in perturbative and $\approx 2.5\%$ in BSSN
simulations. 

\subsection{$B$ Variation Study}
In this study, the grid resolution is fixed at $R/\Delta = 75$, and
only the strength of the initial magnetic field is varied. We perform 
both perturbative and BSSN simulations. The MRI wavelengths in 
these simulation are $\langle \lambda_{\rm MRI}\rangle/\Delta =$3.6,
7.2, and 12.2.  The 
corresponding magnetic field strength parameters are given in
Table~\ref{tab:Bstrengths}.  Note that the last case is the same as
the lowest resolution case in the ``MRI Resolution Study''. 

\begin{figure}
  \begin{center}
    \epsfxsize=2.8in
    \leavevmode
    \hspace{-0.7cm}\epsffile{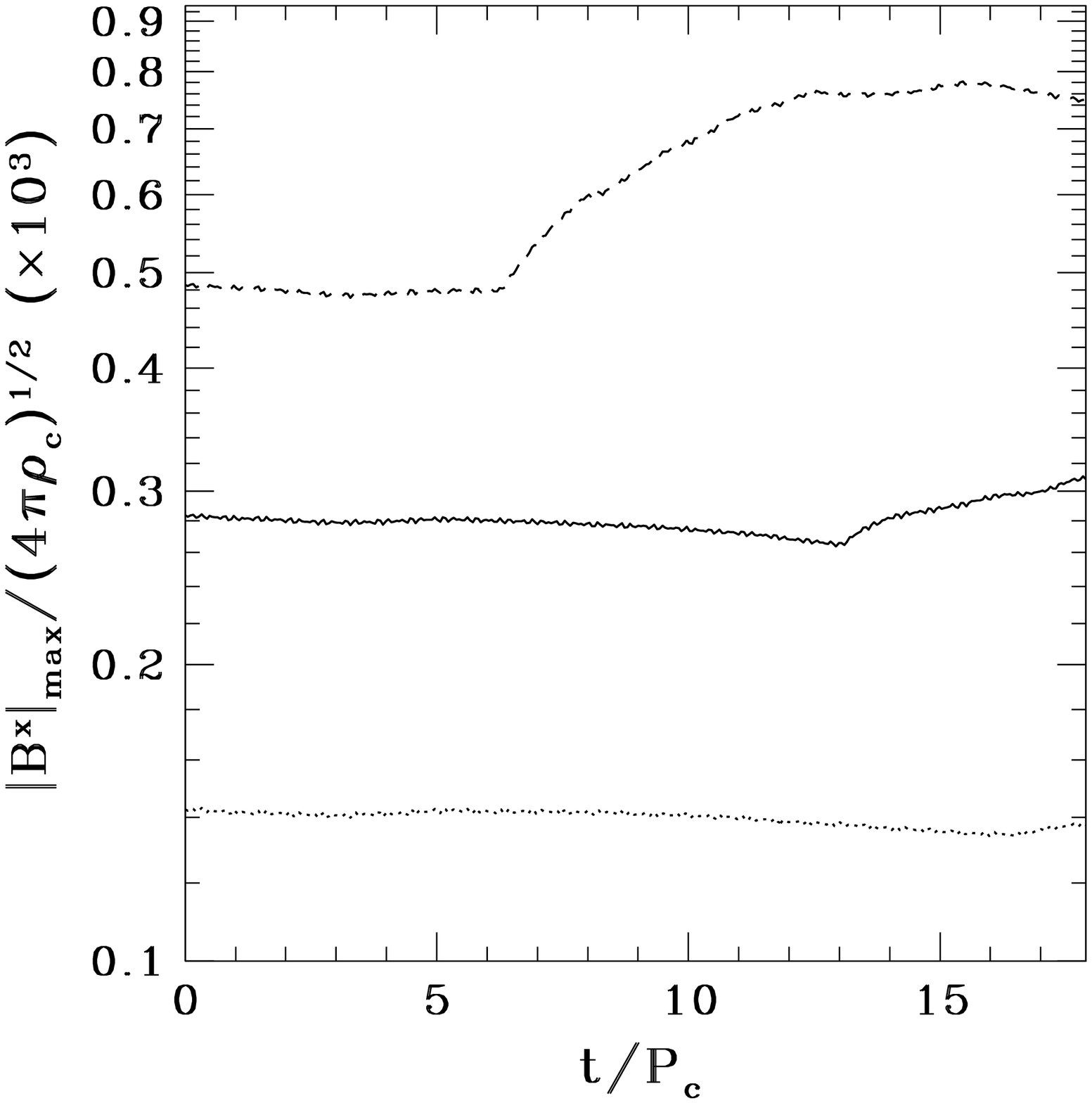}
    \epsfxsize=2.8in
    \leavevmode
    \hspace{-0.2cm}\epsffile{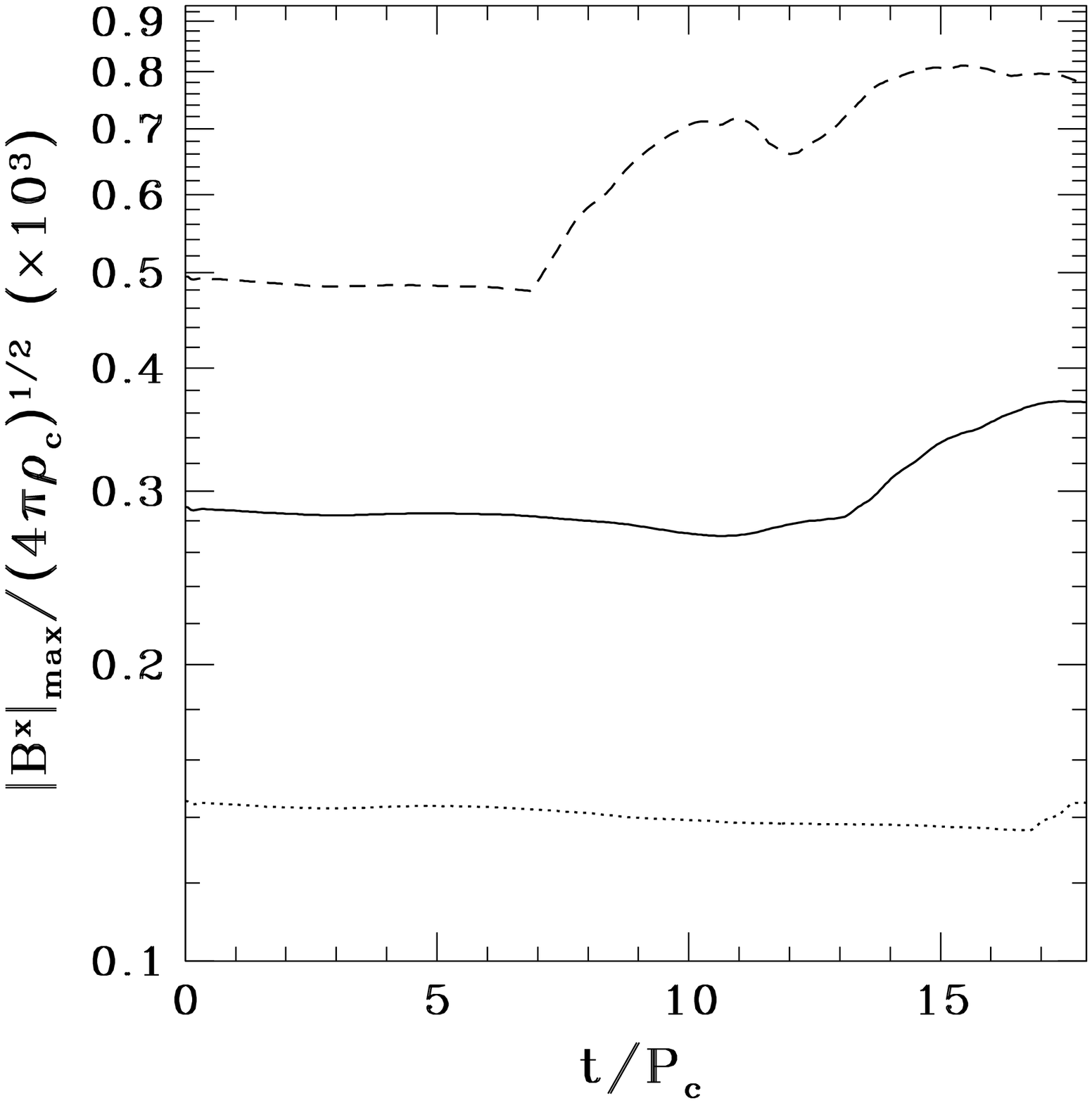}
    \caption{$|B^x|_{\text{max}}$ vs.~time, with different initial
      magnetic field strengths, so that
      $\bkt{\lambda_{\text{MRI}}}/\Delta = 4.0$ (bottom line), 8.0
      (middle line), and 12.2 (top line).  Resolution is fixed at
      $R/\Delta=75$. Data from the perturbative spacetime evolution
      method are shown in the left plot, and BSSN method in the
      right.}
  \label{varyingBstudy1}
  \end{center}
\end{figure}

In Fig.~\ref{varyingBstudy1}, we plot $|B^x(t)|_{\text{max}}$ for 
the three magnetic field strengths.  Although
$\bkt{\lambda_{\text{MRI}}}$ depends on initial magnetic field
strength (Eq.~(\ref{lambMRI2})) , the MRI $e$-folding timescale
$\tau_{\text{MRI}}$ does not (see, e.g.~Eq.~(\ref{tauMRI})).  MRI is
observed only when $\bkt{\lambda_{\text{MRI}}}/\Delta \gtrsim 12$,
which is consistent with the results of~\cite{MHDLett,BigPaper}.

\subsection{Rotation Profile Study}

In this study, we consider a differentially rotating star in which
$\Omega$ increases with cylindrical radius. We expect to see magnetic
braking but not MRI in simulations of this star.

\begin{figure}
  \begin{center}
    \epsfxsize=2.8in
    \leavevmode
    \hspace{-0.7cm}\epsffile{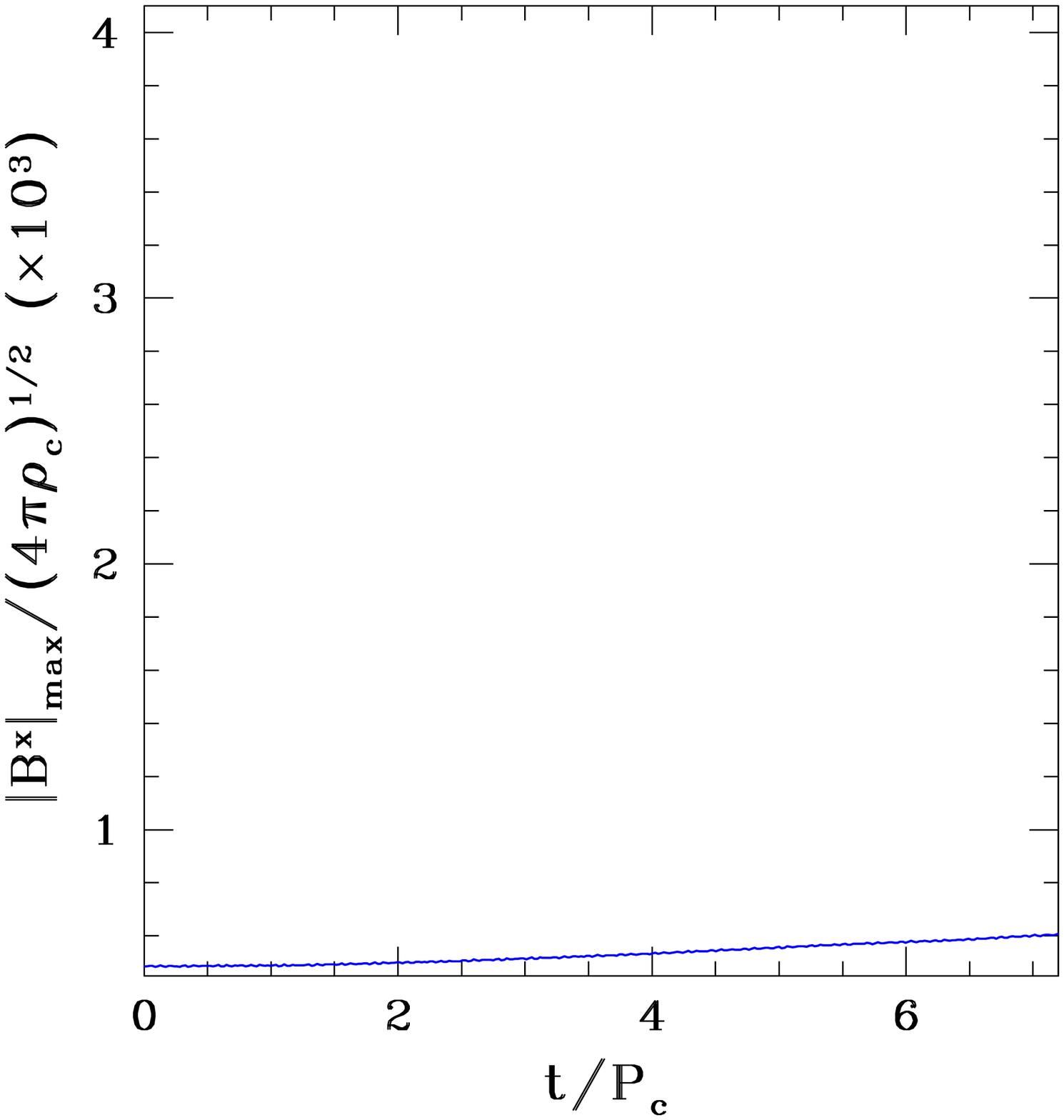}
    \epsfxsize=2.8in
    \leavevmode
    \hspace{-0.7cm}\epsffile{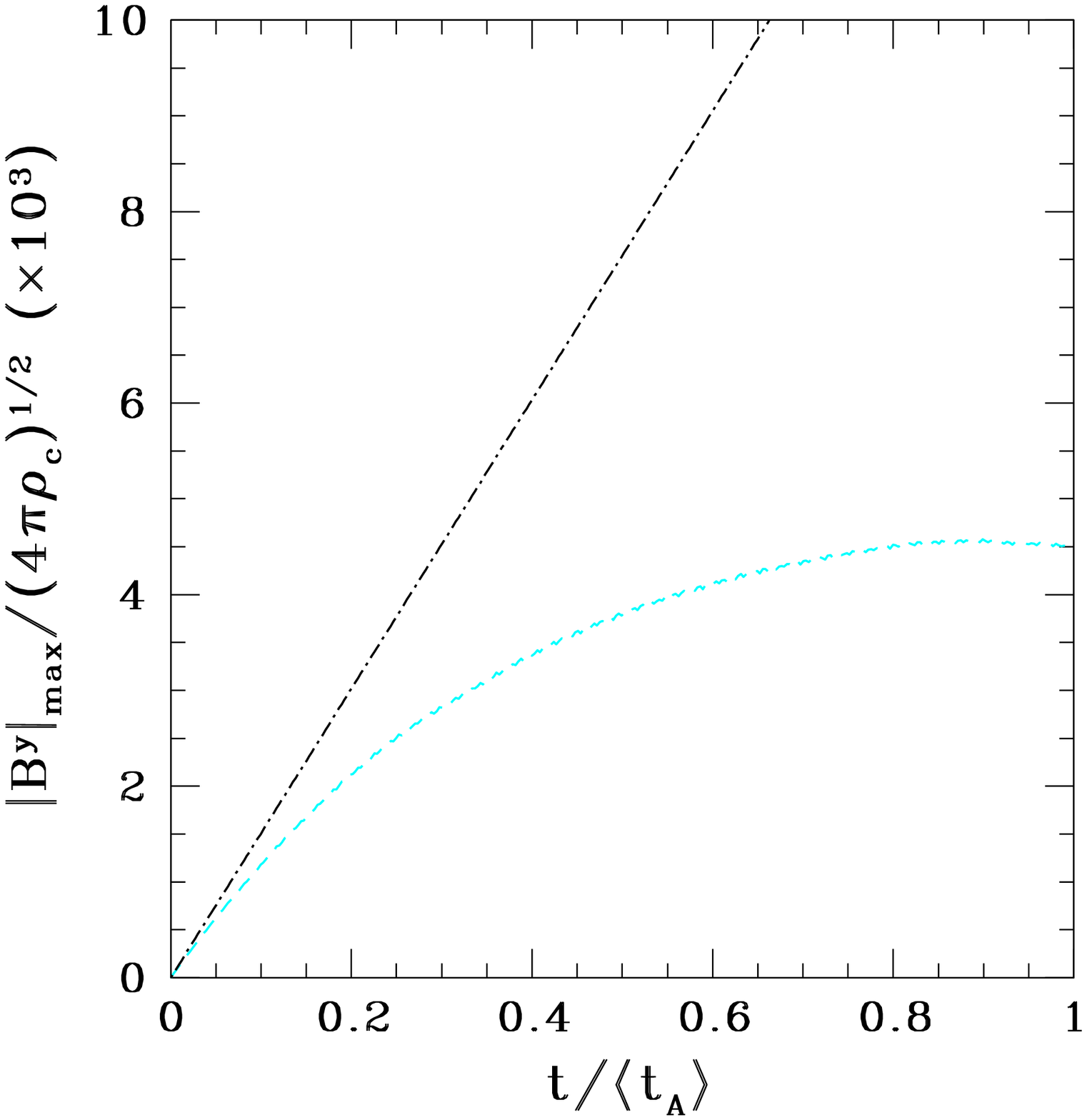} \\
    \leavevmode
    \epsfxsize=1.4in
    \leavevmode
    \hspace{-0.3cm}\epsffile{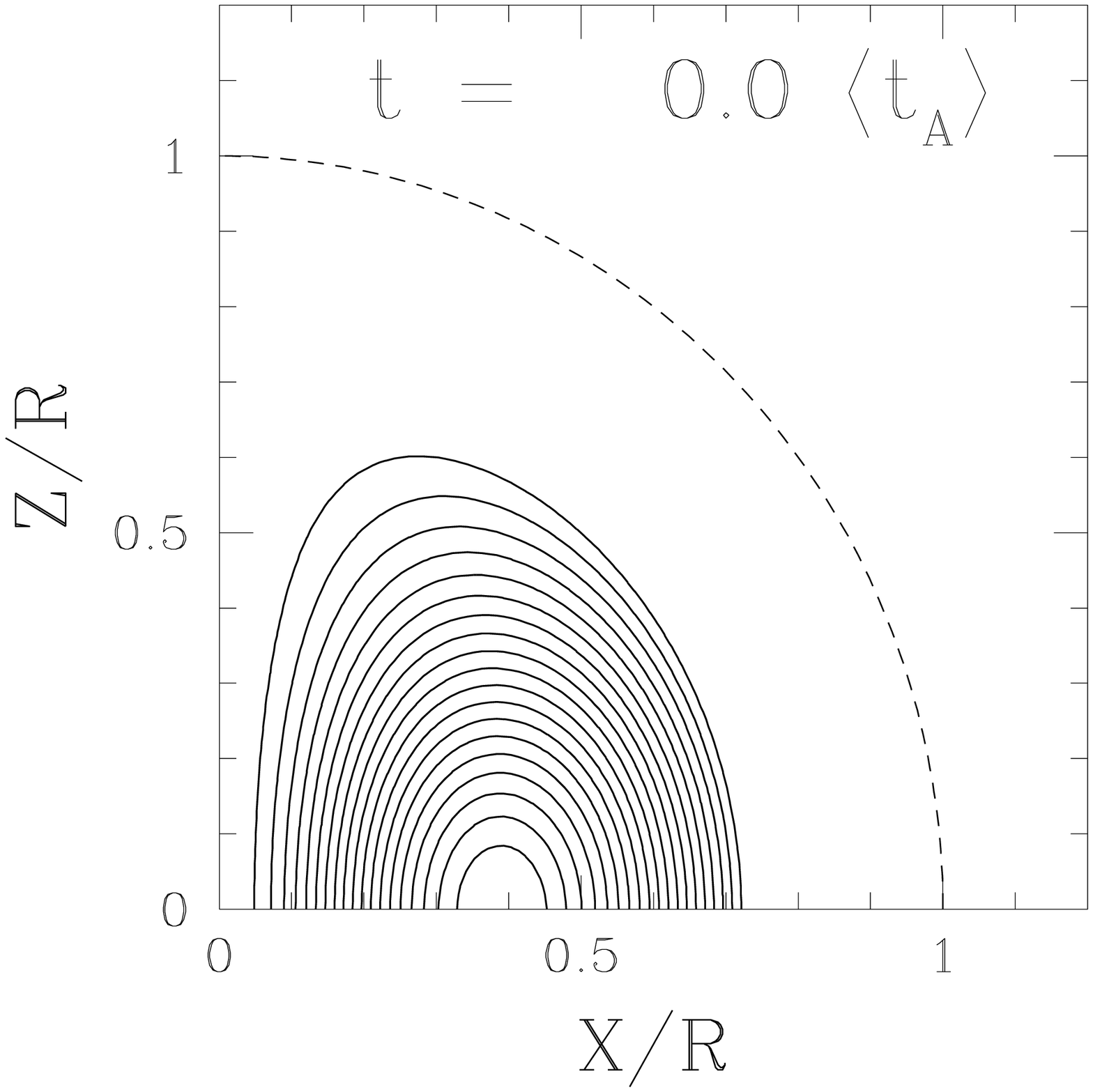}
    \epsfxsize=1.4in
    \leavevmode
    \hspace{-0.3cm}\epsffile{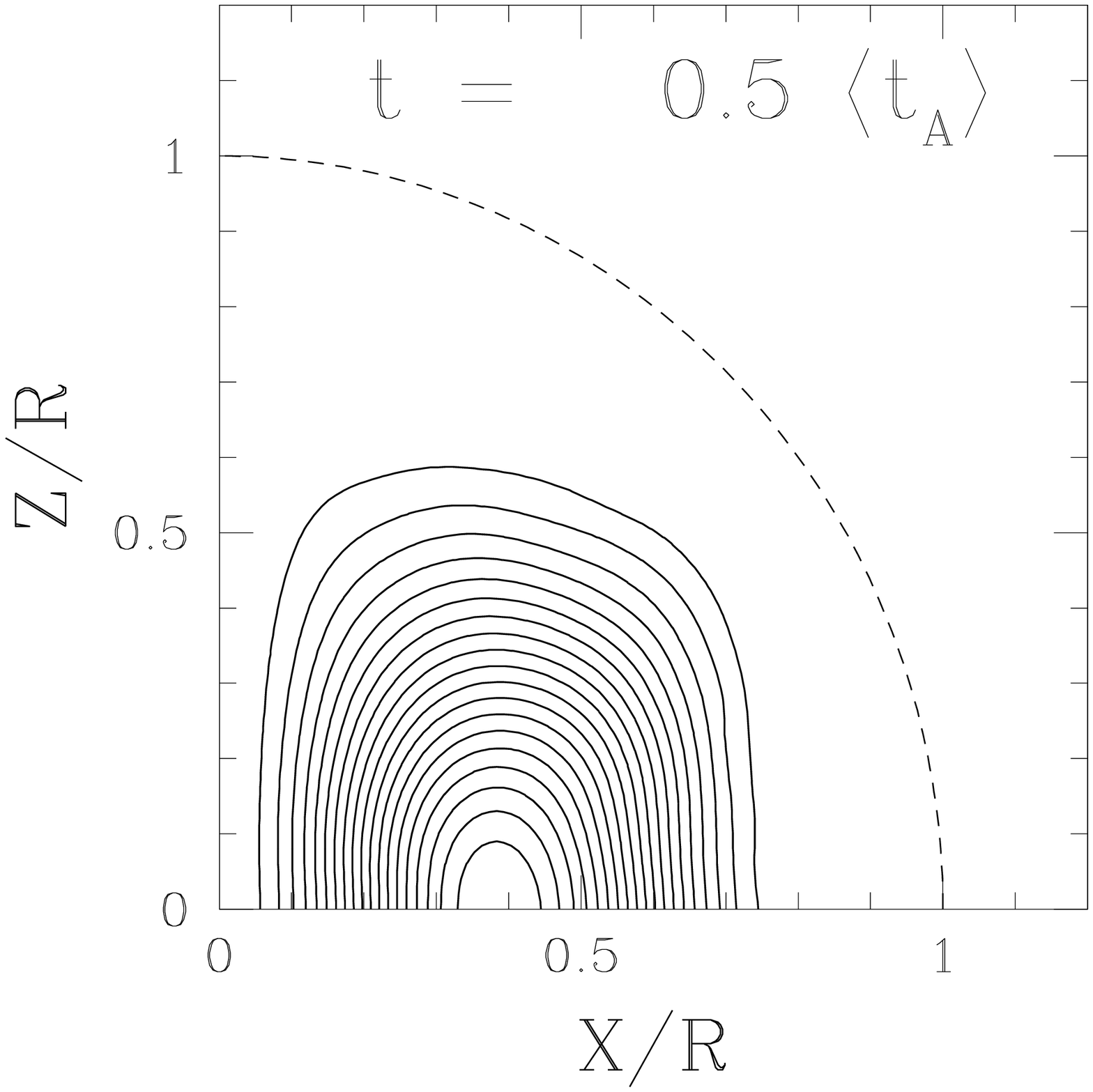}
    \epsfxsize=1.4in
    \leavevmode
    \hspace{-0.3cm}\epsffile{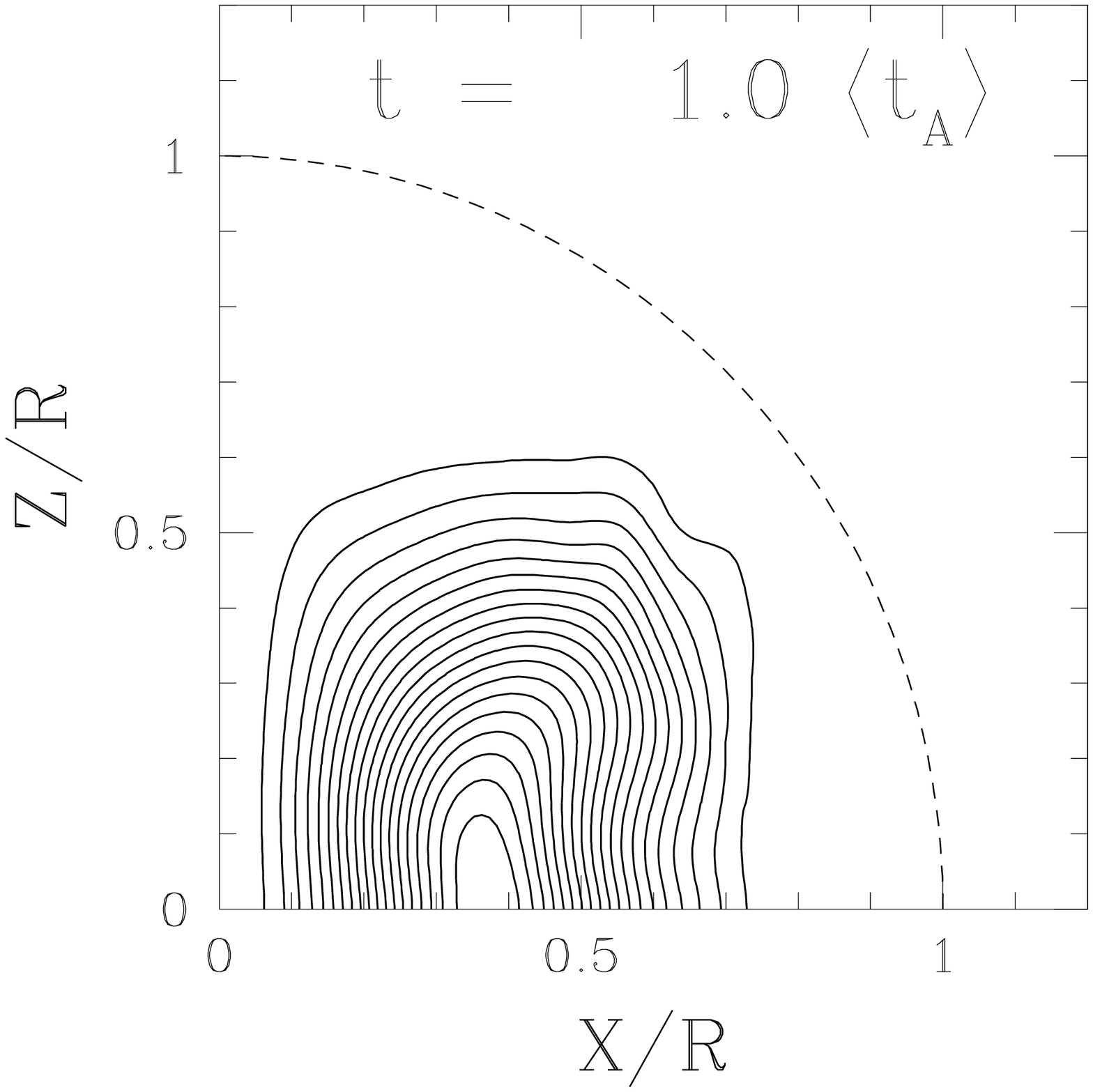} \\
    \epsfxsize=2.8in
    \leavevmode
    \hspace{-0.7cm}\epsffile{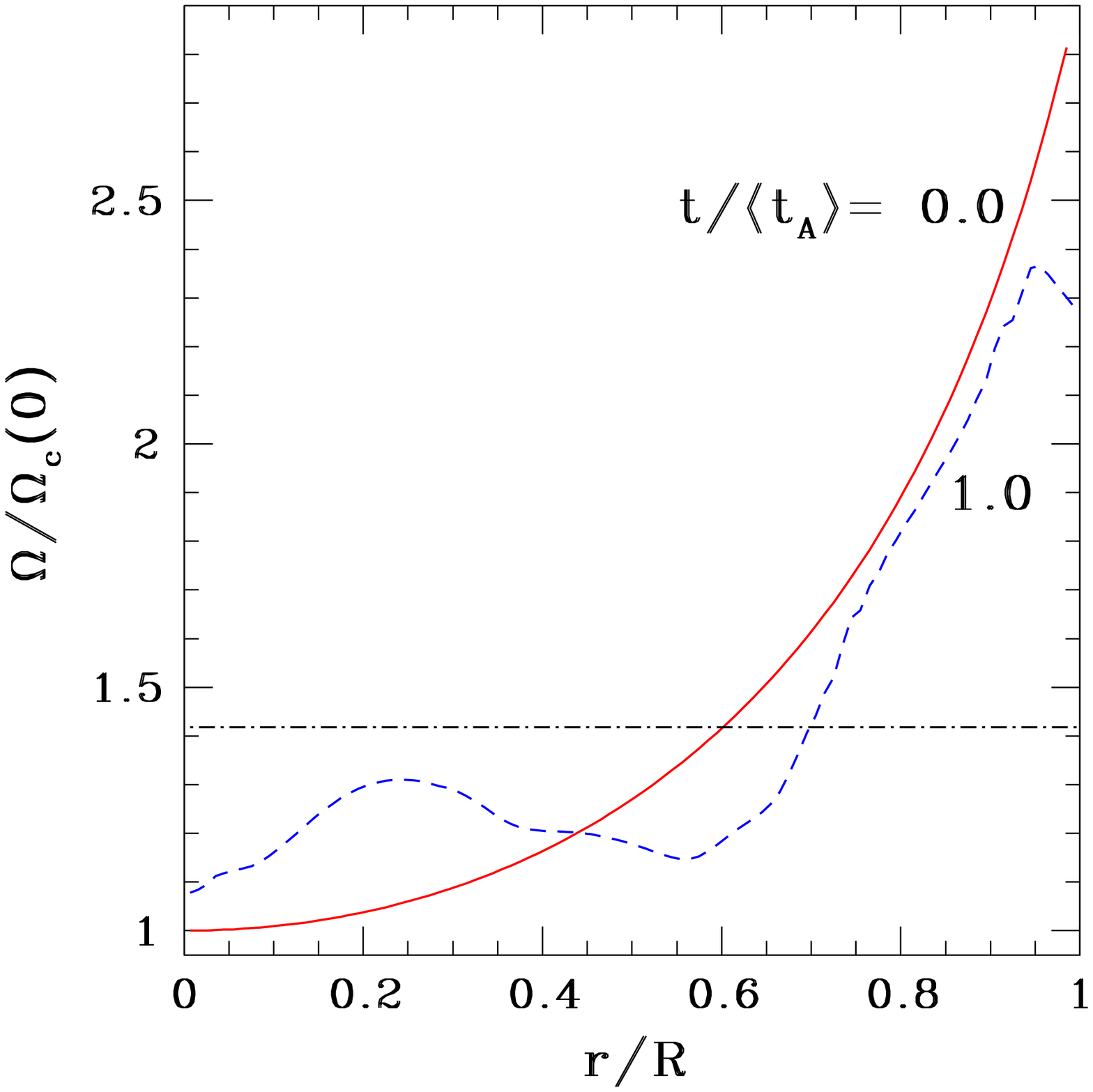}
    \caption{Results from the Rotation Profile Study.  The
    plots from top left to bottom are analogues to 
Figure~\ref{MRIresstudy1}--\ref{MRIresstudy4}.  Top left: $|B^x|_{\rm max}$ 
    vs.\ time, top right: $|B^y|_{\rm max}$ vs.\ time, middle three plots: 
     magnetic field lines
    (contours of $A_{\phi}$), bottom: equatorial rotation profile.} 
  \label{incrotstudy1}
  \end{center}
\end{figure}

Figure~\ref{incrotstudy1} presents the same plots as those in
Figs.~\ref{MRIresstudy1}--\ref{MRIresstudy4} for this star.
The top left plot indicates that, as expected, MRI is absent from
this simulation. Although magnetic braking does appear (top right) as
predicted, the curve is nearly parabolic and does not exhibit a 
sudden increase in slope as observed in Fig.~\ref{MRIresstudy2}. 
This further supports the notion that the sudden increase in slope in
Fig.~\ref{MRIresstudy2} results from an interplay between magnetic
braking and MRI.  Note also that the MRI-induced magnetic field
distortion does not appear in the poloidal plane (middle three
plots). Finally, we see that since the magnetic field-shifting effects
of MRI are absent and the magnetic field is initially confined to the
high density inner region of the star (Fig.~\ref{MRIresstudy2}),
magnetic braking is incapable of flattening the rotation profile
(bottom plot) in the less dense outer layers of the star.

\section{Conclusions}
Our perturbative metric evolution algorithm yields results quite
similar to those produced by the nonperturbative, BSSN-based evolution
scheme.  Because the outer boundary may be moved inward and the metric
updated on a physically relevant timescale, perturbative metric
simulations may be performed at $\sim 1/4$ the computational cost of
BSSN metric simulations in axisymmetry for slowly rotating, weakly
magnetized equilibrium stars.  However, we found that loss of accuracy
did occur after magnetic fields hit the outer boundary. This problem
could be efficiently solved in future work by extending the outer
boundary further from the star. 

The stars we study in this paper are weakly magnetized
($\mathcal{M}/|W| \sim 10^{-6}$) and slowly rotating ($T/|W| \approx
0.005$).  As a result, the Alfv\'en timescale is very long ($\sim
5000M$), and accurate simulations spanning many Alfv\'en times would
be prohibitively expensive. In this paper, we reliably evolved such
stars out to two Alfv\'en times at moderate resolution ($R/\Delta =
75$).  In these simulations, we found that magnetic fields wind and
unwind on the Alfv\'en timescale, resulting in a trade-off between the
kinetic energy in differential rotation and magnetic energy.  
Since we could not perform simulations spanning many Alfv\'en
timescales while accurately resolving MRI, we can only speculate,
based on the $\alpha$-disk model and our previous
work~\cite{StusFirst,JamesCookPaper,MHDLett,LiuShap}, that the
oscillations between the magnetic and kinetic energy will be damped by
MRI-induced dissipative processes over many Alfv\'en times.

In order to observe MRI in our simulations, we find that the spatial
resolution must be set so that $\lambda_{\text{MRI}}/\Delta \gtrsim
10$, in agreement with the results of~\cite{MHDLett,BigPaper}.  We
have also verified that MRI is not present if the star's angular
velocity initially increases with increasing distance from the
rotation axis (i.e.,~the MRI is absent when $\partial_{\varpi} \Omega
>0$).

We find that as resolution is increased, the effects of MRI become
more and more prominent.  Due to the turbulent nature of MRI, we do
not achieve convergence of field amplitude, even if
$\lambda_{\text{MRI}}/\Delta= 32.6$ or $40.7$.  However, we found the
$e$-folding time of MRI, $\tau_{\rm MRI}$ does converge. The
numerically determined value, $\tau_{\rm MRI} \approx  5.0P_c$ is
consistent with the value predicted by the linearized, local Newtonian
analysis ($5.72P_c$). The small difference is due in part to the fact that
our star is relativistic. In addition, the linearized analysis assumes
that $\lambda_{\rm MRI}$ is much smaller than the length scale on
which the magnetic field changes (i.e.\ $\lambda_{\rm MRI} \ll
B/|\nabla B|$), which is not quite satisfied in our magnetic field
configuration. 

Finally, we note that the behavior of the MRI is expected to be 
different in a full 3D calculation because of the effect of
nonaxisymmetric MRI induced by a toroidal magnetic field.
Turbulence may arise and persist more readily in 3D due
to the lack of symmetry.  More specifically, according to the
axisymmetric anti-dynamo theorem~\cite{hgb95,moffatt78}, sustained
growth of the magnetic field energy is not possible through
axisymmetric turbulence.  Thus proper treatment of MRI in
differentially rotating neutron stars requires high resolution
simulations performed in full 3+1 dimensions. The computational cost
of such simulations with existing 3+1 metric evolution schemes has
thus far been prohibitive, but with our perturbative metric solver it
may be possible to perform 3+1 simulations of weakly but realistically
magnetized, slowly rotating stars at a small fraction of the
computational cost. 

\acknowledgments
We gratefully acknowledge useful conversations with C.~Gammie,
M.~Shibata, and B.~Stephens.  Computations presented here were
performed at the National Center for Supercomputing Applications at
the University of Illinois at Urbana-Champaign (UIUC) and on a UIUC
Department of Physics Beowulf cluster with 26 Intel Xeon processors,
each running at 2.4GHz. This work was in part supported by NSF Grants
PHY-0205155 and PHY-0345151, and NASA Grants NNG04GK54G. 

\appendix
\section{Derivation of Shift Equation (Eq.~\ref{finalpde})}
\label{app:shiftEq}
Here we derive the equation for the shift (Eq.~\ref{finalpde}),
starting from the usual momentum constraint equation in the 3+1 ADM
decomposition of Einstein's field equations.  Recall from
Eq.~(\ref{metric}) the perturbative line element is given by 
\beq
ds^2 = -e^{\nu(r)} dt^2 + e^{\lambda(r)} dr^2 + 2
\beta^{\phi}(t,r,\theta) r^2 \sin^2{\theta} dt d\phi + r^2
(\sin^2\theta d\phi^2 + d\theta^2) + \mathcal{O}(\Omega^2) +
\mathcal{O}(B^2). 
\label{app:metric}
\eeq
We begin by writing the momentum constraint equation (Eq.~24 in
\cite{ADM})
\beq
D_j K^j{}_i - D_i K = 8\pi j_i,
\label{mom_const}
\eeq
where $K=K^j{}_j$ and $j_i = -\gamma^b{}_i n_c T^c{}_b$.  
It can easily be shown from the ADM 3-metric evolution equation
(Eq.~35 in \cite{ADM}) $\partial_t \gamma_{ij} = -2e^{\nu/2} K_{ij} +
\mathcal{L}_{\beta} \gamma_{ij}$ that in this approximation $K_{ij}$
is given by
\beq
0 = \partial_t \gamma_{ij} = -2e^{\nu/2} K_{ij} + \beta_{i|j} +
\beta_{j|i}
\label{3metricevol}
\eeq
where $\beta_{i|j}$ denotes the spatial covariant derivative $D_j
\beta^i$, and the first equality reflects the stationarity of the
3-metric in this approximation.  Taking the trace of this equation and
applying the identity $\beta^j{}_{|j} = \frac{1}{\sqrt{\gamma}}
(\sqrt{\gamma} \beta^j)_{,j}$ (where $\gamma=\det{\gamma_{ij}}$)
yields an expression for $K$: 
\beq
  -2 e^{\nu/2} K +
  \frac{2}{\sqrt{\gamma}} (\sqrt{\gamma} \beta^j)_{,j} = 0 .
\eeq
Note that $(\sqrt{\gamma} \beta^j)_{,j} = (\sqrt{\gamma}
\beta^{\phi})_{,\phi} =0$ by axisymmetry, so Eq.~({\ref{mom_const}}) may
be rewritten 
\beq
  K^j{}_{i | j} = 8 \pi j_i.
  \label{mom_const_simp}
\eeq
It follows from the definition of $j_i$ and the metric~(\ref{app:metric}) 
that $j_i = e^{\nu/2} T^0{}_i$. We split the stress-energy tensor 
$T_{\mu \nu}$, as well as $j_i$, into 
a fluid part and an electromagnetic part: 
\beqn
  T_{\mu \nu} &=& T^{\rm F}_{\mu \nu} + T^{\rm EM}_{\mu \nu} \ , \\
  T^{\rm F}_{\mu \nu} &\equiv& \rho_0 h u_{\mu}u_{\nu} + P g_{\mu \nu} \ , \\
  T^{\rm EM}_{\mu \nu} &\equiv& b^2 u_{\mu}u_{\nu} + \frac{b^2}{2} 
g_{\mu \nu} - b_{\mu} b_{\nu} \ , \\ 
  j^{\rm F}_i &=& e^{\nu/2} T^{\rm F\, 0}{}_i \ , \ \ \ \ \ \ \ 
  j^{\rm EM}_i= e^{\nu/2} T^{\rm EM\, 0}{}_i \ .
\eeqn
Our approximation as summarized in Table~\ref{tab:slow_rot_assump} and 
derived in Appendix~\ref{app:slow-rot-der} ensures that $j^{\rm F}_{\phi}$ 
is the dominant component of $j_i$. In this approximation, 
$v^{i} = (0,0,\Omega)$ and 
\beq
j_{\phi} = \rho_o h e^{-\nu/2} (\Omega + \beta^{\phi}) r^2
\sin^2\theta.
\label{jphi}
\eeq
Thus $i=\phi$ remains the only nonzero component of
Eq.~({\ref{mom_const_simp}}) to first order in $\Omega$ and $B$:
\beq
K^j{}_{\phi | j} = 8 \pi \rho_o h e^{-\nu/2} (\Omega + \beta^{\phi}) r^2
\sin^2\theta  \ .
\label{pde_middlestep}
\eeq
Next we expand our expression for $K^j{}_{\phi | j}$:
\beqn
K^j{}_{\phi | j} &=& \frac{1}{\sqrt{\gamma}} (\sqrt{\gamma}\,
K^j{}_\phi)_{,j} - \Gamma^k{}_{\phi m} K^m{}_k \\
&=& \frac{1}{\sqrt{\gamma}} (\sqrt{\gamma}\,
r^2 \sin^2 \theta K^{j\phi})_{,j} - \Gamma^k{}_{\phi m} K^{ml}
\gamma_{kl}.
\label{Kij}
\eeqn
It follows from Eq.~(\ref{3metricevol}) and the identity
$\beta^{i}{}_{|j}=\beta^i{}_{,j} + \Gamma^i{}_{jk} \beta^k$ that
\beq
2e^{\nu/2} K^{ij} = (\beta^i{}_{,l} + \Gamma^i{}_{lk} \beta^k) \gamma^{jl} +
(\beta^j{}_{,l} + \Gamma^j{}_{lk} \beta^k) \gamma^{il}.
\eeq
After computing the necessary Christoffel symbols, we find that 
the nonvanishing components of $K^{ij}$ are:
\beqn
  K^{r\phi} =K^{\phi r}  &=& \frac{1}{2e^{\nu/2}}
  \left(e^{-\lambda} \beta^\phi{}_{,r}\right) \\ 
  K^{\theta \phi} = K^{\phi \theta}  &=& \frac{1}{2e^{\nu/2}}
  \left(\frac{\beta^\phi{}_{,\theta}}{r^2}\right).
\eeqn
Plugging these expressions into Eqs.~({\ref{Kij}}) and then
({\ref{pde_middlestep}}) yields the equation governing the shift:
\beq
\frac{1}{r^4} [ r^4 j(r) \omega_{,r}]_{,r} +
e^{\frac{\lambda-\nu}{2}} \frac{1}{r^2 \sin^3{\theta}} [
  \sin^3{\theta} \omega_{,\theta} ]_{,\theta} +
\frac{4}{r} j'(r) \omega = \frac{4}{r} j'(r) \Omega, 
\label{fin_pde}
\eeq
where $\omega = -\beta^{\phi}$ and $j(r) = e^{-\frac{\lambda(r)+\nu(r)}{2}}$. 

Finally, we perform the angular decompositions as in 
Eq.~({\ref{littomeg_exp}}) and Eq.~({\ref{bigomeg_exp}}) and 
substitute them into
Eq.~({\ref{fin_pde}}) to obtain the equation
\beq
\frac{1}{r^4} \left[r^4 j(r) \omega_l{}'(r)\right]' +
\left[4\frac{j'(r)}{r} - e^{(\lambda-\nu)/2} \frac{l(l+1) -
    2}{r^2}\right] \omega_l(r) = 4 \frac{j'(r)}{r} \Omega_l(r).
\label{finalode2}
\eeq
This equation is the same as Eq.~(30) in~\cite{Hartle2}, which
was derived using a different approach. Each $\omega_l$ must also satisfy 
the boundary conditions at the origin and at infinity (see below), 
as well as the matching conditions [Eqs.~(\ref{match1})
and (\ref{match2})] at the surface of the star.

\subsection{Boundary condition at the origin}
We now determine the boundary condition at the origin by analyzing the
terms of Eq.~({\ref{finalode2}}) in the $r\rightarrow 0$ limit.  

First we analyze $j(r)$ and its derivatives, starting with
$j'(0) = -\frac{1}{2} j(0) [\lambda'(0) + \nu'(0)]$.
From the OV equations, we obtain 
\beqn
\lambda'(0) &=& \left.\frac{2[r m'(r) -
  m(r)]}{(r-2m(r))^2}\right|_{r\to0} = 0 \\
\nu'(0) &=& \left.\frac{2[m(r) + 4\pi r^3 P(r)]} {r[r-2 m(r)]}
\right|_{r\to0} = 0,
\eeqn
where we have used the expression $m(r) = (4/3) \pi r^3 \rho(0)$ as 
$r\to 0$.
Thus $j'(0) = 0$ since $j(r) = e^{-\frac{\lambda+\nu}{2}}$ is regular
at $r=0$.  Next we examine $j'(r)/r$:
\beqn
j'(r)/r|_{r\to 0} &=& j''(0) = (e^{-\frac{\lambda+\nu}{2}})''|_{r\to
  0} \\ 
&=& -\frac{1}{2} j(0) (\lambda''(r) + \nu''(r))|_{r\to0} \\
&=& -\frac{1}{2} j(0) \left.\frac{d}{dr}\left \{ \frac{2[r m'(r) -
  m(r)]}{[r-2m(r)]^2} + \frac{2[m(r) + 4\pi r^3 P(r)]}{r[r-2
      m(r)]} \right \}\right|_{r\to0} \\
&=& -\frac{4}{3} \pi j(0) [4 \rho(0) + 3 P(0)].
\eeqn

We may therefore write Eq.~({\ref{finalode2}}) near
$r\to0$ as follows
\beq
\omega_l''(r) +\frac{4}{r} \omega_l'(r) + 4
\omega_l(r) \frac{j''(0)}{j(0)} -
\frac{l(l+1)-2}{r^2}\omega_l(r) = 4 \Omega_l(0)
\frac{j''(0)}{j(0)}, 
\label{zeroode}
\eeq
with solution given by Eq.~({\ref{rtozero_soln}}). 

\subsection{Boundary condition at infinity}
Outside the star, $P=\rho=0$ and the time independent
(diagonal) metric becomes Schwarzschild, so $e^{\nu} = e^{-\lambda} =
1-2M/r$.  The ODE governing the shift outside the star is therefore
given by
\beq
\frac{1}{r^4}[r^4 \omega_l{}'(r)]' - \frac{1}{1-2M/r}
\frac{l(l+1)-2}{r^2} \omega_l(r)=0.
\label{outereq}
\eeq
Since $2M/r \ll 1$ in the limit $r \to \infty$, we may write
Eq.~(\ref{finalode2}) as 
\beq
\frac{1}{r^4} \frac{d}{dr} (r^4 \frac{d\omega_l}{dr}) -
\frac{l(l+1) - 2}{r^2} \omega_l = 0
\eeq
with solution given by Eq.~({\ref{rtoinf_soln}}).  

Note that the analytic solution for Eq.~(\ref{outereq}) exists and 
is given in terms of the hypergeometric function:
\beq
\omega_l= 
\begin{cases}
 \displaystyle \frac{C_1}{r^3}, & \text{if } l=1, \text{and}\cr \cr 
 \displaystyle \frac{C_l}{r^{l+2}} {_2F_1}(l+2,l-1;2l+2;2M/r), &
 \text{otherwise}. 
\end{cases}
\eeq

\subsection{Rigid rotation case}
In the case of solid body rotation ($\Omega(r,\theta)= 
\text{constant}$), only the $l=1$ mode in Eq.~\ref{finalode}
contributes to $\Omega$.  Thus the right-hand
side of Eq.~(\ref{finalode2}) is zero for $l > 1$.  For
$l>1$, the solution $\omega_l = 0$ satisfies the boundary
conditions at the origin and at infinity [Eqs.~(\ref{rtozero_soln}) 
and (\ref{rtoinf_soln})] and the matching conditions at the stars's
surface [Eqs.~(\ref{match1}) and (\ref{match2})], so $\omega_l = 0$ is
the solution for $l>1$.  This coincides with the result cited
in~\cite{Hartle}.  

\section{Derivation of Slow Rotation Approximation
  Inequalities}
\label{app:slow-rot-der}
In this section, we derive inequalities that must hold in order for
our primary assumption in Appendix~\ref{app:shiftEq} [leading to 
Eq.~({\ref{jphi}})] to be valid. 

We have assumed that the metric can be written in the form~(\ref{metric}) 
at all times, with the shift $\beta^{\phi} = -\omega$ being the only 
non-diagonal component of the metric.
For this to be true, the system has to be (approximately) stationary, 
axisymmetric and the 
stress-energy tensor has to be {\it circular} or 
{\it nonconvective}~\cite{circular}: 
\beqn
  \epsilon_{\alpha} T^{\alpha [\beta} \epsilon^{\gamma} \xi^{\delta]}
&=&0 \ , \\
  \xi_{\alpha} T^{\alpha [\beta} \epsilon^{\gamma} \xi^{\delta]}
&=&0 \ ,
\eeqn
where $\epsilon=\partial/\partial t$ and $\xi = \partial/\partial \phi$ 
are two Killing vector fields associated with stationarity and 
axisymmetry, respectively. These circularity conditions are satisfied 
if the momentum currents in the meridional planes are negligible 
compared with the axial component. Hence we require that 
\beq
|j_{\hat{\phi}}| \gg |j_{\hat{r}}| \ \ \ \ \mbox{and} \ \ \ \ 
|j_{\hat{\phi}}| \gg |j_{\hat{\theta}}| \ ,
\label{cond:circular}
\eeq
where the ``hats'' denote the orthonormal components.

We split the stress-energy tensor into a fluid part and an
electromagnetic part: 
$T^{\mu \nu} = T^{\mu \nu}_{\rm F} + T^{\mu \nu}_{\rm EM}$, where 
\beqn
  T^{\mu \nu}_{\rm F} &=& \rho_0 h u^{\mu} u^{\nu} + Pg^{\mu \nu} \ , \\ 
  T^{\mu \nu}_{\rm EM} &=& b^2 u^{\mu} u^{\nu} + \frac{b^2}{2}g^{\mu \nu} 
  - b^{\mu} b^{\nu} \ .
\eeqn
From this we obtain the following expressions for
$T^{0}{}_i{}^\text{F}$ and $T^{0}{}_i{}^\text{\rm EM}$:
\beqn
T^{0}{}_i{}^\text{F} &=& \rho_0 h u^0 u_i \cr 
 &=& \rho_0 h (u^0)^2\gamma_{ij}(v^j+\beta^j)  \cr
 &\sim & \rho_0 h e^{-\nu} \gamma_{ij} v^j \ , \\
T^{0}{}_i{}^\text{EM} &=& b^2 u^0 u_i - b^0 b_i \cr 
 &\sim & e^{-\nu}\, \frac{v^j}{4\pi}(\gamma_{ij} B^2 - B_i B_j) \ ,
\eeqn
where we have used the fact that $v^j+\beta^j \sim v^j$ and 
$\alpha u^0 = 1 + O(\Omega^2)$ for slowly rotating stars. 
Here the lapse $\alpha = e^{\nu/2}$.

Since $j_{i} = e^{\nu /2} T^{0}{}_{i}$, we may split the momentum
current density in the same way: $j_i = j^{\rm F}_i + j^{\rm EM}_i$.
Using the above expression for $T^{0}{}_i^\text{F}$ and the
metric~(\ref{metric}), we find
\beq
j_{i}^\text{F} = e^{\nu /2} T^{0}{}_{i}{}^\text{F} \implies
\begin{cases}
\displaystyle j_{\hat{\phi}}^\text{F} \sim e^{-\nu/2} \rho_0 h r
\Omega \cr 
\displaystyle j_{\hat{\theta}}^\text{F} \sim \left(
\frac{v^{\hat{\theta}}}{\Omega r} \right) j_{\hat{\phi}}^\text{F}\cr
\displaystyle j_{\hat{r}}^\text{F} \sim e^{-\lambda/2} \left(
\frac{v^{\hat{r}}}{\Omega r} \right) j_{\hat{\phi}}^\text{F} 
\end{cases}  \ .
\label{eq:jF_i}
\eeq
In the absence of magnetic fields, $j_i = j_i^{\rm F}$ and the 
conditions~(\ref{cond:circular}) 
yield $|v^{\hat{\theta}}| \ll |\Omega| r$ and $|v^{\hat{r}}| \ll
|\Omega| r$ [Assuming $e^{-\lambda/2} \sim O(1)$]. 
Applying these inequalities, the electromagnetic part of
$j_i$ becomes
\beq
j_i^\text{EM} = e^{\nu/2} T^0{}_i{}^\text{EM} \implies
\begin{cases}
\displaystyle j_{\hat{\phi}}^\text{EM} \sim \frac{B^2-(B^{\hat{\phi}})^2}
{4\pi\rho_0 h}
\displaystyle j_{\hat{\phi}}^\text{F} \cr
\displaystyle j_{\hat{\theta}}^\text{EM} \sim \frac{B^2}{4\pi\rho_0
  h} \left[\frac{v^{\hat{\theta}} }{\Omega
    r}\left(1-\frac{(B^{\hat{\theta}})^2} {B^2}\right) -
  \frac{B^{\hat{\theta}} B^{\hat{\phi}} }{B^2} \right]
\displaystyle j_{\hat{\phi}}^\text{F} \cr
\displaystyle j_{\hat{r}}^\text{EM} \sim \frac{B^2}{4\pi\rho_0 h} 
e^{-\lambda/2} \left[ \frac{v^{\hat{r}}}{\Omega r} \left( 1-e^{-\lambda} 
\frac{(B^{\hat{r}})^2}{B^2}\right) 
- \frac{B^{\hat{r}} B^{\hat{\phi}} }{B^2} \right]
\displaystyle j_{\hat{\phi}}^\text{F}
\end{cases} \ .
\label{eq:jEM_i}
\eeq
For simplicity, we have ignored the magnetic field terms
when computing the shift in Appendix~\ref{app:shiftEq}. For this 
to be valid, we need to impose an additional condition: 
\beq
|j_{\hat{\phi}}^\text{F}| \gg |j_{\hat{\phi}}^\text{EM}| \ .
\label{cond:add}
\eeq

Equations~(\ref{eq:jF_i}) and~(\ref{eq:jEM_i}) together with the 
conditions~(\ref{cond:circular}) and~(\ref{cond:add}) yield 
the following inequalities which must be satisfied for the shift 
equations in Appendix~\ref{app:shiftEq} to be valid:
\begin{align}
  \left| \frac{v^{\hat{r}} }{\Omega r} \right| \ll 1 \ , \ \ \ \ \
  \left| \frac{v^{\hat{\theta}} }{\Omega r}\right| \ll 1 \ , \ \ \ \ \
  \frac{ (B^{\hat{r}})^2}{4\pi \rho_0 h} \ll 1 \ , \ \ \ \ \
  \frac{ (B^{\hat{\theta}})^2}{4\pi \rho_0 h} \ll 1 \ ,\text{ and} \ \ \ \ \
  \frac{ (B^{\hat{\phi}})^2}{4\pi \rho_0 h} \lesssim 1 \ .
\end{align}
For information on how well these inequalities are satisfied in our
simulations, see Table~I.


\begin{thebibliography}{99}
\bibitem{MRI0} 
  V. P. Velikhov, Soc. Phys. JETP {\bf 36}, 995 (1959);
  S. Chandrasekhar, Proc.~Natl.~Acad.~Sci.~USA {\bf 46}, 253 (1960).

\bibitem{MRI} 
  S. A. Balbus and J. F. Hawley, Astrophys.~J. {\bf 376},
  214 (1991).

\bibitem{BinMerg-Rasio}
  F. A. Rasio and S. L. Shapiro, Astrophys.~J. {\bf 432}, 242 (1994);
  Class.~Quant.~Grav. {\bf 16} R1 (1999).
  
\bibitem{BinMerg-ShapShib}
  T. W. Baumgarte, S. L. Shapiro, and M. Shibata, 
  Astrophys.~J.~Lett. {\bf 528}, L29 (2000).
  
\bibitem{BinMerg-Shib}
  M. Shibata and K. Uryu, Phys.~Rev.~D {\bf 61}, 064001 (2000);
  M. Shibata and K. Uryu, Prog.~Theor.~Phys. {\bf 107}, 265 (2002);
  M. Shibata, K. Taniguchi, and K. Uryu, Phys.~Rev.~D {\bf 68},
  084020, (2003); M. Shibata and K. Taniguchi, Phys.~Rev.~D {\bf
  73}, 064027 (2006).

\bibitem{Zwerg}
  T. Zwerger and E. M\"uller, Astron.~Astrophys. {\bf 320}, 209 (1997);
  M. Ruffert and H.-T. Janka, Astron.~Astrophys. {\bf 344}, 573 (1999).

\bibitem{liu01} 
  Y.\ T.\ Liu and L.\ Lindblom, Mon.~Not.~R.~Astron.~Soc. {\bf 324},
  1063 (2001);
  Y.\ T.\ Liu, Phys.~Rev.~D {\bf 65}, 124003 (2002).

\bibitem{StusFirst}
  S. L. Shapiro, Astrophys.~J. {\bf 544}, 397 (2000).

\bibitem{mouschpaleo}
  T. C. Mouschovias and E. V. Paleologou, Astrophys.~J. {\bf 230}, 204
  (1979); T. C. Mouschovias and E. V. Paleologou, Astrophys.~J.  {\bf
  237}, 877 (1980).

\bibitem{JamesCookPaper}
  J. N. Cook, S. L. Shapiro, and B. C. Stephens, Astrophys.~J. {\bf
  599}, 1272 (2003).

\bibitem{LiuShap}
  Y. T. Liu and S. L. Shapiro, Phys.~Rev.~D {\bf 69}, 044009 (2004).

\bibitem{spruit99} 
  H. C. Spruit, Astron.~Astrophys. \textbf{349}, 189 (1999).

\bibitem{BigPaper}
  M. D. Duez, Y. T. Liu, S. L. Shapiro, M. Shibata, and B. C. Stephens,
  Phys.\ Rev.\ D {\bf 73}, 104015 (2006).

\bibitem{MHDLett}
  M. D. Duez, Y. T. Liu, S. L. Shapiro, M. Shibata, and
  B. C. Stephens, Phys.~Rev.~Lett. {\bf 96}, 031101 (2006).

\bibitem{GRB2} M. Shibata, M. D. Duez, Y. T. Liu, S. L. Shapiro, and
  B. C. Stephens, Phys.~Rev.~Lett. {\bf 96}, 031102 (2006).

\bibitem{MHDIntro}
  M. D. Duez, Y. T. Liu, S. L. Shapiro, and B. C. Stephens,
  Phys.~Rev.~D {\bf 72}, 024028 (2005) 

\bibitem{SS} M. Shibata and Y.-I. Sekiguchi,
  Phys.~Rev.~D {\bf 72}, 044014 (2005).
 
\bibitem{Hartle}
  J. B. Hartle, Astrophys.~J. {\bf 150}, 1005 (1967).

\bibitem{Hartle2}
  J. B. Hartle, Astrophys.~J. {\bf 161}, 111 (1970).

\bibitem{BSSN}
  M. Shibata and T. Nakamura, Phys.~Rev.~D {\bf 52}, 5428 (1995);
  T. W. Baumgarte and S. L. Shapiro, Phys.~Rev.~D {\bf 59}, 024007
  (1998). 

\bibitem{MRIrev} 
  S. A. Balbus and J. F. Hawley, Rev.~Mod.~Phys. {\bf 70}, 1 (1998). 

\bibitem{BSS} 
  T. W. Baumgarte, S. L. Shapiro, and M. Shibata,
  Astrophys.~J.~Lett. {\bf 528}, L29 (2000).
 
\bibitem{TOV}
  J. R. Oppenheimer and G. Volkoff, Phys.~Rev. {\bf 55}, 374 (1939).

\bibitem{ADM} 
  J. W.\ York, in {\it Sources of Gravitational Radiation},
  edited by L.\ Smarr (Cambridge University Press, 1979).

\bibitem{abpst01} 
  M. Alcubierre, B. Br\"ugmann, D. Pollney, E. Seidel, and
  R. Takahashi, Phys.~Rev.~D {\bf 64}, 061501(R) (2001).

\bibitem{HJYo}
  M. D. Duez, S. L. Shapiro, and H.-J. Yo, Phys.~Rev.~D {\bf 69}
  104016 (2004).

\bibitem{cartoon} 
  M.\ Alcubierre, S.\ Brandt, B.\ Br\"ugmann,
  D.\ Holz, E.\ Seidel, R.\ Takahashi, and J.\ Thornburg,
  Int.~J.~Mod.~Phys. {\bf D10}, 273 (2001).

\bibitem{PPM} 
  P.\ Colella and P.\ R.\ Woodward, J.~Comput.~Phys., 
  {\bf 54}, 174 (1984).

\bibitem{HLL} 
  A.\ Harten, P.\ D.\ Lax, and B.\ van Leer {\it On Upstream
  Differencing and Godunov-Type Schemes for Hyperbolic Conservation
  Laws}, SIAM Rev., 25, 35-61, (1983). 

\bibitem{CookPaper}
  G. B. Cook, S. L. Shapiro, and S. A. Teukolsky, Astrophys.~J. {\bf
  398}, 203 (1992).

\bibitem{hgb95} 
  J. F. Hawley, C. F. Gammie, and S. A. Balbus, Astrophys.~J. {\bf
  440}, 742 (1995); J. F. Hawley, Astrophys.~J. {\bf 528}, 462
  (2000). 

\bibitem{moffatt78} 
  H. K. Moffatt, {\it Magnetic Field Generation in Electrically
  Conducting Fluids} (Cambridge Univ.\ Press, 1978).

\bibitem{circular} 
  B.\ Carter, in {\it Black Holes -- Les Houches 1972},
  edited by C.\ DeWitt and B.\ S.\ DeWitt (Gordon and Breach, New
  York, 1973); A.\ Papapetrou, Ann.~Inst.~H.~Poincar\'e {\bf A 4},
  83 (1966); B.\ Carter, J.~Math.~Phys. {\bf 10}, 70 (1969); E.\
  Gourgoulhon, and S.\ Bonazzola, Phys.~Rev.~D {\bf 48}, 2635
  (1993).

\end{thebibliography}
\end{document}